\documentclass{aa} 
\usepackage[utf8]{inputenc} 
\usepackage{threeparttable}
\usepackage[normalem]{ulem}
\usepackage{txfonts}
\usepackage{hyperref}
\usepackage{amsmath}

\newcommand{\Gyr}{\mathrm{Gyr}}
\newcommand{\Myr}{\mathrm{Myr}}
\newcommand{\yr}{\mathrm{yr}}
\newcommand{\dex}{\mathrm{dex}}
\newcommand{\Msun}{\mathrm{M}_\odot}
\newcommand{\Zsun}{\mathrm{Z}_\odot}

\newcommand{\A}{\mathring{\mathrm{A}}}
\newcommand{\m}{\mathrm{m}}
\newcommand{\kms}{\mathrm{km}/\mathrm{s}}
\newcommand{\kmsMpc}{\mathrm{km}\,\mathrm{s}^{-1}\,\mathrm{Mpc}^{-1}}
\newcommand{\vs}{~~\mathrm{vs}~~}

\newcommand{\dn}{$\rm D4000_n$}
\newcommand{\hb}{H$\beta$}

\newcommand{\hdg}{H$\delta_{\rm A}$+H$\gamma_{\rm A}$}
\newcommand{\ha}{H$\alpha$}
\newcommand{\mgfep}{$\rm [MgFe]^\prime$}
\newcommand{\mgtwofe}{$\rm [Mg_2Fe]$}
\newcommand{\mli}[1]{\mathit{#1}}

\defcitealias{Gallazzi2005}{G05}
\defcitealias{Gallazzi2021}{G21}

\begin{document} 
\title{Re-assessing the stellar population scaling relations of the galaxies in the Local Universe}
    \subtitle{}

   \author{D. Mattolini,
          \inst{1,2}
          S. Zibetti,
          \inst{2,3}
          A. R. Gallazzi,
          \inst{2}
          L. Scholz-Díaz
          \inst{2}
          \and
          J. Pratesi
          \inst{2}
          }
\authorrunning{Mattolini et al.}
   \institute{Dipartimento di Fisica ,
            Universit\`a di Trento, Via Sommarive 14, I-38123 Povo (TN), Italy\\
            \email{daniele.mattolini@unitn.it}\\
            \email{daniele.mattolini@inaf.it}
        \and
            INAF-Osservatorio Astrofisico di Arcetri, Largo Enrico Fermi 5, I-50125 Firenze, Italy
        \and
             Dipartimento di Fisica e Astronomia, Universit\`a degli Studi di Firenze, Via G. Sansone 1, I-50019 Sesto Fiorentino, Italy
             }

\date{Received 02 April 2025. Accepted 13 August 2025}

\abstract
{Local galaxies follow scaling relations between mass and stellar population properties, such as age and metallicity. These relations encode fundamental information about the past evolutionary history of galaxies.}
{We want to revise stellar population scaling relations of local galaxies leveraging the largest spectroscopic dataset provided by the SDSS DR7 ($0.005<z<0.22$) and using improved Stellar Population Synthesis (SPS) methods and novel SDSS aperture bias corrections.}
{We apply statistical weights to account for selection biases and implement corrections for SDSS fibre aperture limitations. In a Bayesian framework, we estimate stellar masses, mean stellar ages, and mean stellar metallicities by comparing spectral indices and photometry with composite stellar population models, adopting state-of-the-art ingredients and updated prescriptions to better capture the complexity of galaxies star formation and chemical enrichment histories. We also test different models and priors.}
{We estimate light-weighted mean stellar ages for 354,977 galaxies ($\mathit{SNR}\ge10$) and metallicities for 89,852 galaxies ($\mathit{SNR}\ge20$) and study their dependence on stellar mass. Key findings include:
i) A revised bimodal distribution in the mass-age plane, with a young sequence (dominant at low masses) and an old sequence (dominant at high masses), partly overlapping in mass and transitioning at $M_\mathrm{tr}=10^{10.80}\,\mathrm{M_\odot}$.
ii) A Mass-Metallicity Relation (MZR) shifted $\sim0.2\,\dex$ higher than previous studies. Our aperture corrections produce mass-dependent reductions in masses, ages, and metallicities, enhancing the young sequence and steepening the MZR at low masses.
iii) Using H$\alpha$-based SFR classification, we found that while star-forming/young and quiescent/old correspondences generally hold, exceptions exist for many galaxies. Quiescent galaxies show flatter, less scattered MZR than star-forming ones, converging at high masses.
iv) Different SPS modelling assumptions significantly impact results, with star formation and chemical enrichment histories having the strongest effects.}
{These revised relations provide new benchmarks for galaxy evolution studies and simulations. Systematic effects of $0.1-0.2\,\dex$ can arise from uncorrected aperture biases and different SPS modelling choices. Consistent assumptions should be used when comparing observations and models.}    

\maketitle
\section{Introduction}\label{sec:Introduction}
The observed properties of galaxies in the Local Universe are connected through well-known scaling relations, which are fundamental tools to investigate how galaxies form and evolve.
The scaling relations that connect the stellar mass of the galaxies with the ages and metallicities of the stellar population that compose them are of particular relevance, as stellar populations retain a wealth of ``archaeological'' information about the past star-formation and chemical-enrichments histories.
Traditionally the study of these relations was accessible only for quiescent elliptical galaxies \citep[e.g.,][]{Tinsley1976a, Tinsley1980a, BruzualA.1983a, Gonzalez1993, Kauffmann1996}.
With the advent of more refined models for Stellar Population Synthesis (SPS) \citep[e.g.,][]{Bruzual2003} these relations have been characterised for galaxies with any star formation activity.
The advent of higher resolution spectroscopic survey such as SDSS \citep{York2000, Stoughton2002} allowed the estimation of stellar ages and metallicities, and their relations with galaxy stellar mass, for a large number of galaxies in the Local Universe \citep[e.g.,][]{Kauffmann2003, Kauffmann2003a, Panter2003, Panter2008, Tremonti2004, Gallazzi2005, Gallazzi2008, Gallazzi2021, Cid-Fernandes2007, Asari2009, Mannucci2010, Pasquali2010, Peng2015, Trussler2020, Sanchez2018, Sanchez2020}.
These relations showed the existence of a bimodal mass-age relation composed by two distinct sequences of young low-mass and old high-mass galaxies \citep[see also][]{Mateus2006}.
A smooth transition between the two sequences occurs at a stellar mass of $M_\star\sim10^{10.5}\,\Msun\sim3\times10^{10}\,\Msun$ \citep{Kauffmann2003, Gallazzi2005, Moustakas2013, Haines2017}.
In addition to the mass-age relation, galaxies also exhibit a strong connection between stellar mass and stellar metallicity, known as the mass-(stellar) metallicity relation (MZR).
The stellar MZR of galaxies in the Local Universe has a unimodal shape with increasing stellar metallicities for increasing masses, flattening in the high mass end.
These relations have been used in the local Universe to characterize galaxy evolution within a downsizing scenario \citep{Gallazzi2005, Panter2008, Fontanot2009, Thomas2010}.
In this context higher-mass galaxies -- on average older and more metal-rich -- efficiently formed stars at earlier cosmic epochs, resulting in shorter formation timescales and higher enrichment efficiencies.
On the contrary lower-mass galaxies reached the peak of their star formation activity more recently, resulting in longer formation timescales and lower chemical enrichment efficiencies.

Over the last decades, models of galaxy formation and evolution in a cosmological context have become capable of realistically reproducing several observational properties of Local Universe galaxies \citep[see the review by][and references therein]{Somerville2015}.
They not only allow for realistic representation of Local Universe galaxies, but also are able to predict the properties of galaxies at higher redshifts.
Local Universe scaling relations are fundamental observational constraints to the treatment of the different astrophysical mechanisms governing galaxy growth in theoretical models of galaxy formation.

Modern models of galaxy formation and evolution are built upon a number of scaling relations, which have been characterised in the Local Universe mainly thanks to extensive spectroscopic surveys. 
The SDSS is the largest statistical database to-date used to study galaxy scaling relations.
However, we know that SDSS spectroscopic measurements suffer from an observational bias due to the limited aperture of the optical fibres used to extract galaxies spectra, which is generally referred to as ``aperture effects'' \citep[e.g.,][]{Kauffmann2003, Gallazzi2005, Gallazzi2008, Peng2015, Trussler2020, Gallazzi2021}.
In fact, galaxy spectra in the SDSS are extracted using optical fibres centred on the centre of the target galaxy and collecting light from a circular region with a diameter of $3"$.
Since a galaxy's apparent size depends on both its physical dimensions and distance, the fibre coverage varies across different galaxies.
This leads to a non-uniform loss of light from the outer regions of observed galaxies.
In addition to this bias, models and inference methods may introduce systematic biases in the estimates of stellar population parameters.
The combination of observational and systematic biases can result in uncertainties in the scaling relations which, in turn, may give rise to systematics when comparing with model predictions, or with different datasets.

In recent years the advent of IFU surveys such as MaNGA \citep{Bundy2014a} and CALIFA \citep{Sanchez2012} has enabled the study of galaxies in the Local Universe on spatially resolved scales \citep[see ][for a review]{Sanchez2020}, quantifying spatial variations and gradients of stellar population properties \citep[e.g.][]{Perez2013, Sanchez-Blazquez2014, GonzalezDelgado2014, GonzalezDelgado2015, Goddard2017, Zibetti2017, Zibetti2020, Zibetti2022, Martin-Navarro2018, Li2018, Johnston2022}.
Recently \cite{Zibetti2025} have leveraged the spatial information and extensive coverage of the CALIFA dataset to statistically characterize and correct the aperture biases in the SDSS spectroscopy.

Taking advantage of these corrections, as well as of the advancements in population synthesis models, we build upon the work of \cite{Gallazzi2005} to revise the mass-age and mass-stellar metallicity scaling relations in the Local Universe.
Specifically, we improve on previous work in the following ways: on the observation side we apply correction factors accounting for the aperture effects in the absorption indices estimates; on the modelling and stellar population parameter side, we use updated Simple Stellar Populations (SSP) obtained with state-of-the-art modelling \citep[2019 version of][SSPs]{Bruzual2003}, and introduce more refined prescriptions for Star Formation Histories (SFH) \citep{Sandage1986, Asari2007, Pacifici2016}, Chemical Enrichment Histories (CEH) \citep{Erb2008, Zibetti2017, Camps-Farina2021, Camps-Farina2022}, and dust absorption \citep{Charlot2000}.
Furthermore we also implement statistical weights accounting for Malmquist \citep{Schmidt1968} and sample selection biases.
As a result, we present volume-weighted aperture-corrected scaling relations obtained with state-of-the-art stellar population synthesis models, exploiting the large statistics provided by the SDSS DR7 sample.

Our revised scaling relations not only provide a new reference which can be used by theoretical models to constrain the processes shaping galaxy evolution \citep[such as galaxy quenching, e.g.,][]{Peng2015, Casado2015, Spitoni2017, Trussler2020, Bluck2020a, Bluck2020b, Corcho-Caballero2023a, Corcho-Caballero2023b, Looser2024} and fine-tune the corresponding parameters, but will also represent the local benchmark for higher redshift scaling relations \citep[e.g.,][]{Gallazzi2014, Cappellari2023, Nersesian2024, Gallazzi2025}.
This is particularly meaningful given the advent of the next generation surveys such as WEAVE-StePS \citep{Iovino2023a} and 4MOST-StePS \citep{Iovino2023b} which will provide deep spectroscopy of large near-mass selected samples, bridging the redshift gap between SDSS at $z\sim0$ and LEGA-C \citep{vanderWel2016} at $0.6<z<1$, enabling a uniform characterization of stellar populations.

The paper is organized as follows. 
In section \ref{sec:Data_analysis} we present the dataset used throughout this work and describe the methodology used to estimate the statistical weights accounting for observational and selection biases.
In section \ref{sec:Methodology&Inference} we describe the procedure to retrieve the galaxy properties from the observed spectra, focusing also on the analysis of the individual ingredients of the Stellar Population Synthesis modelling.
In section \ref{sec:Results} we present our revised mass-age and mass-stellar metallicity relations for galaxies in the Local Universe, both for the general population and considering passive and star-forming galaxies separately.
In section \ref{sec:Models_comparison} we analyse the systematics induced in the scaling relations by different choices for the ingredients of the SPS models, with a detailed analysis of the differences with respect to previous works from our group, and specifically to \cite{Gallazzi2005} and \cite{Gallazzi2021}.
We discuss the importance of these revised and bias-corrected relations in the context of galaxy evolution in section \ref{sec:Discussion}.
We summarise our results and conclusions in section \ref{sec:Summary&Conclusions}.
Throughout the paper, we assume a $\Lambda\text{CDM}$ cosmology with $H_0=70\,\kmsMpc$, $\Omega_\mathrm{m}=0.3$, and $\Omega_\Lambda=0.7$.

\section{Data analysis}\label{sec:Data_analysis}
In this section we describe the sample and dataset used in this work. We also describe the process followed to obtain statistical corrections for biases due to sample selection and aperture effects.
\subsection{Observational dataset\label{subsec:observational_dataset}}
The dataset used in this work is composed of both photometric and spectroscopic observations drawn from the Sloan Digital Sky Survey Data Release 7 \citep[][SDSS DR7 hereafter]{Abazajian2009}. 
The SDSS \citep{York2000, Stoughton2002} is an optical photometric and spectroscopic survey conducted using a $2.5~\m$ telescope at Apache Point Observatory (APO), in Southern New Mexico. 
The photometry is obtained in the five bands $ugriz$ \citep{Fukugita1996}, using a mosaic CCD camera \citep{Gunn1998}.  
Spectroscopic observations are carried out using 640 optical fibres with $3"$ diameter apertures. 
The spectra cover the wavelength range $3800-9200~\A$, with an average spectral resolution $R=\lambda/\Delta\lambda\sim2000$.
The DR7 delivers photometry of 375 million objects and spectroscopy of $\sim1.6$ million objects \citep{Abazajian2009}, including $\sim930,000$ galaxies.

The observational features we used to estimate galaxy properties are a set of absorption indices, which can be extracted from the spectra of the galaxies \citep{Worthey1994, Worthey1997}, and the SDSS $ugriz$ petrosian photometry.
Given the studies that have characterised the dependence of the absorption indices on the ages and metallicities of galaxies stellar populations \citep{Worthey1994, Worthey1997, Thomas2003, Bruzual2003}, we followed \cite{Gallazzi2005} and used the break at $4000~\A$ (\dn), as defined in \cite{Balogh1999}, a set of three Balmer indices \hb, and \hdg, and two composite magnesium and iron indices \mgfep, and \mgtwofe:
\begin{equation}\label{eq:Metal_indices}
        \left\{
    \begin{array}{lr}
        \rm[MgFe]^\prime=\sqrt{\rm Mgb\,(0.72\,Fe5270+0.28\,Fe5335)}\\
        \\
        \rm[Mg_2Fe]=0.6\,Mg_2+0.4\,\log(Fe4531+Fe5015)\
    \end{array}
    \right.~~~,
\end{equation}
as defined by \cite{Thomas2003} and \cite{Bruzual2003}.
Such a choice for the absorption indices is driven by the need to reduce the age-metallicity degeneracy.
This is achieved by using mainly age-sensitive indices such as high-order Balmer lines, and the \dn~break, in synergy with indices that are mainly metallicity-sensitive, such as \mgfep~and \mgtwofe.
These composite magnesium and iron indices are designed to be good tracers of the overall metallicity and, at the same time, have only a little dependence on $\rm[\alpha/Fe]$ \citep{Thomas2003, Bruzual2003}.
The index measures used in this work have been obtained from the emission-line-subtracted spectra by \cite{Brinchmann2004} and are available from the ``MPA-JHU catalogues''
\footnote{
    The catalogues are available at the following link:
    \url{https://wwwmpa.mpa-garching.mpg.de/SDSS/DR7/raw_data.html}}.

The MPA-JHU catalogues contain photometric and spectroscopic properties of $927~552$ galaxies, some of which are associated to repeated observations of the same galaxy.
From this sample we rejected $48~599$ observations due to low quality redshift and velocity dispersion measurement\footnote{We rejected galaxies with invalid (negative) estimates for redshift and velocity dispersion uncertainties. We also rejected galaxies with the \texttt{z\_warning} flag greater than zero.}.
We combined the repeated observations and created a catalog of unique observations for each galaxy identified as unique by the photometric pipeline \citep{Lupton1999a}.
The repeated spectroscopic observations were combined by calculating the error-weighted averages of the galaxy properties that we used in our analysis, namely: redshift, velocity dispersion $\sigma_\star$ (used to broaden the spectral features of the models to match the observed ones), Star Formation Rate (SFR, used to separate passive and star-forming galaxies), and index measures (for \dn, \hb, H$\gamma_{\rm A}$, H$\delta_{\rm A}$, and the indices needed to compute  \mgfep, and \mgtwofe).
In the combination of the repeated absorption index measures, we used only those associated to valid (positive) uncertainties.
The photometric information are drawn from the NYU Value-Added Galaxy Catalog by \citet[NYU-VAGC hereafter\footnote{We used the kcorrect.none.petro.z0.00.fits and kcorrect.none.petro.z0.10.fits catalogues, which may be found at \url{https://sdss.physics.nyu.edu/vagc-dr7/vagc2/kcorrect/}}]{Blanton2005}, which we matched with the MPA-JHU after the combination of the duplicated observations, using the nearest neighbour based on the sky coordinates.
We thus obtained a catalog of science-grade and unique observations for the full SDSS DR7 galaxy sample, containing $825~263$ galaxies.

We then selected those galaxies that constitute a suitable sample for the stellar population analysis.
We selected only galaxies that are part of the Main Galaxy Sample (MGS), hence with $14.5\leq r_\mathrm{petro,~dered}\leq17.77$~mag \citep{Strauss2002}\footnote{We relaxed the selection on Petrosian half-light surface brightness $\mu_{50}\leq24.5~\mathrm{mag~arcsec^{-2}}$, as this requirement is automatically matched for the SNR requested by stellar population analysis.}.
We also restricted the sample to galaxies in the redshift range $0.005<z\leq0.22$.
Galaxies with $z\leq0.005$ are excluded in order to avoid bad distance determinations due to deviations from the Hubble flow; galaxies with $z>0.22$ are excluded because of severe mass incompleteness given the apparent magnitude limits of the SDSS.
We also rejected galaxies with velocity dispersion exceeding $375~\kms$ to exclude possible non-physical measurements (possibly large errors). Rare extreme or peculiar objects may be rejected as well, yet this may be considered acceptable given the statistical nature of this study.
We also used the SDSS photometric flags to reject $253~037$ galaxies with bad quality photometric measurements\footnote{
This was done following the recommendations in \url{https://www.sdss3.org/dr9/tutorials/flags.php\#clean}}.
As at faint absolute magnitudes the representativeness of our sample drops dramatically (see next Sec. \ref{subsec:Correction_statistical_biases}), we rejected galaxies with absolute r-band magnitude $M_r>-17.5$\footnote{With this selection we rejected $822$ galaxies with $\mathrm{SNR\geq10}$, and $121$ galaxies with $\mathrm{SNR\geq20}$}.
\cite{Gallazzi2005} showed that a signal-to-noise ratio (SNR) per pixel of at least $10\text{ and }20$ are needed to get reliable estimates of ages and metallicities, respectively \citep[see also ][]{Rossi2025}.
For this reason we also imposed SNR cuts to select suitable galaxies to perform the analysis on mass-age and mass-metallicity scaling relations separately.
After the combination of the duplicated observations and the sample selections we are left with $354~977$ galaxies with $\mathrm{SNR}\geq10$, and $89~852$ galaxies with $\mathrm{SNR}\geq20$.

\subsection{Corrections for statistical biases}\label{subsec:Correction_statistical_biases}
The apparent-magnitude cut that defines our sample implies that our sample is not
volume-complete. 
We corrected for such incompleteness by calculating statistical volume weights following \cite{Schmidt1968}.
Based on the Petrosian magnitudes in the NYU-VAGC catalog, we computed the minimum and maximum redshifts, $z_\mathrm{min}$ and $z_\mathrm{max}$, at which each galaxy enters the selection cuts, taking into account the K-corrections computed with the Kcorrect code by \cite{Blanton2007}. The ratio between the comoving volume $\Delta V_\mathrm{surv}=V(z=0.22)-V(z=0.005)$ within the redshift boundaries of our sample selection and the comoving volume $\Delta V_\mathrm{max}=V(z_{max})-V(z_{min})$ within $z_\mathrm{min}$ and $z_\mathrm{max}$ gives the statistical weight of each galaxy, in order to account for the Malmquist bias.

Incompleteness due to fibre allocation, data quality rejection, and SNR selection can also induce biases in how the selected sample is representative of the underlying parent population.
To correct for such effects, we calculated statistical weights based on the unbiased photometric NYU-VAGC catalogue.
We calculated completeness corrections for each SNR selection by comparing the 2D distribution of the complete NYU-VAGC sample and of our samples in the $\left(^{0.1}M_\mathrm g-\,^{0.1}M_\mathrm r\right)\vs ^{0.1}M_\mathrm r$ plane, obtained using the absolute magnitudes K-corrected at redshift $z=0.1$.
We estimated the completeness fractions $\kappa$ calculating the ratio between the NYU-VAGC number-density distribution and the one obtained with our sample.
Adopting for each galaxy the statistical weight $W\equiv \frac{\Delta V_\mathrm{surv}}{\Delta V_\mathrm{max}}\kappa^{-1}$ we can thus obtain a volume-complete statistical representation of the galaxy population. 

\begin{figure}
\centering
\includegraphics[width=\hsize]{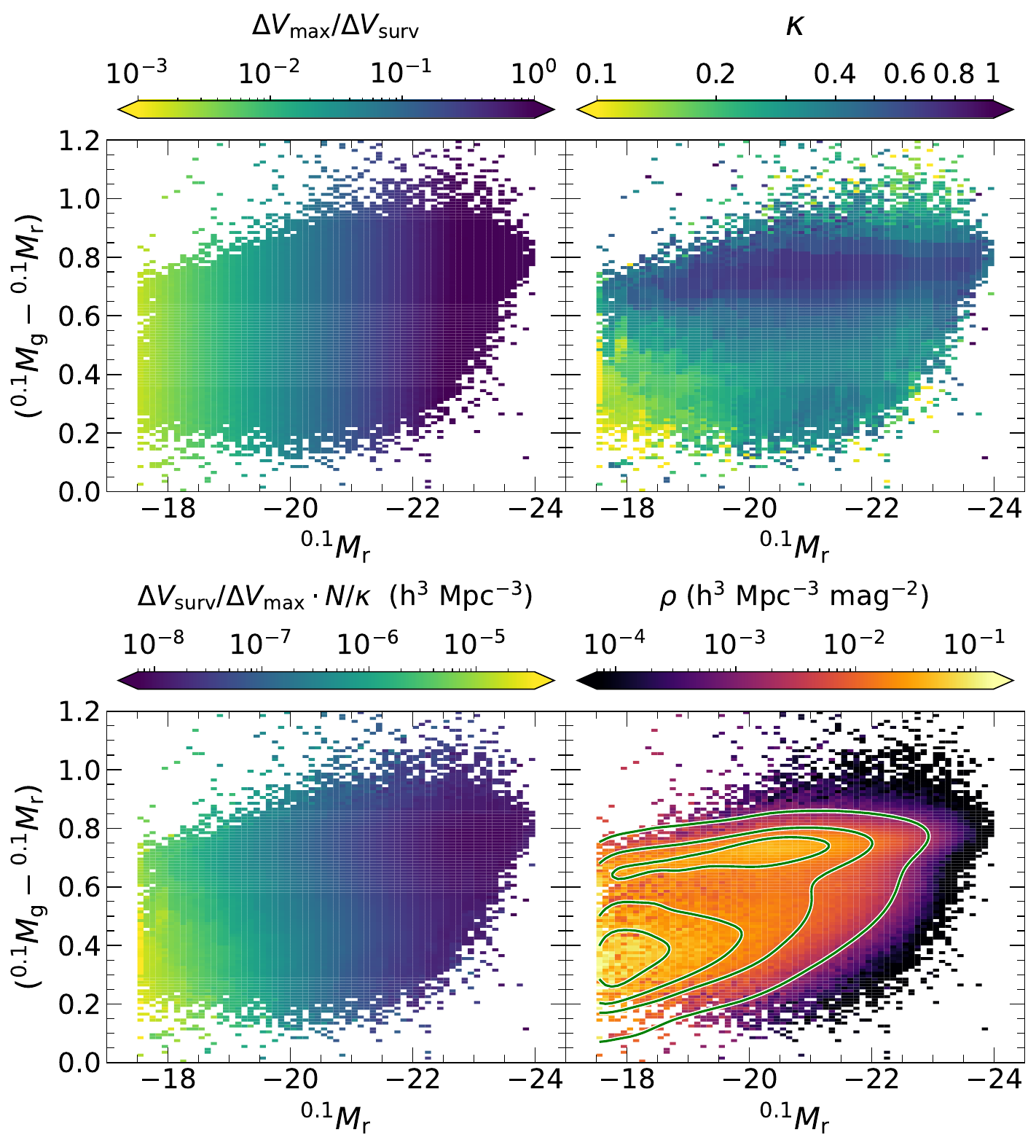}
  \caption{Colour Magnitude Diagram (CMD) for the galaxy sample with $\mathrm{SNR}\geq10$. The \emph{top left panel} displays the volume completeness (i.e. the inverse of the volume-weights). 
  The \emph{top right panel} displays the completeness fractions (i.e. the inverse of the completeness-weights). 
  The \emph{bottom left panel} displays the statistical weights obtained combining the volume and completeness statistical weights, multiplied by the factor $N=4.89\times10^{-9}~h^3\,\mathrm{Mpc}^{-3}$, to convert number counts to the number volume-density of galaxies. The \emph{bottom right panel} displays the number density of galaxies in the CMD applying the completeness corrections. The green contours enclose $16\%$, $50\%$, $84\%$, and $97.5\%$ of the galaxy sample, and have been obtained on the distribution smoothed with a Gaussian KDE. Similar plots are applicable to the $\mathrm{SNR}\geq20$ sample.}
    \label{fig:Corrections}
\end{figure}

In order to reproduce the volume density of galaxies 
we multiply the statistical weights by the factor $N=4.89\times10^{-9}~h^3\mathrm{Mpc}^{-3}$ by which we can exactly reproduce the \cite{Blanton2003} luminosity function in the magnitude range $-22<~^{0.1}M_\mathrm{r}<-21$.
Figure \ref{fig:Corrections} illustrates the statistical corrections both for the volume and completeness effects, as well as the resulting distribution in the colour-magnitude diagram
for the galaxies with $\mathrm{SNR}\geq10$ (similar plots are obtained for $\rm SNR\geq20$). The top panels show the 2D maps of $\frac{\Delta V_\mathrm{max}}{\Delta V_\mathrm{surv}}$ and $\kappa$ (left and right panels, respectively).
Volume corrections depend essentially on the absolute magnitude alone, resulting in a gradient along the x-axis, with fainter galaxies having the strongest corrections.
As we can see from Fig. \ref{fig:Corrections}, moving towards faint absolute magnitudes (and blue colours), the overall representativeness drops below $10^{-3}$. In order to avoid our analysis to be plagued by galaxies with very high statistical noise, we limit the study of scaling relations to galaxies with $M_r<-17.5$, as already mentioned.

On the other hand, the completeness fractions depend both on the magnitude and the colour of the galaxies.
The highest completeness fraction $\kappa\sim0.65$ is reached by galaxies with $\left(^{0.1}M_g-\,^{0.1}M_r\right)\sim0.75$.
Bluer galaxies have progressively lower completeness fractions reaching $\kappa\sim0.1$ at the faint end and bluest colour.
The bottom left panel shows the 2D map of the overall statistical corrections (in units of $\rm h^3\mathrm{Mpc}^{-3}$) obtained by multiplying the completeness correction by the volume correction times the conversion factor $N$. Note the dependence on both colour and magnitude, inherited from the completeness corrections.
The bottom right panel shows the number density of galaxies when implementing the overall statistical weights.
The green contours identify the density levels enclosing $16\%$, $50\%$, $84\%$, and $97.5\%$ of the complete galaxy sample.
Such contours were derived by smoothing the distribution using a gaussian Kernel Density Estimator (KDE).

As a sanity check on the robustness of our statistical weights, throughout the work we have verified that the weighted distributions obtained from the $\mathrm{SNR}\geq10$ sample are consistent with the ones obtained from the more restricted sample at $\mathrm{SNR}\geq20$, except for the noise due to the different number statistics. The mass functions shown in Fig. \ref{fig:Mass_function} illustrate this.
A posteriori we can claim that our galaxy sample is representative for galaxy masses as low as $\sim10^9~\Msun$ overall, and down to $\sim10^{9.5}~\Msun$ for passive/old galaxies (which are a minority at the low-mass end). These limits can be graphically identified in Fig. \ref{fig:Mass_function} as the masses at which the mass functions display a sharp down-bending, also indicated by the shaded areas.

\subsection{Corrections to indices for fibre aperture bias}\label{sec:Aperture_effects}
\begin{figure}
\centering
\includegraphics[width=\hsize]{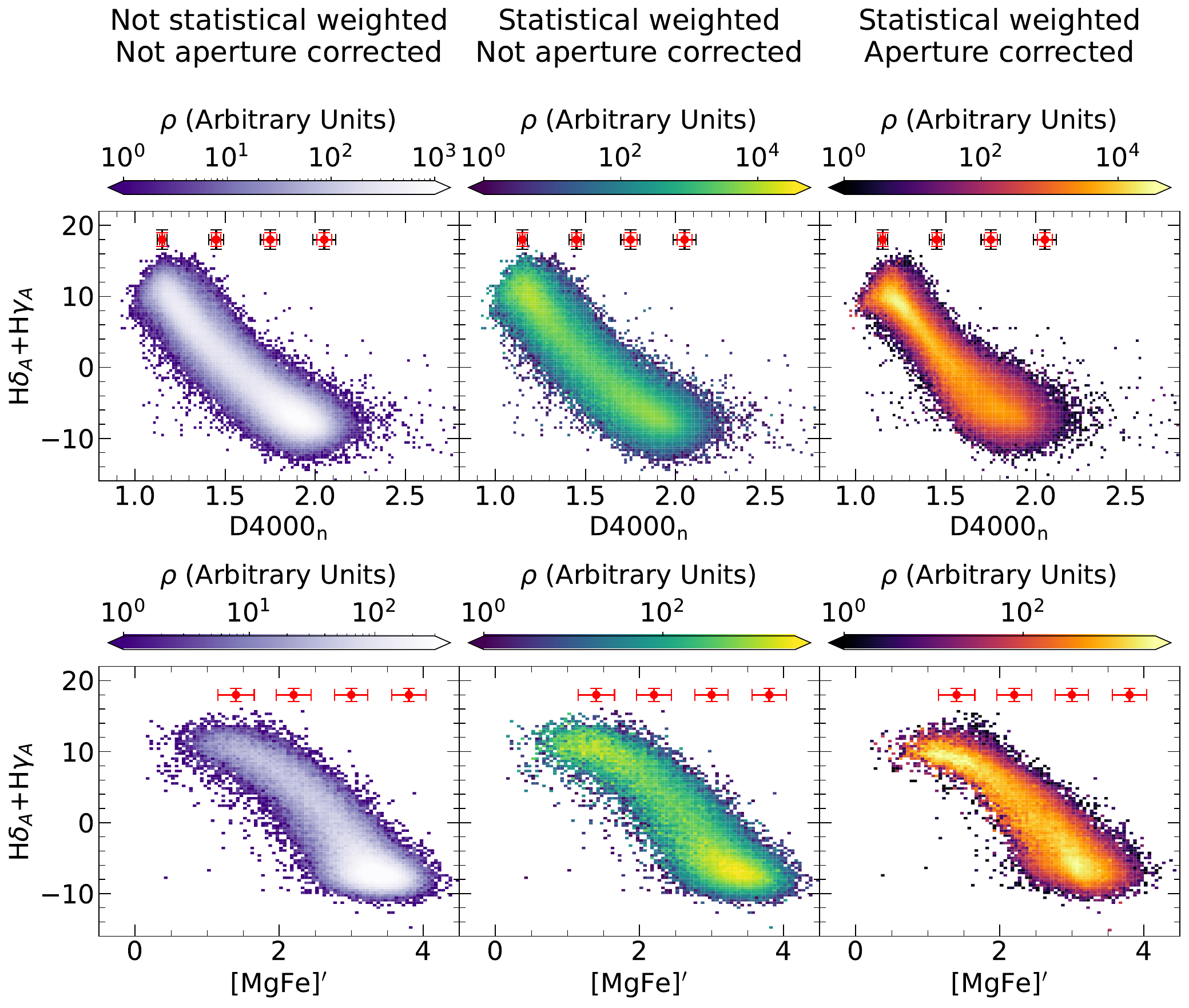}
  \caption{Distribution of the galaxies in the  H$\delta_{\rm A}$+H$\gamma_{\rm A}$ vs \dn~ and H$\delta_{\rm A}$+H$\gamma_{\rm A}$ vs \mgfep~ planes before and after the implementation of the statistical weights and the aperture corrections.
  The \emph{left panels} display the distribution of the original data, i.e. simple number counts for uncorrected index values.
  The \emph{central panels} display the distributions of galaxies weighted by the volume and completeness statistical corrections, for uncorrected index values.
  The \emph{right panels} display the distributions obtained implementing both the statistical weights and the corrections for aperture effects.
  The error bars correspond to the median uncertainties calculated in 
  four bins of \dn~and \mgfep, for samples with $\mathrm{SNR}\geq20$ (red) and $\mathrm{SNR}\geq10$ (black, only \emph{top panels}).}
    \label{fig:Aperture_correction_comparison}
\end{figure}
The SDSS optical fibres were positioned on the centre of the target galaxies and collected light from a circular region with a diameter of $3''$, hence only from a central portion of a galaxy's extent.
Since a galaxy's apparent size depends both on its physical dimensions and on distance, the fibre coverage varies across different galaxies.
This leads to a non-uniform loss of light from their outer regions.
Several studies showed that galaxies may present significant radial gradients in their stellar populations properties \citep[e.g., based on IFU observations,][]{GonzalezDelgado2014, GonzalezDelgado2015, Goddard2017, Li2018, Zibetti2017, Zibetti2020, Zibetti2022}.
Also, galaxies physical properties may change in different galaxy components.
For instance it is known that bulges are typically composed mainly of old and metal-rich stellar populations, with low-level or absent star formation.
On the contrary, discs display a variety of populations both in age and metallicity.
The strength of stellar population gradients in galaxies thus depends on the properties and dominance of each component, hence on morphology and, in turn, on mass.
The combination of the non-uniform galaxy coverage from SDSS fibres, together with the presence of radial gradients in stellar population properties, causes a systematic observational bias, which we will refer to as ``aperture effects'' in the following.

\cite{Zibetti2025} calculated the corrections to the absorption indices measured on SDSS spectra needed to obtain the indices relative to the full integrated light of the galaxy.
The authors used integral field observations from CALIFA to simulate the effect of a fixed $3''$ aperture on the measurement of absorption indices from galaxies with different masses and morphologies.
The aperture-corrections are divided into first and second order offsets, which are assigned to each galaxy based on its position in the index-strength $\mathrm{vs}~~(g-r)$ and $r_{50}\vs M_\mathrm{r,petro}$ planes.
We used the indices measures and the half-light radii (with the respective errors) obtained from the MPA-JHU catalog (converting the $r_{50}$ from arcsecond to kpc using the angular diameter distance estimated with the MPA-JHU redshift).
The absolute magnitudes $M_{\rm petro}$ and $(g-r)$ colours are obtained from the NYU-VAGC, using the petrosian values.

The effects of the aperture corrections and of the statistical weights are illustrated in the distributions of the observed galaxies in the H$\delta_{\rm A}$+H$\gamma_{\rm A}$ vs \dn~ and H$\delta_{\rm A}$+H$\gamma_{\rm A}$ vs \mgfep~ planes, in Fig.  \ref{fig:Aperture_correction_comparison} (top and bottom row, respectively). These index-index planes  are diagnostic for the stellar population age and metallicity, respectively \citep{Gallazzi2005, Gallazzi2014}. 
The left panels display the distribution prior to applying either statistical weights or aperture corrections.
The central panels display the galaxy density distribution obtained applying the statistical corrections for volume and completeness.
The right panels present the fully corrected distributions, after applying the aperture corrections to the index measurements and the statistical weights.
The aperture corrections substantially enhance the galaxies populating the parts of the planes associated to high values of \hdg~and low values of both \dn~and \mgfep.
Since \dn~and \mgfep~are proxies for ages and metallicities respectively, this results in an enhancement of the galaxies with young and metal poor stellar populations.
Furthermore, the aperture corrections reduce the scatter of the 2D distributions in both H$\delta_{\rm A}$+H$\gamma_{\rm A}$ vs \dn~ and H$\delta_{\rm A}$+H$\gamma_{\rm A}$ vs \mgfep~ planes, as analysed in more detail in \cite{Zibetti2025}.

\section{Methodology }\label{sec:Methodology&Inference}
In this section we describe the methodology used to perform the inference of the stellar population parameters, detailing the ingredients used in the Stellar Population Synthesis.

\subsection{Bayesian inference of stellar population properties}\label{subsec:Inference_method}
We estimated the galaxy stellar population properties by means of the ``Bayesian Stellar population Analysis'' (BaStA)  tool, developed by A. Gallazzi and S. Zibetti and already presented in several works by our group \citep[e.g.,][]{Gallazzi2005,Zibetti2017}. 
Here we provide a brief overview of its operating principles.
The key idea is to derive the posterior Probability Distribution Function (PDF) of a given stellar population parameter using a Bayesian approach in which the prior PDF is defined by a pre-computed library of SPS models  \citep[following the approach of][]{Kauffmann2003, Gallazzi2005, daCunha2008}.
In our case the prior distribution is composed by a library of $500~000$
Composite Stellar Populations (CSPs).
To each model spectrum, generated by prior assumptions in SFH, CEH and dust (see sec. \ref{subsec:Stellar_Population_Synthesis}), we can associate physical parameters (e.g. mean stellar age and metallicity) and observational parameters (e.g. photometry and absorption indices measurements).
Therefore, given an observed dataset for a galaxy, for each SPS model we define the likelihood given the data as:
\begin{equation}\label{Likelyhood}
	\mathcal{L}\propto{\exp}\left(-\frac{\chi^2}{2}\right)~~,
        ~~~~~~~~~~~~~~~~~~
        \chi^2=\left\|\frac{\rm O-M}{\sigma}\right\|^2~~,
\end{equation}
where O and $\sigma$ are the measurements and uncertainties of the observed quantities, and M are the corresponding quantities predicted by the given model.
For each physical parameter we then obtain its posterior PDF by marginalising over all other properties.
From this we estimate the fiducial value of the parameter as the median, and its uncertainty given by half the range between the $16^\mathrm{th}$ and $84^\mathrm{th}$ percentiles.

As observational constraints we use the five indices and the five photometric fluxes defined in Sec. \ref{subsec:observational_dataset}. The observed indices are confronted with the model indices computed for a velocity dispersion broadening consistent with the observed one. \footnote{Note that the velocity dispersion $\sigma_\star$ is not corrected for the aperture effects, hence the aperture-corrected indices are compared with model indices at the $\sigma_\star$ as measured within the fibre. This is not expected to cause significant biases, because the dependence of indices on $\sigma_\star$ is weak and the aperture effects on $\sigma_\star$ are typically small.}

We apply this Bayesian approach to infer the light-weighted and (present-day) mass-weighted ages and metallicities, as defined in equation \ref{eq:light_weighted} and \ref{eq:mass_weighted} respectively:

\begin{equation}\label{eq:light_weighted}
        \left\{
    \begin{array}{lr}
        \mathit{Age}_\mathrm{wr}=\frac{\int_{t=0}^{t=t_0}dt~(t_0-t)~\text{SFR}(t)~\mathcal{L}'(t)}{\int_{t=0}^{t=t_0}dt~\text{SFR}(t)~\mathcal{L}'(t)}\\
        \\
        Z_\mathrm{wr}=\frac{\int_{t=0}^{t=t_0}dt~Z_\star (t)~\text{SFR}(t)~\mathcal{L}'(t)}{\int_{t=0}^{t=t_0}dt~\text{SFR}(t)~\mathcal{L}'(t)}
    \end{array}
    \right.
\end{equation}
\begin{equation}\label{eq:mass_weighted}
        \left\{
    \begin{array}{lr}
        \mathit{Age}_\mathrm{wm}=\frac{\int_{t=0}^{t=t_0}dt~(t_0-t)~\text{SFR}(t)~(1-R(t))}{\int_{t=0}^{t=t_0}dt~\text{SFR}(t)~(1-R(t))}\\
        \\
        Z_\mathrm{mw}=\frac{\int_{t=0}^{t=t_0}dtZ_\star (t)~\text{SFR}(t)~(1-R(t))}{\int_{t=0}^{t=t_0}dt~\text{SFR}(t)~(1-R(t))}
    \end{array}
    \right.
\end{equation}
where $t$ measures the time along the SFH, $t_0$ is the time elapsed since the start of the SFH until the epoch of observation, $\text{SFR}(t)$ is the star formation rate as a function of time, $\mathcal{L}'(t)$ is the luminosity arising  per unit of formed stellar mass from the SSPs of age $t_0-t$, and $R(t)$ is the fraction of stellar mass returned to the ISM.

Since mass is an extensive property, it cannot be directly extracted from the models. For each model we compute the normalization factor that has to be applied to the model photometric fluxes in order to match the observed ones. This is done by minimizing the $\chi^2$ corresponding to the magnitudes in the five SDSS bands. Since by construction all models are normalized to $1~\Msun$ in present day stellar mass (including living stars and stellar remnants), this normalization factor directly gives the present day stellar mass in units of solar masses for a given model. This parameter is then treated as the other physical parameters in the marginalization of the posterior PDF.

\subsection{The stellar population model library}\label{subsec:Stellar_Population_Synthesis}
In this section we provide an essential description of how the library of CSPs that constitute the prior for our Bayesian inference was constructed. More technical details can be found in Appendix \ref{App:Model_library}.

In this work we adopted the grid of SSPs from the 2019 version of the \citet[CB19 hereafter]{Bruzual2003}\footnote{The CB19 SSPs are available at this link: \url{https://www.bruzual.org/CB19/}}.
Such SSPs assume a \cite{Chabrier2003} Initial Mass Function (IMF) with a maximum $M_\star=100~\Msun$, use the MILES spectral libraries \citep{Sanchez-Blazquez2006}, and model stellar evolution via the PARSEC \citep{Bressan2012, Marigo2013a, Chen2015} evolutionary tracks \citep[for a more detailed description of the CB19 ingredients see Appendix A of ][]{Sanchez2022a}.

We synthesised the CSPs combining SSPs based on parametric Star Formation Histories and Chemical Enrichment Histories. The emergent spectra are then attenuated according to a dust attenuation model.
We used SFHs composed by a continuum parametric component superimposed with random bursts of star formation, to incorporate stochasticity.
For the continuum component we implemented a delayed Gaussian SFH, following \cite{Sandage1986} and \cite{Gavazzi2002a}: 

\begin{equation}\label{eq:Sandage_profile}
    \mli{SFR}_\tau(t)\propto\frac{(t-t_{\mathrm{form}})}{\tau}\exp\left(-\frac{(t-t_{\mathrm{form}})^2}{2\tau^2}\right)
\end{equation}
with timescale $\tau$.
This formalism allows more flexibility in the description of a galaxy SFH, with respect to a decaying exponential law \citep[adopted in several studies, including, e.g.][]{Gallazzi2005}, by including an increasing phase, which may reveal more appropriate for the early phase of galaxy evolution or even the present-day evolution of low-mass galaxies \citep{Asari2007, Pacifici2016}. 
The look-back time of the beginning of the SFH $t_\mathrm{form,lb}\equiv t_0-t_\mathrm{form}$ ($t_0$ being cosmic time of the observation) is chosen randomly between $500~\Myr$ and  $20~\Gyr$
\footnote{Note that for the CB19 SSPs the maximum available age is $14~\Gyr$, therefore we render the oldest part of the SFH by modifying the weight for the $14$-$\Gyr$ old component (see Sec. \ref{subsubsec:Evolutionary_tracks}).}
, from a uniform logarithmic distribution.
The parameter $\tau$ (time of the maximum SFR since the beginning of the SFH and time-scale of the Gaussian decay) is randomly generated ranging from $1/20$ to $2$ times $t_\mathrm{form,lb}$. The smallest $\tau/t_\mathrm{form,lb}$ correspond very closely to SSPs of age equal to $t_\mathrm{form,lb}$ \citep[see also][]{Zibetti2024a}.

The burst-like component is composed by one or more bursts of star formation (up to six), and affects two thirds of the models. 
The probability to have $N_\mathrm{burst}$ is proportional to $\exp^{-N_\mathrm{burst}}$. Each burst can form between $10^{-3}$ and $2$ times the total mass formed in the continuos component. The age of each burst can vary from $10^5~\yr$ to $t_\mathrm{form,lb}$. Although not explicitly modelled, different combinations of a secular component with $\tau\ll t_\mathrm{form,lb}$ and recent burst(s) can reproduce post-starburst galaxies or truncations of the SFH.

Our CSPs were constructed using the parametric CEHs defined in \cite{Zibetti2017}:
\begin{equation}\label{eq:Chemical_Enrichment_History}
    Z_\star(t)=Z_\star(M(t))=Z_\mathrm{\star~final}-\left(Z_\mathrm{\star~final}-Z_{\star~0}\right)\left(1-\frac{M(t)}{M_\mathrm{final}}\right)^\alpha,
\end{equation}
where $M(t)$ and $M_\mathrm{final}$ are the stellar mass formed at time $t$ and the total stellar mass formed until the end of the SFH respectively, and $\alpha\geq0$ is a parameter that describes how fast the chemical enrichment proceeds relative to the formation of stars and can vary between $0.25$ and $19$.
In case of the presence of bursts in the SFH, at each burst is assigned a metallicity $Z_{\star~burst}$ equal to the metallicity of stars formed in the continuous mode at time of the burst, plus a random offset taken from a log-normal distribution with $\sigma=0.2~\dex$.

It should be noted that the leaky-box model CEH defined in eq. \eqref{eq:Chemical_Enrichment_History} \citep[see also][]{Erb2008} does not allow possible decreases in $Z_\star$ due to, e.g., the infall of metal poor gas from the IGM, and the consequent dilution of the gas reservoir of the galaxy. This kind of phenomena is rendered, at least in part, by the down-scatter in the $Z_{\star~burst}$ with respect to the secular component.
Note that the increasing behaviour of the stellar metallicity over time is empirically confirmed to be predominant on average among galaxies \cite[e.g.][]{Camps-Farina2021, Camps-Farina2022}.

We implemented dust attenuation via the \cite{Charlot2000} two component model.
This model considers different attenuation laws for the diffuse Inter Stellar Medium (ISM) and the dust in the Birth Clouds (BC), with attenuation proportional to $\lambda^{-0.7}$ and $\lambda^{-1.3}$ respectively.
Since young stars (i.e. with $Age\leq10^7~\yr$) reside in the BCs, they suffer attenuation by both components.
Young stars are affected by a total optical depth in the V-band of $\tau_V$, with a fraction $\mu$ attributed to the diffuse ISM, and a fraction $1-\mu$ attributed to the BC.
Older stars are effectively attenuated only by the diffuse ISM, hence with a V-band optical depth of $\mu\tau_V$. $\tau_V$ and $\mu$ are generated using the same prior PDF as in \citet{daCunha2008} \citep[see also][]{Zibetti2017}.

As detailed in Appendix \ref{App:Model_library}, the final model library is produced after applying an equalization in (H$\delta_{\rm A}$+H$\gamma_{\rm A}$, \dn), in order to make the model density constant in this plane of observable quantities. Moreover, in the actual procedure of posterior marginalization, we assign zero weight to models with $t_\mathrm{form,lb}<1.5~\Gyr$, which appear to be unrealistic for galaxies with $M_\star \gtrsim 10^9\,\Msun$ in the local Universe. This leaves us with $400~481$ effective models.

\section{Results}\label{sec:Results}
\begin{figure*}
\centering
\includegraphics[height=0.6\textwidth]{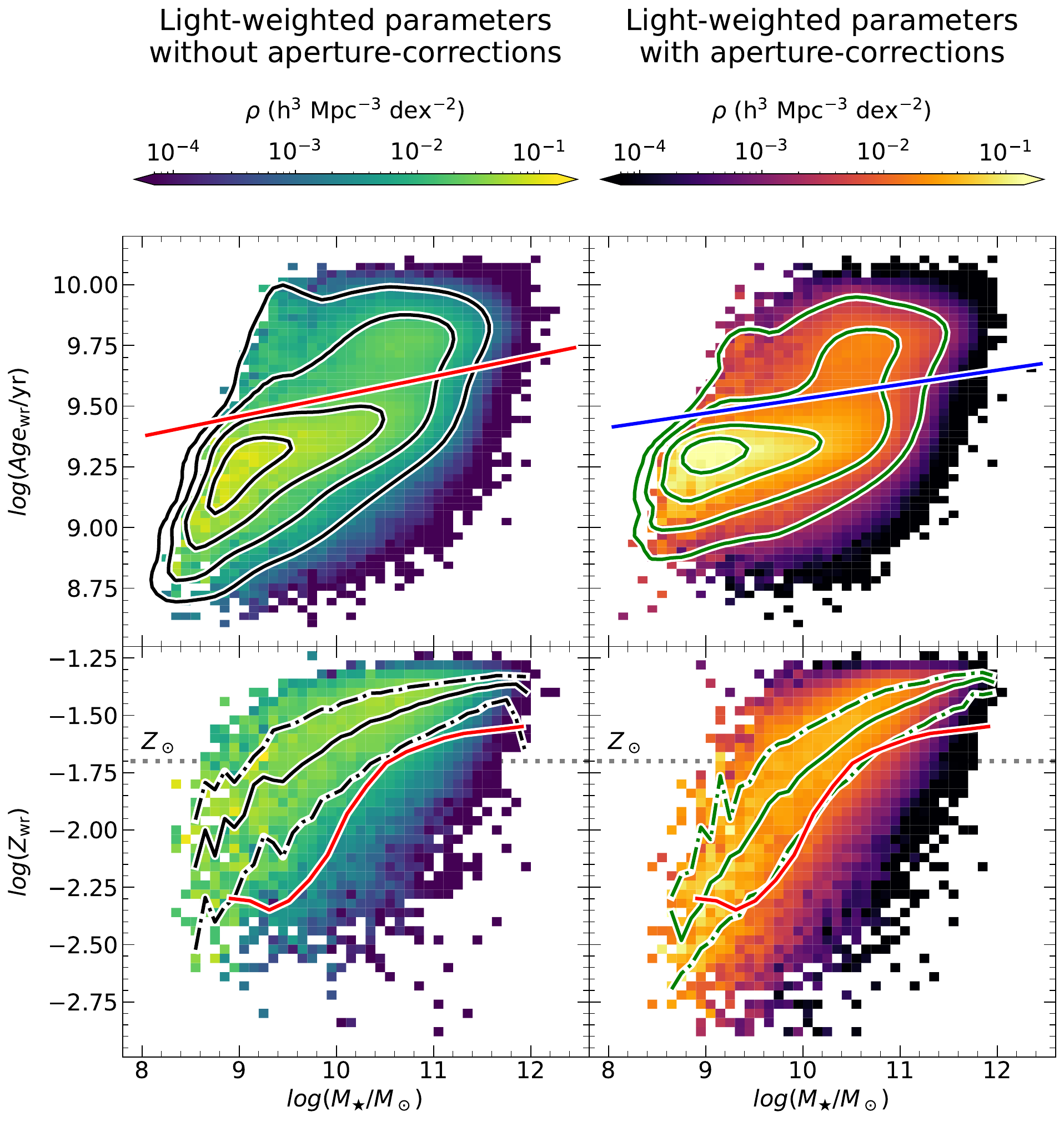}
\includegraphics[height=0.6\textwidth]{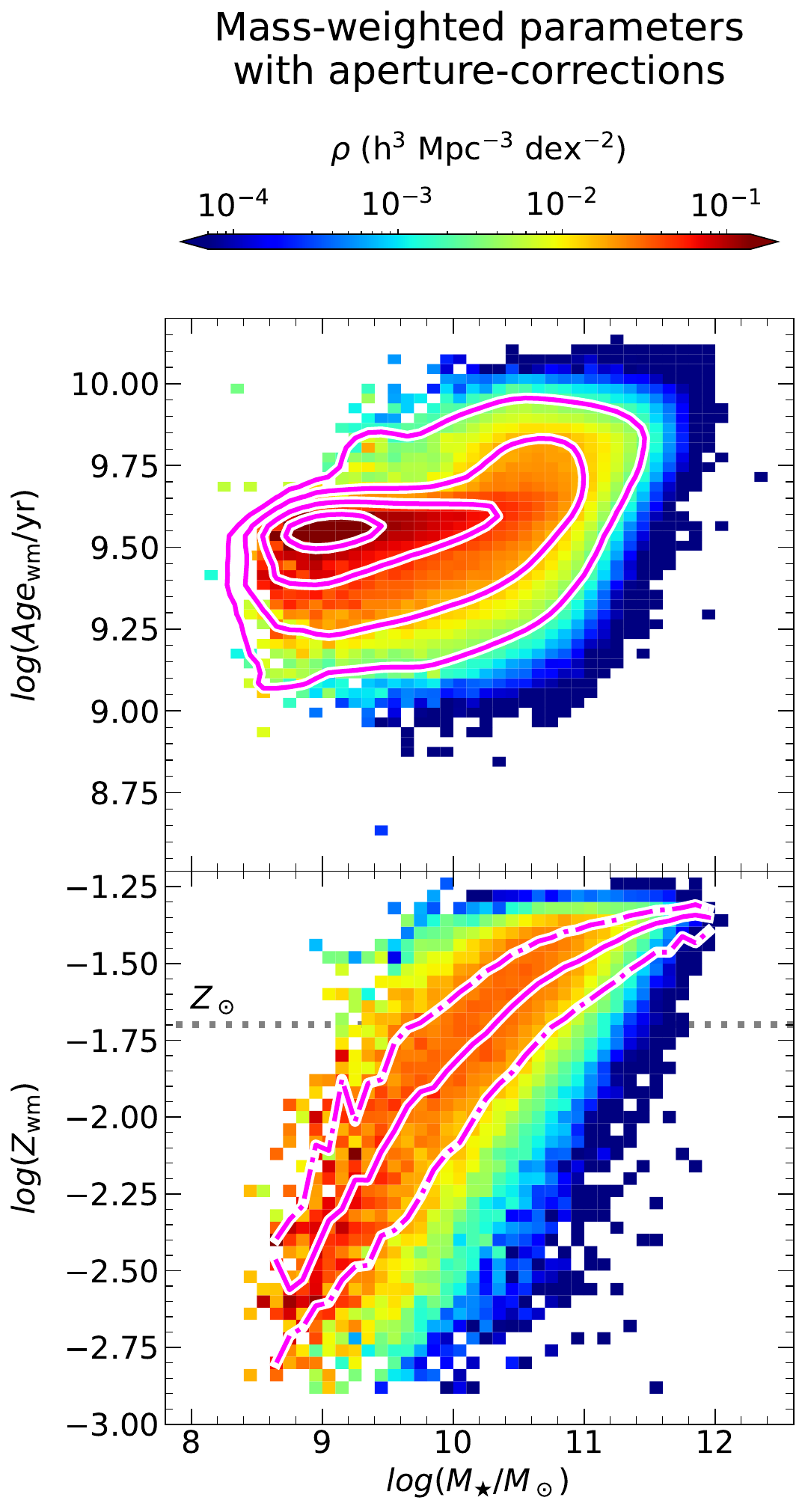}
  \caption{Number density of galaxies for the SDSS DR7 in the mass-age and mass-metallicity planes, obtained using both the volume and completeness statistical weights.
  The \emph{left and central panels} display the distributions obtained with light-weighted mean ages and metallicities, without the implementation of aperture corrections (\emph{left panels}) and implementing aperture corrections (\emph{central panels}).
  The \emph{right panels} display the distributions obtained with mass-weighted mean ages and metallicities for the aperture-corrected sample.
  In the mass-age planes, the coloured black, green, and fuchsia lines identify the density levels enclosing $16\%$, $50\%$, $84\%$, and $97.5\%$, obtained smoothing the corresponding distribution using a Gaussian KDE.
  The red and blue lines, overplotted on the top of the mass-age distributions in the \emph{left and central panels}, for uncorrected and aperture-corrected estimates, respectively, identify the divisions between young and old galaxies as defined in Sec. \ref{subsec:Transition_mass}.
  In the mass-metallicity planes, the coloured black and green solid and dashed lines identify the median and $16^\mathrm{th}$ and $84^\mathrm{th}$ percentiles of the corresponding distributions in bins of mass.
  The red lines overplotted on top of the mass-metallicity distribution identify the median relation from \cite{Gallazzi2005}.
  The dotted horizontal lines identifies the solar metallicity $Z_\odot\equiv0.02$.}
    \label{fig:Scaling_relations_comparison}
\end{figure*}
Using the inference method described in sec. \ref{sec:Methodology&Inference} we estimated properties for the galaxy sample both implementing and not-implementing the aperture corrections.
In this section, we present and analyse the distributions of galaxies in the mass-age and mass-stellar metallicity planes, taking into account volume and completeness statistical corrections (as calculated in section \ref{subsec:Correction_statistical_biases}).
We characterise the stellar populations scaling relations that emerge from these distributions. 
By comparing the relations obtained from the fiducial aperture-corrected parameters with those without the implementation of aperture corrections we show that these cause significant changes in the scaling relations.
Finally, based on existent aperture-corrected estimates of the SFR, we analyse the scaling relations for passive and star-forming galaxies separately.
\subsection{Local Universe mass-age and mass-metallicity scaling relations}\label{subsec:scaling_relations_results}
The central panels of figure \ref{fig:Scaling_relations_comparison} display the number density distributions of galaxies in the local Universe in the planes defined by light-weighted mean stellar age ($Age_{\rm wr}$) vs stellar mass (top row) and by light-weighted mean stellar metallicity ($Z_{\rm wr}$) vs stellar mass (bottom row), based on the parameters obtained with the aperture-corrected indices.
The left panels display the same relations obtained without the implementation of the aperture correction.
The right panels display the scaling relations for the aperture-corrected \emph{mass-weighted} mean ages and metallicities, which will be analysed at the end of this section. If not specified otherwise, age and metallicity are meant as light-weighted means in the following.
In each mass-age plane the contours identify the isodensity levels enclosing $16\%$, $50\%$, $84\%$, and $97.5\%$ of the complete galaxy sample, calculated from the respective distribution smoothed with a gaussian KDE.
In each mass-metallicity plane the solid and dashed lines represent the median and $16^\mathrm{th}$ and $84^\mathrm{th}$ percentiles of the metallicity distribution as a function of mass, respectively.
The red and blue lines in the light-weighed mass-age planes identify the separation between young and old galaxies, which will be defined in section \ref{subsec:Transition_mass}.

\begin{figure}
\centering
\includegraphics[width=\hsize]{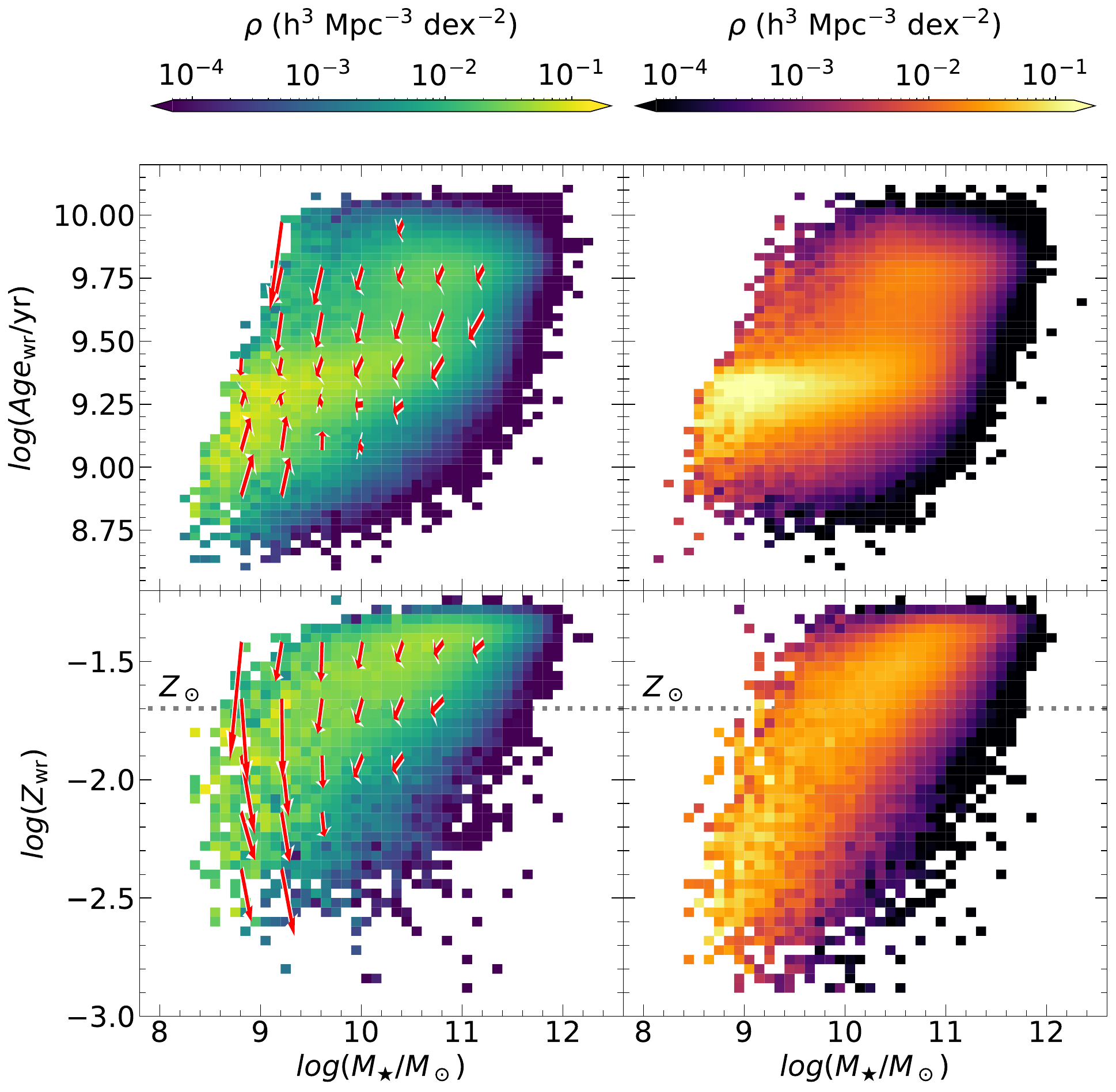}
  \caption{Number density of galaxies for the SDSS DR7 in the mass-age and mass-metallicity planes, obtained using both the volume and completeness statistical corrections. 
  Following the same colour-code of Fig. \ref{fig:Scaling_relations_comparison}, the \emph{left panels} display the distributions obtained without aperture corrections, and the \emph{right panels} display the distributions obtained from aperture-corrected indices.
  The red arrows in the \emph{left panels} represent the median shift of galaxies in each bin due to the implementation of aperture corrections.
  The dotted horizontal lines identify the solar metallicity $Z_\odot\equiv0.02$.}
    \label{fig:Scaling_relations_comparison_fluxes}
\end{figure}
\begin{figure*}
\centering
\includegraphics[width=0.8\textwidth]{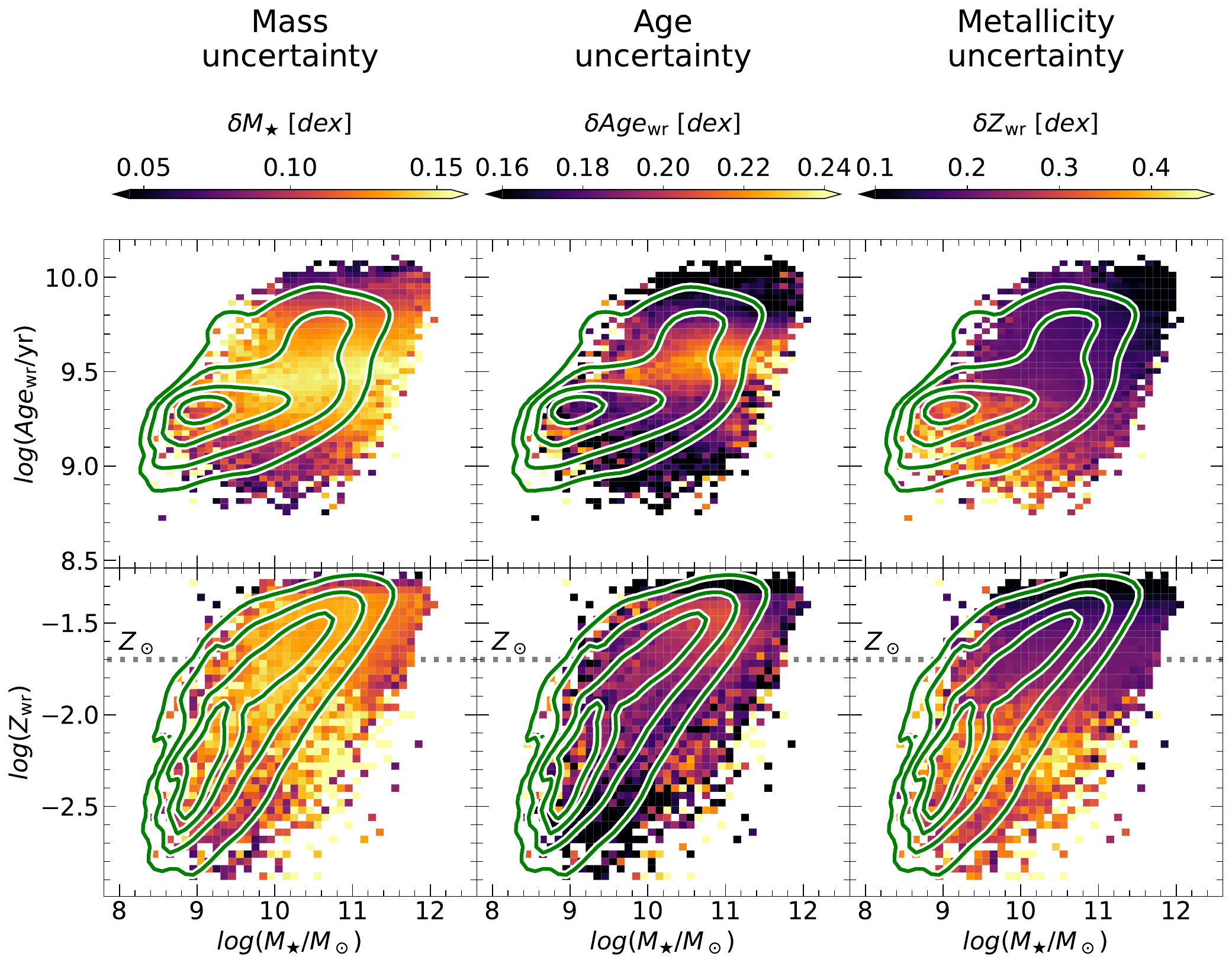}
  \caption{Mass-age (\emph{top row}) and mass-metallicity (\emph{bottom row}) distributions of galaxies with $\mathrm{SNR}\geq20$, colour-coded by the uncertainties $\delta$ in mass (\emph{left panels}), light-weighted mean age (\emph{central panels}), and light-weighted mean metallicity (\emph{right panels}) estimates.
  The green lines identify the number density levels enclosing $16\%$, $50\%$, $84\%$,  and $97.5\%$ of the statistically-weighted distributions.
  The dotted horizontal lines identifies the solar metallicity $Z_\odot\equiv0.02$.}
    \label{fig:Mass_age_metallicitiy_uncertainties_maps}
\end{figure*}
\begin{figure}
\centering
\includegraphics[width=\hsize]{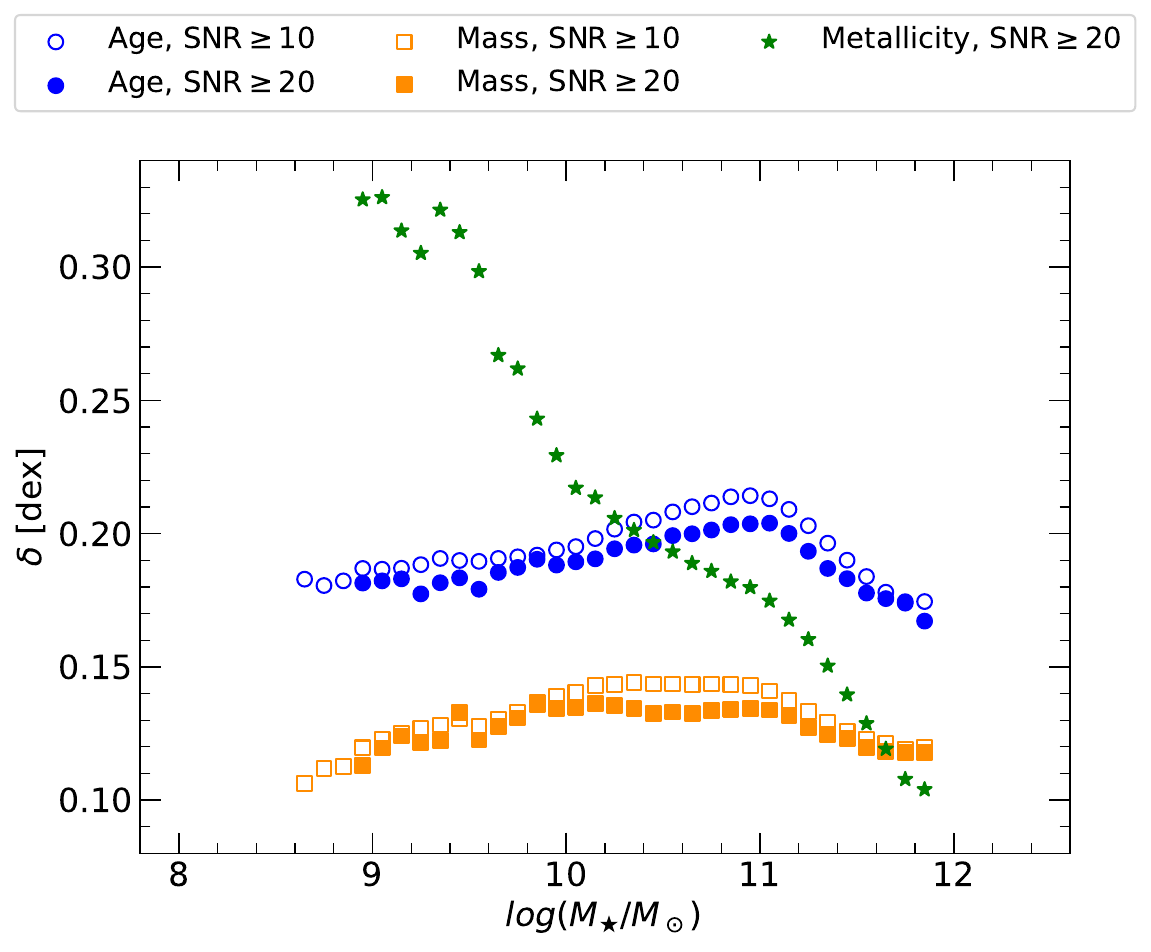}
  \caption{Uncertainties in masses (squares), light-weighted mean stellar ages (circles), and light-weighted mean stellar metallicities (stars) estimates.
  Empty symbols represent the uncertainties for the galaxies with $\mathrm{SNR}\geq10$, while the filled ones are for $\mathrm{SNR}\geq20$. Note that for the metallicity, estimates are considered only for $\mathrm{SNR}\geq20$.}
    \label{fig:Mass_age_metallicitiy_uncertainties}
\end{figure}
\begin{figure}
\centering
\includegraphics[width=1\hsize]{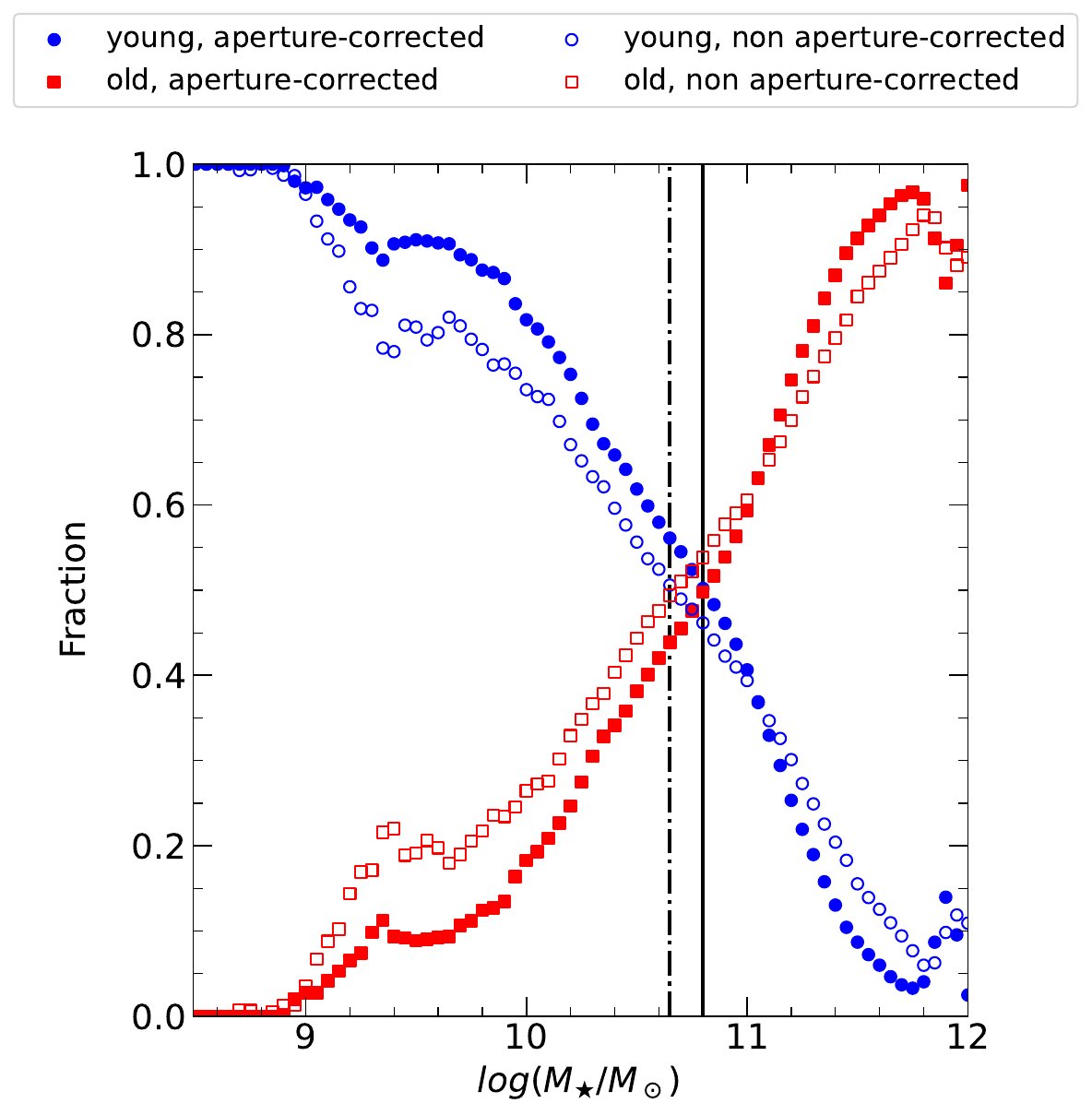}
  \caption{Fractions of young (blue) and old (red) galaxies, as defined in Sec. \ref{subsec:Transition_mass}, in $0.1~\dex$-wide mass bins. 
  Filled and empty symbols refer to fractions derived from aperture-corrected estimates and uncorrected estimates, respectively.
  The solid and dashed vertical lines identify the transition masses based on the aperture corrected ($M_\mathrm{tr}^\mathrm{ac}=10^{10.80}~\Msun$) and uncorrected ($M_\mathrm{tr}=10^{10.65}~\Msun$) estimates, respectively.}
    \label{fig:Transition_mass_statistics}
\end{figure}

\subsubsection{Light-weighted scaling relations}
In the aperture-corrected age vs. mass plane (upper central panel) we see a general trend of increasing light-weighted ages with increasing galaxy mass.
For masses below $10^{10}~\Msun$ we note the presence of a prominent population of galaxies with ages around $10^{9.3}~\yr$, and a scarcely populated distribution with $Age_\mathrm{wr}>10^{10.5}~\yr$.
With increasing galaxy mass this sequence of young galaxies becomes progressively less populated, while a sequence of galaxies with light-weighted ages around $10^{9.75}~\yr$ arises.
At masses $M_\star\gtrsim10^{11}~\Msun$ the young sequence vanishes, leaving a population of old and massive galaxies.
These results confirm the presence of a mass-age relation as already observed in several studies in the literature \citep[e.g.,][]{Kauffmann2003a, Gallazzi2005,Gallazzi2021, Pasquali2010, Trussler2020, Zibetti2022}.
We note, however, that our mass-age relation displays a more prominent bimodality, whereby the young and old sequences of galaxies coexist in the mass range $10^{10}-10^{11}~\Msun$, as also highlighted by the green isodensity contours.
In the aperture-corrected metallicity vs. mass plane (bottom central panel) galaxies exhibit a unimodal MZR of increasing stellar metallicity with increasing galaxy mass.
We visualize this relation and its scatter by overplotting the median relation and $16^\mathrm{th}$ and $84^\mathrm{th}$ percentiles of the metallicity distribution as a function of $M_\star$, identified by the green solid and dashed lines respectively, over the 2D distribution.
The MZR relation is steeper and has a wider spread at lower masses, while becoming progressively flatter and narrower with increasing galaxy mass.

When comparing our aperture-corrected mass-age relation with the uncorrected one (upper left panel) we see significant differences.
The most striking difference concerns the low mass end, where the uncorrected relation shows a much steeper decrease of the light-weighted ages with decreasing galaxy mass below $10^9~\Msun$.
For masses $M_\star\gtrsim10^9~\Msun$, the implementation of aperture corrections result in an enhancement of the population of the young sequence with respect to the uncorrected distribution.
In the mass-stellar metallicity plane we see that the aperture-corrected MZR is characterised by systematically lower stellar metallicities.
We note that this difference is mass dependent, with progressively stronger deviations between the two MZRs with decreasing stellar mass, resulting in a steeper slope for the aperture correction estimates.

In conclusion, aperture corrections significantly influence both the mass-age and the mass-stellar metallicity relations. 
The comparisons indicate that the effect of such corrections is mass dependent.
We analyse this dependence in more details in Fig. \ref{fig:Scaling_relations_comparison_fluxes}, where we show the median shift of galaxies in the mass-age and mass-metallicity planes due to aperture corrections.
Following the same colour-code as in figure \ref{fig:Scaling_relations_comparison}, the left panels show the uncorrected distributions and the right panels show the aperture-corrected ones.
Over the uncorrected distributions we plotted the median variations in mass, age and metallicity in each 2D bin, identified by the red arrows.
Looking at the mass-age distribution, we see that the aperture corrections cause the largest variations in age to the intermediate-age ($\mathit{Age}_\mathrm{wr} \sim 10^{9.6}$\,yr) galaxies, which reduce their light weighted age by up to $\sim0.15~\dex$. In general, age variations are negative by a few $0.01$ up to $0.1~\dex$ over most of the age-mass plane.
The majority of galaxies also suffer of a reduction of their mass estimate by up to $\sim0.1~\dex$.
The only exception is represented by the galaxies with $M_\star<10^{10}~\Msun$ and $Age_\mathrm{wr}\lesssim10^{9.3}~yr$, which display an increase both of the mass and light-weighted age estimates by up to $\sim0.1~\dex$ and $\sim0.15~\dex$, respectively.
In the mass-metallicity plane we find a mass-dependent decrease of the stellar metallicity, with negligible effects in the high mass end, and shifts down to $\sim-0.4~\dex$ in the low-mass end.

\subsubsection{Uncertainties of stellar population parameters}
Figure \ref{fig:Mass_age_metallicitiy_uncertainties_maps} displays the aperture-corrected mass-age and mass-metallicity planes, colour-coded by the median uncertainty $\delta$ in the mass (left panels), age (central panels), and metallicity (right panels) estimates, respectively, for the $\rm SNR\geq20$ galaxies in each bin on the plane.
The green contours identify the isodensity levels as calculated in Fig. \ref{fig:Scaling_relations_comparison}.
The mass-age plane (top panels) shows that mass uncertainties range from $\sim0.07~\dex$ to $\sim0.17~\dex$, age uncertainties range from $\sim0.15~\dex$ to $\sim0.25~\dex$, and metallicity uncertainties (bottom panels) range from $\sim0.1~\dex$ to $\sim0.4~\dex$.
We see that both mass and age uncertainties mostly depend on age, while metallicity uncertainties significantly depend both on metallicity and age.
The highest mass and age uncertainties pertain to the range $10^{9.3}-10^{9.8}~\yr$.
As we can see in the mass-metallicity planes, mass and age uncertainties depend only marginally on metallicity.
On the contrary, metallicity estimates display larger uncertainties for more metal-poor and younger galaxies. In the mass-age diagram we observe a mild trend for less massive galaxies to have larger metallicity uncertainties, which could be the result of the interplay between the mass-age and the mass-metallicity relations.
Interestingly, the largest uncertainties in mass and age are found for galaxies with intermediate light-weighted ages (the so-called ``green valley'' galaxies). This transition range is typically represented by complex SFHs, which are affected by larger degeneracies in the spectral properties. These degeneracies would then be reflected in larger uncertainties on age and, as a consequence, on mass-to-light ratio, hence on stellar mass. 
Remarkably, metallicity estimates do not appear to be affected by SFH degeneracies.
The larger metallicity uncertainties in the metal poor regime derive from the weakness of the absorption features at low metallicity, but are also related to the younger ages that characterize the low-mass metal-poor galaxies.

Figure \ref{fig:Mass_age_metallicitiy_uncertainties} summarises the information about the mass (squares), age (circles), and metallicity (stars) uncertainties as a function of stellar mass.
The median errors in mass bins are shown for the three quantities, both for the galaxies with $\text{SNR}\geq10$ (empty symbols) and $\text{SNR}\geq20$ (filled symbols).
We find that age and mass uncertainties attain values of $\sim0.19~\dex$ and $\sim0.12~\dex$ respectively, not changing significantly when using the sample with higher SNR, as already shown by \cite{Gallazzi2005}. This indicates that the reliability of our age and mass estimates is not limited by SNR as long as it is larger than $10$, rather by the intrinsic degeneracies of CSP. In fact, we have verified that the distributions in the mass-age planes (weighted by the appropriate statistical factors computed in Sec. \ref{subsec:Correction_statistical_biases}) derived for the $\mathrm{SNR}\geq10$ and the $\mathrm{SNR}\geq20$ samples are statistically consistent, the only significant difference being the noise due to the different number statistics.
When comparing to \citet[][see their Fig. 8]{Gallazzi2005} we find that our age uncertainties are larger by $\sim0.05~\dex$. This is likely due to the greater variety (and degeneracies) in our CSP models, providing a more extended coverage of the physical parameter space, due to the changes that we implemented in the SPS models (see sec. \ref{sec:Models_comparison} for more details).

It is interesting to compare these age uncertainties with the width of the young and old sequences. From the mass-binned analysis of the age distribution we can estimate the width of both sequences in $\sim0.15~\dex$, which are smaller than the uncertainties showed in Fig. \ref{fig:Mass_age_metallicitiy_uncertainties}.
On the one hand, this implies that the two sequences are intrinsically very narrow, below the resolution limits given by the parameter uncertainties. On the other hand, this also indicates that part of the uncertainty budget is systematic, thus does not contribute to the broadening of the distributions. 

Metallicity uncertainties (for $\rm SNR\geq20$ only) reduce significantly with increasing mass, from $\sim0.3~\dex$ at $M_\star=10^9~\Msun$ to $\sim0.1~\dex$ at $M_\star\sim~10^{12}~\Msun$.
As analysed with the uncertainty distributions in mass-age and mass-metallicity planes (Fig. \ref{fig:Mass_age_metallicitiy_uncertainties_maps}), this mass dependence of the metallicity uncertainties descends from the MZR and the primary dependence of metallicity uncertainties on the metallicity itself.
Binning galaxies in $0.1~\dex$ wide mass-bins, we find that for low mass galaxies the spread of the MZR is $\sim0.35~\dex$, hence comparable with the uncertainty on the stellar metallicity.
With increasing galaxy mass, the width of the distribution gradually decreases, reaching $0.07~\dex$, thus smaller than the uncertainties in the high-mass end ($0.1-0.15~\dex$). Similar considerations as above about the possible systematic component in the estimate uncertainties apply here \citep[see also discussion in section 4 of][]{Zibetti2020}.

\subsubsection{Mass-weighted scaling relations}
When considering the mass-weighted properties (right panels of Fig. \ref{fig:Scaling_relations_comparison}) we find significant differences in the mass-age plane, and negligible differences in the mass-metallicity plane.
In the mass-age plane we note that the galaxies in the young sequence have mass-weighted ages $\sim0.3~\dex$ higher with respect to the light-weighted ones.
On the other hand, galaxies in the old sequence have a much smaller difference between the ages estimates, with variations up to $\sim0.1~\dex$.
Using mass-weighted quantities the young sequence is located at ages around $Age_\mathrm{mw}\sim10^{9.6}~\yr$, for stellar masses $M_\star\lesssim10^{10.5}~\Msun$.
The reduced separation between the young and old sequences makes the bimodality found in the mass range $10^{10}-10^{11}~\Msun$ less marked, resulting in a smooth transition between the two.
These differences between mass- and light-weighted ages are indeed expected considering the variation of the mass-to-light ratio as a function of age for SSPs, with young stellar populations being overluminous with respect to older ones for a given stellar mass. This is a well known effect resulting in light-weighted mean ages being biased towards younger populations with respect to mass-weighted mean ages \citep[e.g.][]{Zibetti2017}. This bias progressively reduces moving to overall older ages.

In the mass-metallicity plane, we observe only minor discrepancies between the light-weighted and mass-weighted MZRs. This may be interpreted as a consequence of metal absorption indices being mainly determined by the bulk of the old stellar population which contribute cold stars, thus resulting in a reduction of the ``outshining'' bias that more heavily affects the age estimates \citep[e.g.][]{Serra2007a}.

\subsection{A mass-dependent age bimodality}\label{subsec:Transition_mass}
As previously mentioned, in the mass-age plane we can identify a young sequence dominating at low masses ($\lesssim10^{10}~\Msun$) and an old sequence dominating at large masses ($\gtrsim10^{11}~\Msun$). A transition between these sequences occurs for masses in the range $10^{10}-10^{11}~\Msun$.
In this paragraph we analyse this transition in detail.
The existence of a transition mass in the stellar populations scaling relations was already pointed out in several studies \citep[e.g.,][]{Kauffmann2003a, Gallazzi2005, Peng2010, Haines2017, Gallazzi2021} and is generally interpreted as linked to the efficiency of feedback mechanisms which cause galaxy quenching.
Hence a correct estimation of such transition mass is of vital importance to further investigate and constrain such processes.
Interestingly, as shown in figure \ref{fig:Scaling_relations_comparison_fluxes}, aperture effects shift galaxies between the two sequences, thus, by implementing careful corrections for these effects, our work constitutes a key contribution in unravelling the issues about a reliable determination of the transition mass.

To calculate the transition mass we define a separation between the two sequences.
For the sake of simplicity we adopt a linear cut running all along the green valley, i.e. the locus of the minima of the age distributions as a function of mass.
Since the minima are often ill-determined due to the unequal populations of the two age peaks, we proceed as follows.
We identify the peak of the age distribution for the galaxies populating the young and old sequences by fitting a double Gaussian to the age distribution for galaxies in $0.1~\dex$-wide mass bins in the range $10^{9.2}-10^{11.8}~\Msun$.
We perform a linear fit between $\log \mathit{Age}_\mathrm{wr,~young~peak}$ and $\log \mathit{M}_\star$, and $\log \mathit{Age}_\mathrm{wr,~old~peak}$ and $\log \mathit{M}_\star$.
The mean of the slopes of these lines is taken as the slope of the dividing line.
The intercept is determined by imposing the line to go through the minimum of the galaxy density distribution, projected along the direction of the dividing line.
We obtain separation lines described by a linear relation of the form:
\begin{equation}\label{eq:Separation_line}
    \log(\mathit{Age}_\mathrm{cut}/\mathrm{yr})=m\cdot\log(\mathrm{M}_\star/10^{10.5}\,M_\odot)+\log(\mathit{Age}_\mathrm{cut,10.5}/\mathrm{yr})
\end{equation}
where $m$ is the slope, and $\log(\mathit{Age}_\mathrm{cut,10.5}/\mathrm{yr})$ is the intercept of the dividing line calculated at $M_\star=10^{10.5}~\Msun$.
Without the implementation of aperture corrections we obtain $m=0.082\pm0.033$ and $\log(\mathit{Age}_\mathrm{cut,10.5}/\mathrm{yr})=9.58 \pm 0.34$ while, implementing aperture corrections, we obtain $m^\mathrm{ap}=0.059\pm0.019$ and $\log(\mathit{Age}_\mathrm{cut,10.5}/\mathrm{yr})^\mathrm{ap}=9.56 \pm 0.21$
\footnote{
Errors on these parameters are obtained from a Monte Carlo realisation in which the slope is varied following a normal distribution centred on the mean slope with $\sigma$ corresponding to half of the difference between the maximum and the minimum of the slope.}
.

The divisions between the two sequences are shown by the blue and red lines in the mass-age planes of figure \ref{fig:Scaling_relations_comparison} and \ref{fig:Mass_age_passive_starforming}, with and without aperture corrections, respectively.
We identified galaxies as young when they lay below the lines, and old otherwise.
We computed the fraction of young and old galaxies as a function of $M_\star$ in bins of $0.1~\dex$-wide bins, with a step of $0.05~\dex$.
Figure \ref{fig:Transition_mass_statistics} displays fractions of young (blue) and old (red) galaxies both for non aperture-corrected (empty symbols) and  aperture-corrected (filled symbols) galaxies.
The old sequence is not sampled below $M_\star=10^9~\Msun$, hence we limit the analysis to the mass range $M_\star>10^9~\Msun$.
With increasing galaxy mass we observe a gradual decrease of the fraction of young galaxies, with a faster decrease for stellar masses above $10^{10}~\Msun$.
Therefore, in the mass range $10^{10}-10^{11}~\Msun$ the galaxy distribution becomes increasingly dominated by old galaxies.
We defined the transition mass as the central value of the mass bin where the two fractions cross over.
We find $\log(M^\mathrm{ac}_\mathrm{tr}/\Msun)=10.80 \pm 0.05~\dex$ (solid line) for the aperture-corrected sample, and $\log(M_\mathrm{tr}/\Msun)=10.65 \pm 0.05~\dex$ (dashed line) for the non aperture-corrected.
Since the primary effect of aperture corrections is to systematically move intermediate-age galaxies to lower ages, thus making many of them transition from the old to the young sequence (as shown in Fig. \ref{fig:Scaling_relations_comparison_fluxes}), for a given galaxy mass the fraction of young (old) galaxies is higher (lower) in the case of aperture-corrected estimates than for uncorrected ones.
This results in a shift of the transition mass towards higher masses by $0.15~\dex$ when implementing aperture corrections.

\subsection{The scaling relations of star-forming and passive galaxies}\label{subsec:Passive_vs_starforming}
\begin{figure}
\centering
\includegraphics[width=\hsize]{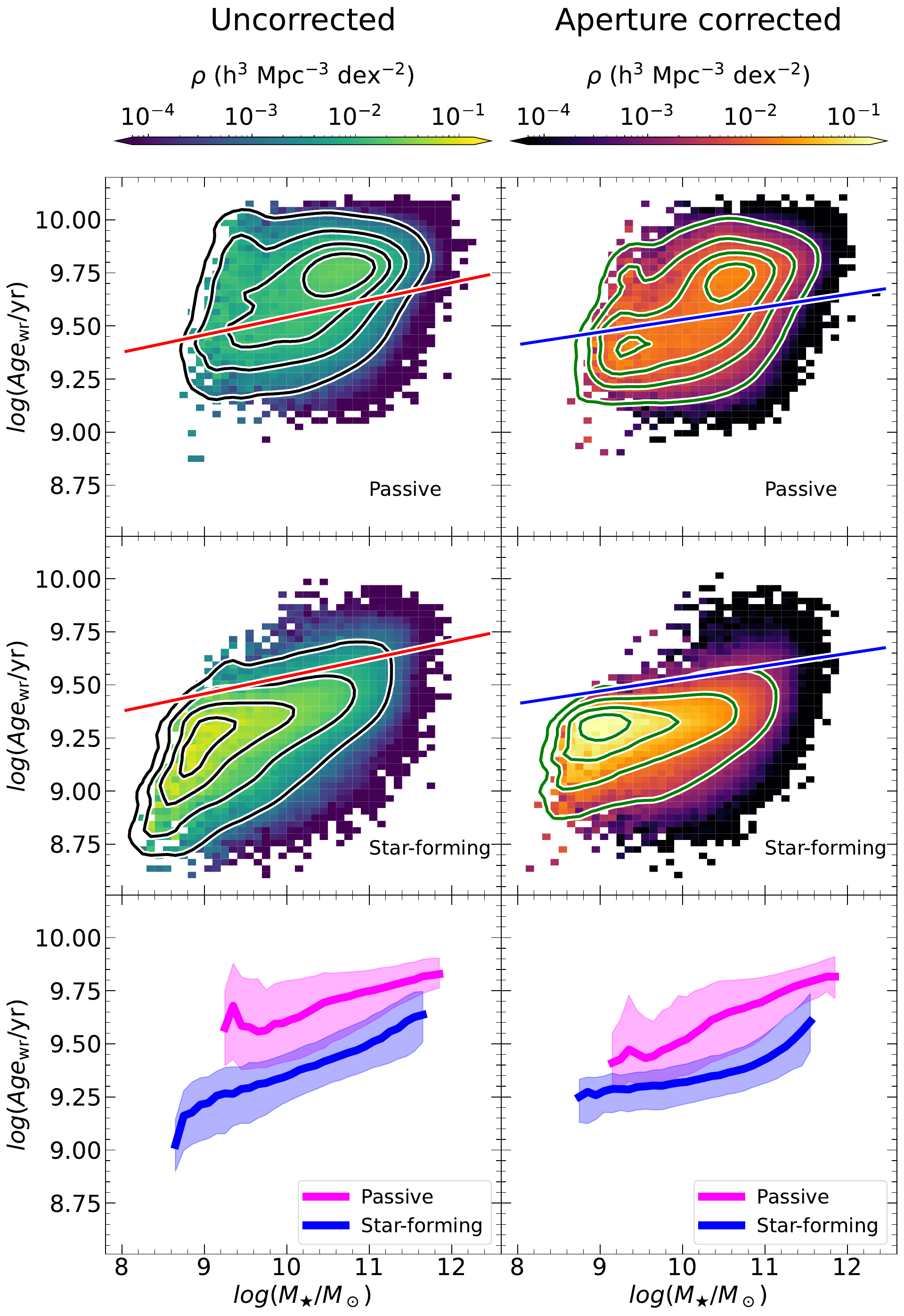}
  \caption{Number density of passive (\emph{top row}) and star-forming (\emph{middle row}) galaxies in the mass-age plane, obtained using both the volume and completeness statistical corrections. 
  Following the same colour-code of Fig. \ref{fig:Scaling_relations_comparison}, the \emph{left panels} show the distributions obtained without the use of the aperture corrections, and the \emph{right panels} show the distributions obtained implementing aperture corrections.
  The contours identify the levels containing $16\%$, $50\%$, $84\%$, and $97.5\%$ of the total integral of the distributions.
  The red and blue lines identify the divisions between young and old galaxies as defined in Sec. \ref{subsec:Transition_mass}.
  The bottom panels display the median relations (solid lines) and the respective $16^\mathrm{th}$ and $84^\mathrm{th}$ percentiles (shaded area) as a function of mass, for passive (red) and star-forming (blue) galaxies.}
    \label{fig:Mass_age_passive_starforming}
\end{figure}
\begin{figure}
\centering
\includegraphics[width=\hsize]{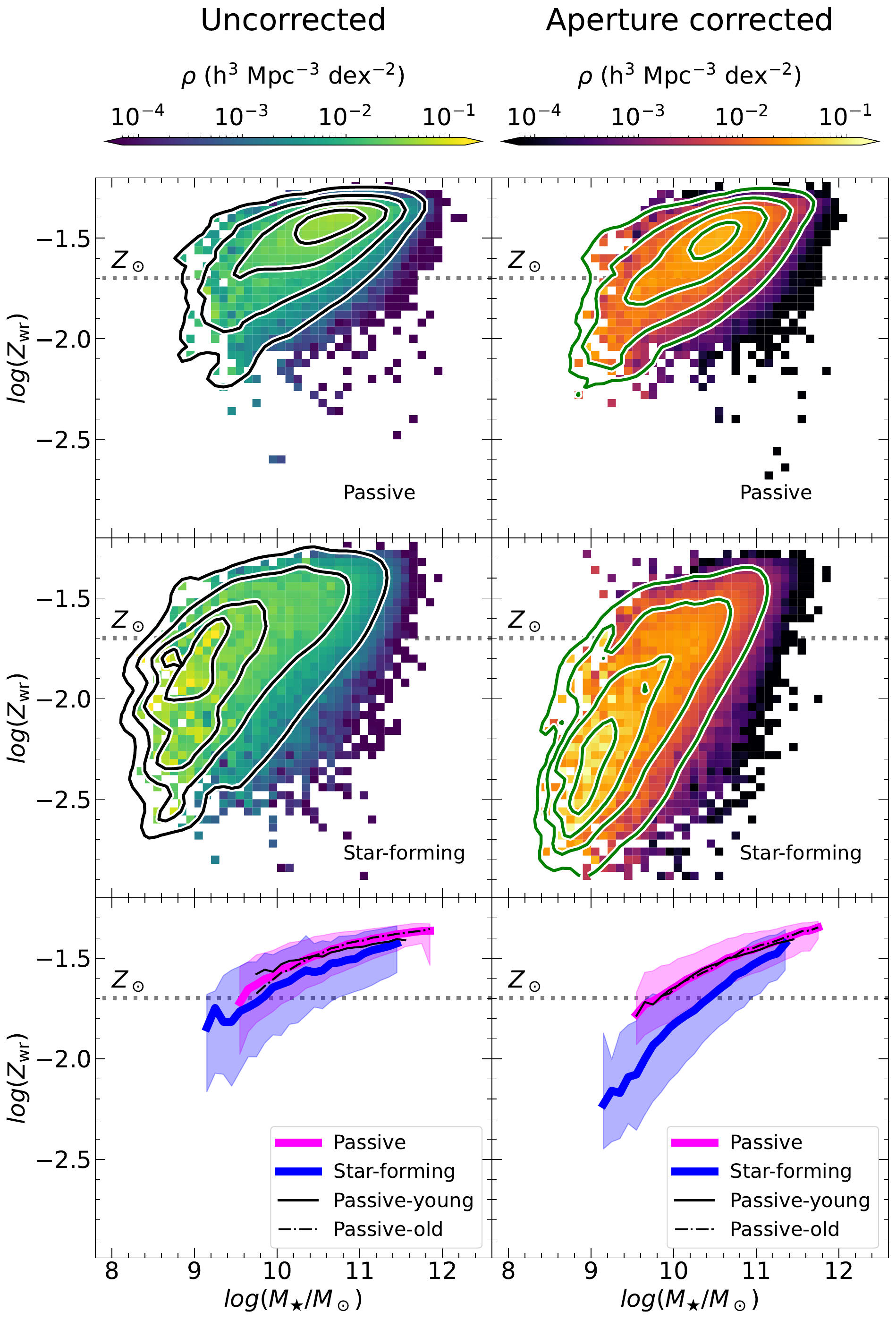}
  \caption{Same as Fig. \ref{fig:Mass_age_passive_starforming}, but for the metallicity-mass distributions.
  The solid and dot-dashed black lines in the bottom panels identify the median MZRs obtained with the subsamples of passive--young and passive--old galaxies, respectively.
  The dotted horizontal lines identify the solar metallicity $Z_\odot\equiv0.02$.}
    \label{fig:Mass_metallicity_passive_starforming}
\end{figure}
Previous studies have shown that passive and star-forming galaxies obey different scaling relations \citep{Mateus2006, GonzalezDelgado2015, Schawinski2014, Peng2015, Trussler2020, Looser2024}.
We study the scaling relations of SDSS DR7 galaxies separately for star-forming and passive, based on the aperture-corrected star formation rate estimates by \cite{Brinchmann2004}
\footnote{
Aperture-corrected star formation rate estimates have been estimated by \cite{Brinchmann2004} based on galaxy colours.
Corrections for galaxies with low levels of star formations were calculated building upon \cite{Salim2007}, as described by the MPA-JHU catalogues documentation at:
    \url{https://wwwmpa.mpa-garching.mpg.de/SDSS/DR7/sfrs.html}
}
, as available from the MPA-JHU catalogues.
More specifically, we adopt a cut based on the deviation from the star-forming main sequence as determined in \cite{Gallazzi2021}:
\begin{equation}\label{eq:Gallazzi_sfr_selection}
    \log(\text{sSFR}/\mathrm{yr}^{-1})=-0.15\log(M_\star/M_\odot)-8.46~~~.
\end{equation}
This relation is characterized by a dispersion $\sigma=0.3$. We classified galaxies as passive if they fall below the relation by more than $2\sigma$, and star-forming otherwise.

Figure \ref{fig:Mass_age_passive_starforming} and figure \ref{fig:Mass_metallicity_passive_starforming} present the comparison of the mass-age and mass-metallicity relations for passive (top panels) and starforming (central panels) galaxies, respectively.
The bottom panels display the median (solid line) and $16^\mathrm{th}$-$84^\mathrm{th}$ percentile range (shaded area) of the relations for star-forming (blue) and passive (red) galaxies.
The left panels display the uncorrected distributions and the right panels display the aperture-corrected ones.
In both figures the black and green contours identify the isodensity levels.
In Fig. \ref{fig:Mass_age_passive_starforming} the blue and red lines identify the separations between young and old galaxies as defined in sec. \ref{subsec:Transition_mass}.
From the mass-age plane we see that star-forming and passive galaxies follow different relations \citep[as already observed by][]{Schawinski2014, Peng2015, Trussler2020}.
As expected, star-forming and passive galaxies are mainly associated to the young and old sequences identified in Fig. \ref{fig:Scaling_relations_comparison}, respectively.
However, the mass-age distribution of passive galaxies presents a significant tail of young galaxies, which is visible at all masses below $\sim 10^{11}\,\mathrm{M_\odot}$, and becomes particularly conspicuous below $\sim 10^{9.5}\,\mathrm{M_\odot}$, especially in the aperture-corrected version. These low-mass passive galaxies are indeed characterised by young stellar populations, with ages below the dividing line ($\sim10^{9.4}~\yr$), consistent with the ages of starforming galaxies of similar mass.
On the other hand, high-mass star-forming galaxies reach light-weighted ages up to $\sim10^{9.6}~\yr$, very similar to their passive counterparts.
Star-forming galaxies reach lower ages in the low-mass end when aperture-corrections are taken into account, as already seen in the left panel of Fig. \ref{fig:Scaling_relations_comparison}.
Moreover, passive low-mass galaxies display both lower ages and a narrower distribution when aperture corrections are implemented.

\begin{figure}
\centering
\includegraphics[width=\hsize]{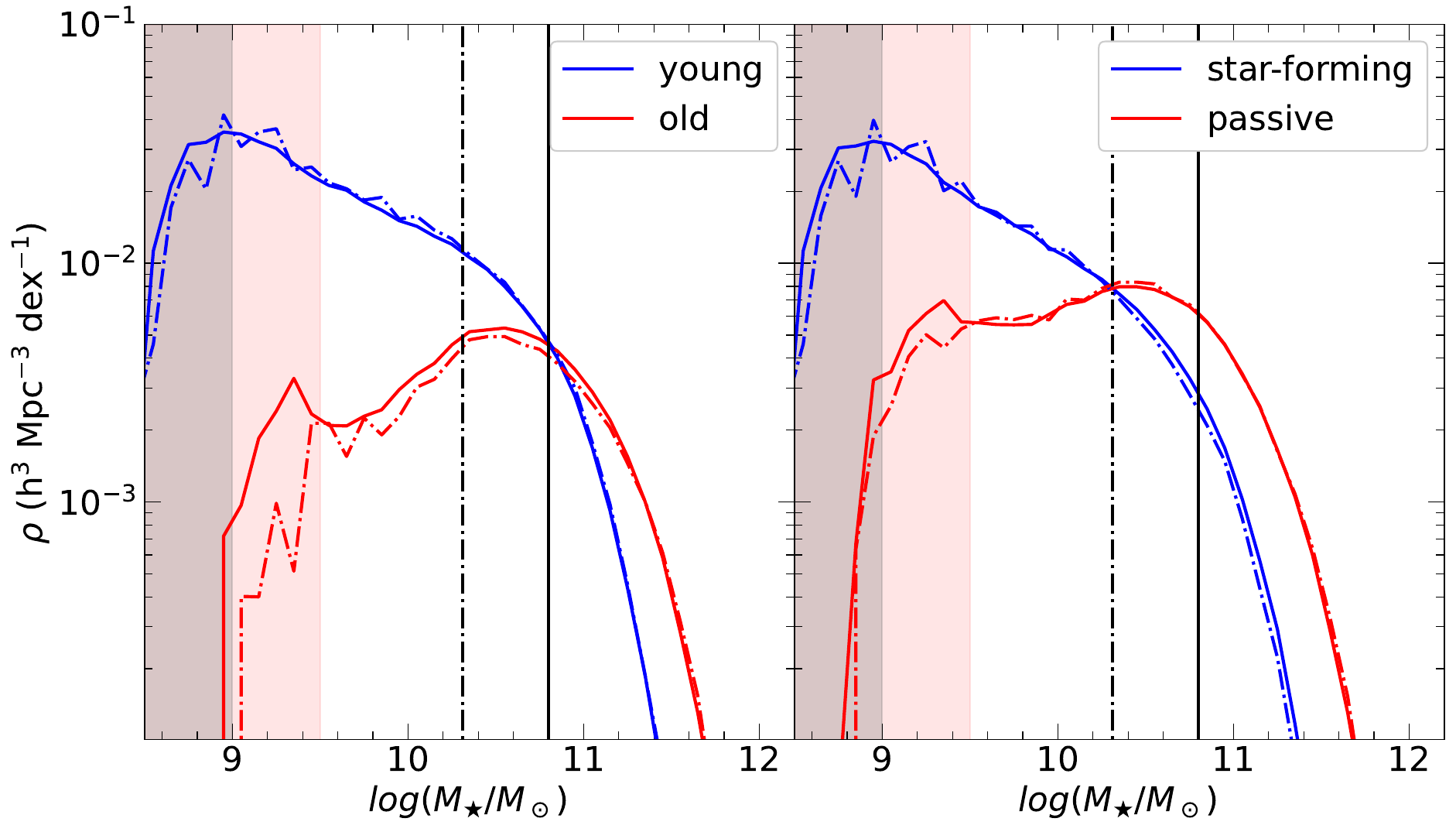}
  \caption{Mass function of the $\rm SNR\geq10$ (solid lines) and $\rm SNR\geq20$ (dot-dashed lines) galaxy samples, including aperture corrections. In the \emph{left panel} galaxies are divided into young (blue) and old (red), following the division described in sec. \ref{subsec:Transition_mass}. In the \emph{right panel} galaxies are divided into star-forming (blue) and passive (red), following the \cite{Gallazzi2021} separation (see eq. \ref{eq:Gallazzi_sfr_selection}). The solid and dot-dashed black lines identify the transition masses obtained from the mass-functions, for the separations based on the age and SFR, respectively. The shaded areas identify the mass ranges where our sample is not representative, namely $\lesssim M_\star\sim10^{9.5}~\Msun$ for passive/old galaxies (red shading), $\lesssim M_\star\sim10^9~\Msun$ for star-forming/young galaxies (grey shading). Notably, the statistical weights ensure that we recover consistent mass functions irrespective of the SNR cut.}
    \label{fig:Mass_function}
\end{figure}

The differences in the classifications according to sSFR (passive vs. star-forming) or according to age (old vs. young) are apparent when comparing their mass functions. In figure \ref{fig:Mass_function} we show the mass-functions of the galaxy sample (with aperture corrections implemented) obtained when separating galaxies based on the age (left panel) and on the star formation activity (right panel). Solid and dot-dashed lines represent the samples with $\rm SNR\geq10$ and $\rm SNR\geq20$, respectively, which, thanks to the statistical weights computed in Sec. \ref{subsec:Correction_statistical_biases}, behave in a very consistent way.
We find that the mass-functions of passive and old galaxies (red lines) have an overall similar shape, in accordance to a general correspondence between light-weighted ages and star formation activity.
However, the mass-function of old galaxies is overall less populated and particularly suppressed at masses $\lesssim10^{10.3}~\Msun$, reflecting the fact that the population of passive galaxies is characterised by light-weighted ages younger than the separation defined in sec. \ref{subsec:Transition_mass}, especially at the low-mass end.
This, in turn, leads to lower transition mass between star-forming and passive galaxies (dot-dashed vertical lines in Fig. \ref{fig:Mass_function}, $M_\mathrm{tr}\sim10^{10.3}~\Msun$) with respect to the transition mass between young and old galaxies (solid vertical lines in Fig. \ref{fig:Mass_function}, $M_\mathrm{tr}=10^{10.80}~\Msun$), corresponding to a difference of $0.5~\dex$, i.e. a factor $\sim3$.

From the mass-metallicity relations of Fig. \ref{fig:Mass_metallicity_passive_starforming} we see that star-forming and passive galaxies follow distinct MZRs \citep[as already observed by][]{Peng2015, Trussler2020, Gallazzi2021, Looser2024}.
At any given $M_\star$ passive galaxies are located at higher median metallicities.
Furthermore, when considering aperture-corrections, they follow a flatter and narrower relation with respect to star-forming galaxies.
Although passive galaxies are located at higher median metallicities, they significantly overlap with the distribution of star-forming galaxies (bottom panels of Fig. \ref{fig:Mass_metallicity_passive_starforming}), resulting in a broad yet unimodal MZR when the whole sample is considered (bottom panels of Fig. \ref{fig:Scaling_relations_comparison}).
Analysing the uncorrected MZRs for passive and star-forming galaxies we see that they have only a marginal separation, as opposed to the very clear difference observed in the aperture-corrected ones.
Furthermore, not implementing aperture corrections affects also the slopes of both MZRs. 

In appendix \ref{App:MZR_young_old} we show that the partial MZRs obtained by splitting the sample into young and old galaxies are in substantial agreement with the ones obtained based on the splitting on the SFR.
However, looking at the MZRs of young passive galaxies (solid black line, in the bottom panels of Fig. \ref{fig:Mass_metallicity_passive_starforming}), i.e. of those galaxies that have a discordant classification based on age and on SFR, we see that they follow essentially the same relation as the passive old ones (dotted black line, in the bottom panels of Fig. \ref{fig:Mass_metallicity_passive_starforming}). This is indicative, somewhat surprisingly, of a stronger link of the stellar MZR with the current SFR (on $10^7~\yr$ timescale), rather than with the mean age, which characterizes the full extent of the SFH of a galaxy.

It is worth emphasizing again that aperture corrections are fundamental to correctly characterise the differences between the scaling relations of star-forming, passive, young, and old galaxies.
In particular, in the mass-age relation the corrections significantly affect passive low-mass galaxies, which turn out to have very similar ages to equally massive star forming galaxies. 
Moreover, in the mass-metallicity plane they are critical in assessing whether passive and star-forming galaxies obey different MZRs, and correctly characterise their slopes. Such a differentiation has been used in several works to derive constraints on the mechanisms producing galaxy quenching \citep{Peng2015, Trussler2020, Looser2024} and test chemical evolution models \citep[e.g.][]{Spitoni2017}.

\section{Assessment of the systematic biases and uncertainties}\label{sec:Models_comparison}

As already mentioned in previous sections, in order to obtain our new determination of the mass-age and mass-metallicity scaling relations, are based on SPS models with a number of improvements with respect to previous studies, in particular \cite{Gallazzi2005} and \cite{Gallazzi2021} \citepalias[][hereafter]{Gallazzi2005, Gallazzi2021}.
In this section we analyse in detail the effect of the changes in the SPS modelling and parameter estimation.
To do so we compare our results with the benchmark of \citepalias{Gallazzi2005}, which are the closest relations to ours in terms of the methodology for properties estimation.
\citetalias{Gallazzi2005} employed the same Bayesian inference method used in this work (described in Section \ref{subsec:Inference_method}) to derive the full posterior PDF of the physical parameters of observed galaxies. However, their inference differs from ours in several aspects, including the dataset of observational constraints, the SPS models adopted to generate the prior CSP model library, and the treatment of dust attenuation.

\citetalias{Gallazzi2005} analysed the SDSS DR2 \citep{Abazajian2004}, which differs from the DR7 both for the more limited statistics, and for the data reduction pipeline, which directly affects the measurements of the absorption indices from the galaxies spectra.
Considering the modelling, we introduced several changes to the SPS models, including updates to the stellar spectral library and 2D evolutionary tracks, more complex star formation and chemical enrichment histories prescriptions, and an explicit modelling of the dust content (see Table \ref{tab:SPS_models2} for a summary of these changes, first and last rows for \citetalias{Gallazzi2005}/\citetalias{Gallazzi2021} and this work, respectively).
Another fundamental difference with respect to \citetalias{Gallazzi2005} resides in the method used to model and infer dust attenuation and its effect on the mass determination (see Sec. \ref{subsubsec:Dust_treatment} for more details).

Since in this section we want to isolate and characterise the importance of these changes for the scaling relations, we will consider estimates obtained \emph{without} the implementation of aperture corrections, for consistency with previous works.

\subsection{Cumulative effects of the changes in SPS modelling}\label{subsec:Cumulative_improvements_Stellar_Population_Synthesis}
The bottom panels (\emph{i}, \emph{j}) of figure \ref{fig:Comparison_models} show the distributions in the mass-age and mass-metallicity planes for the SDSS DR7 sample studied in this work.
Differently from Fig. \ref{fig:Scaling_relations_comparison}, we show the distributions normalised in each bin of mass
\footnote{This is done similarly to figure 8 of \citetalias{Gallazzi2005}. Note, however, that the distribution in figure 8 of \citetalias{Gallazzi2005} shows the stack of the PDFs of the physical parameters, while in this work we show the distributions of the fiducial estimates of each galaxy as given by the PDF medians.}.
The black lines identify the median (solid) and $16^\mathrm{th}$ and $84^\mathrm{th}$ percentiles (dashed) of the galaxy distribution in bins of mass.
Additionally we plot the median relations of \citetalias{Gallazzi2005} (red lines).

The bottom panels (\emph{i}, \emph{j}) of Fig. \ref{fig:Comparison_models} illustrate the cumulative effects on the scaling relations due to the changes in the SPS modelling and to the observational dataset, going from \citetalias{Gallazzi2005} to the present work.
To isolate the effect of the improvement in the data from SDSS DR2 to SDSS DR7 we also plot the \citetalias{Gallazzi2021} relations (yellow lines).
The latter were obtained with the same stellar population synthesis model and inference method of \citetalias{Gallazzi2005}, but based on the SDSS DR7.
The differences between the median relations of \citetalias{Gallazzi2005} and \citetalias{Gallazzi2021} are negligible at the high-mass end for both mass-age and mass-metallicity scaling relations.
For $\mathrm{M}_\star\lesssim10^{10}~\Msun$ \citetalias{Gallazzi2021} estimates higher median ages and metallicities in comparison to \citetalias{Gallazzi2005}, for a given galaxy mass.
For masses below $\sim10^{9.2}~\Msun$ \citetalias{Gallazzi2005} has flatter relations in both mass-age and mass-metallicity with respect to \citetalias{Gallazzi2021}.

Nonetheless, comparing \citetalias{Gallazzi2021} results with our relation, we find significant differences, which cannot be due to anything else than the different modelling.
We see that our mass-age relation is flatter over the whole mass range.
Furthermore, for masses below $\sim10^{9.6}~\Msun$ we obtain significantly higher light-weighted ages, with differences up to $\sim0.25~\dex$. 
For higher stellar masses, our median relation predicts lower light-weighted ages, with differences up to $\sim0.2~\dex$.
The differences obtained for masses in the range $10^{10}-10^{11}~\Msun$ are connected to the bimodality analysed in section \ref{subsec:scaling_relations_results}.
In the mass-metallicity plane our relation has a shape similar to the \citetalias{Gallazzi2021}, but shifted by $\sim0.25~\dex$ towards higher metallicities.
In summary, the bottom panels of Fig. \ref{fig:Comparison_models} show that the changes in the modelling (Table \ref{tab:SPS_models2}, first and last rows), are the main reason for the changes in the scaling relations between this work and \citetalias{Gallazzi2005}, while the improvement due to the different SDSS data releases has only a minor effect.

\subsection{Relevance of the changes in the ingredients of stellar population synthesis models}\label{subsec:Model_changes}
\begin{table*}
\caption{Ingredients of the Stellar Population Synthesis models used in the comparison for the assessment of the systematics.}             
\label{tab:SPS_models2}     
\centering        
\tiny
\begin{threeparttable}
\begin{tabular}{p{0.18\textwidth}|p{0.09\textwidth}|p{0.09\textwidth}|p{0.18\textwidth}|p{0.12\textwidth}|p{0.20\textwidth}}
\hline
\noalign{\vspace{3pt}}
Model name / ref. work & \multicolumn{2}{c|}{SSP} & SFH & CEH & Dust \\
 & Stellar library & Evolutionary tracks &  &  &  \\
\noalign{\vspace{3pt}}
\hline
\hline
\noalign{\vspace{3pt}}
G05 \& G21& STELIB\tnote{a} & Padova94\tnote{c} & {Exponential+bursts} & $Z_\star$ constant & {Dust-free / fit colour excess} \\
\noalign{\vspace{3pt}}
\hline
\noalign{\vspace{3pt}}
Exp\_FixZ\_BC03\_STELIB& STELIB\tnote{a} & Padova94\tnote{c} & {Exponential+bursts} & $Z_\star$ constant & {two components (ISM \& BC)\tnote{g} / fit photometry} \\
\noalign{\vspace{3pt}}
\hline
\noalign{\vspace{3pt}}
Exp\_FixZ\_BC03\_MILES& MILES\tnote{b} & Padova94\tnote{c} & {Exponential+bursts} & $Z_\star$ constant & {two components (ISM \& BC)\tnote{g} / fit photometry} \\
\noalign{\vspace{3pt}}
\hline
\noalign{\vspace{3pt}}
San\_FixZ\_BC03\_MILES& MILES\tnote{b} & Padova94\tnote{c} & {Delayed Gaussian\tnote{e} + bursts} & $Z_\star$ constant & {two components (ISM \& BC)\tnote{g} / fit photometry} \\
\noalign{\vspace{3pt}}
\hline
\noalign{\vspace{3pt}}
San\_VarZ\_BC03\_MILES& MILES\tnote{b} & Padova94\tnote{c} & {Delayed Gaussian\tnote{e} + bursts} & {time increasing $Z_\star$\tnote{f}} & {two components (ISM \& BC)\tnote{g} / fit photometry} \\
\noalign{\vspace{3pt}}
\hline
\noalign{\vspace{3pt}}
San\_VarZ\_CB19\_MILES (this work fiducial)& MILES\tnote{b} & PARSEC\tnote{d} & {Delayed gaussian\tnote{e} + bursts} & {time increasing $Z_\star$\tnote{f}} & {two components (ISM \& BC)\tnote{g} / fit photometry} \\
\noalign{\vspace{3pt}}
\hline
\end{tabular}
\begin{tablenotes}
\footnotesize
\item[a] \cite{LeBorgne2003}
\item[b] \cite{Sanchez-Blazquez2006}
\item[c] \cite{Alongi1993a, Bressan1993, Fagotto1994, Fagotto1994a, Girardi1996}
\item[d] \cite{Bressan2012, Marigo2013a, Chen2015}
\item[e] The delayed Gaussian law is defined by eq. \ref{eq:Sandage_profile}.
\item[f] The time increasing $Z_\star$ is defined by Eq. \ref{eq:Chemical_Enrichment_History}.
\item[g] \cite{Charlot2000}.
\end{tablenotes}
\end{threeparttable}
\end{table*}

In this section we analyse the impact of the changes introduced in the main ingredients of the SPS models.
To achieve this, we analyse the scaling relations derived from SPS models that progressively transition, one ingredient at a time, from the \citetalias{Gallazzi2005} and \citetalias{Gallazzi2021} models to the one adopted in this work.
The changes will be introduced sequentially, following the order in which they were implemented in the models.
Since we already analysed the effects of the improvement in the dataset, the following comparisons will be performed only with the \citetalias{Gallazzi2021} results.
The different versions of the models, which we use in the comparisons, are listed in table \ref{tab:SPS_models2}.
We start by comparing the \citetalias{Gallazzi2021} results with the ones we obtain when introducing an explicit dust treatment in the SPS models (which directly impacts the colours, but not the dust insensitive indices)  and a different inference for the dust absorption (Sec. \ref{subsubsec:Dust_treatment}).
Next we show the effect of updating the stellar spectral library, on which the SPS spectra are based, from STELIB \citep{LeBorgne2003} to MILES, which has a direct impact on indices strengths and colours at fixed isochrone (Sec. \ref{subsubsec:Stellar_libraries}).
Subsequently, we change the SFH (delayed Gaussian + bursts instead of a decaying exponential + bursts), allowing for a phase with a rising SFR (Sec. \ref{subsubsec:Star_formation_history}).
In the next step we implement a time dependent CEH (instead of fixed metallicity), allowing for a monotonic rise of $Z_\star$ (Sec. \ref{subsubsec:Chemical_enrichment_history}).
Finally, we consider the effect of updating the stellar evolutionary tracks \citep[PARSEC instead of Padova94][]{Alongi1993a, Bressan1993, Fagotto1994, Fagotto1994a, Girardi1996}, which modifies the isochrones, hence impacting index strengths and colours, and the way they map age and metallicity (Sec. \ref{subsubsec:Evolutionary_tracks}).
Additionally, the new isoschrones come with an extension of the metallicity range to higher values.

\begin{figure*}
\sidecaption
  \includegraphics[width=12cm]{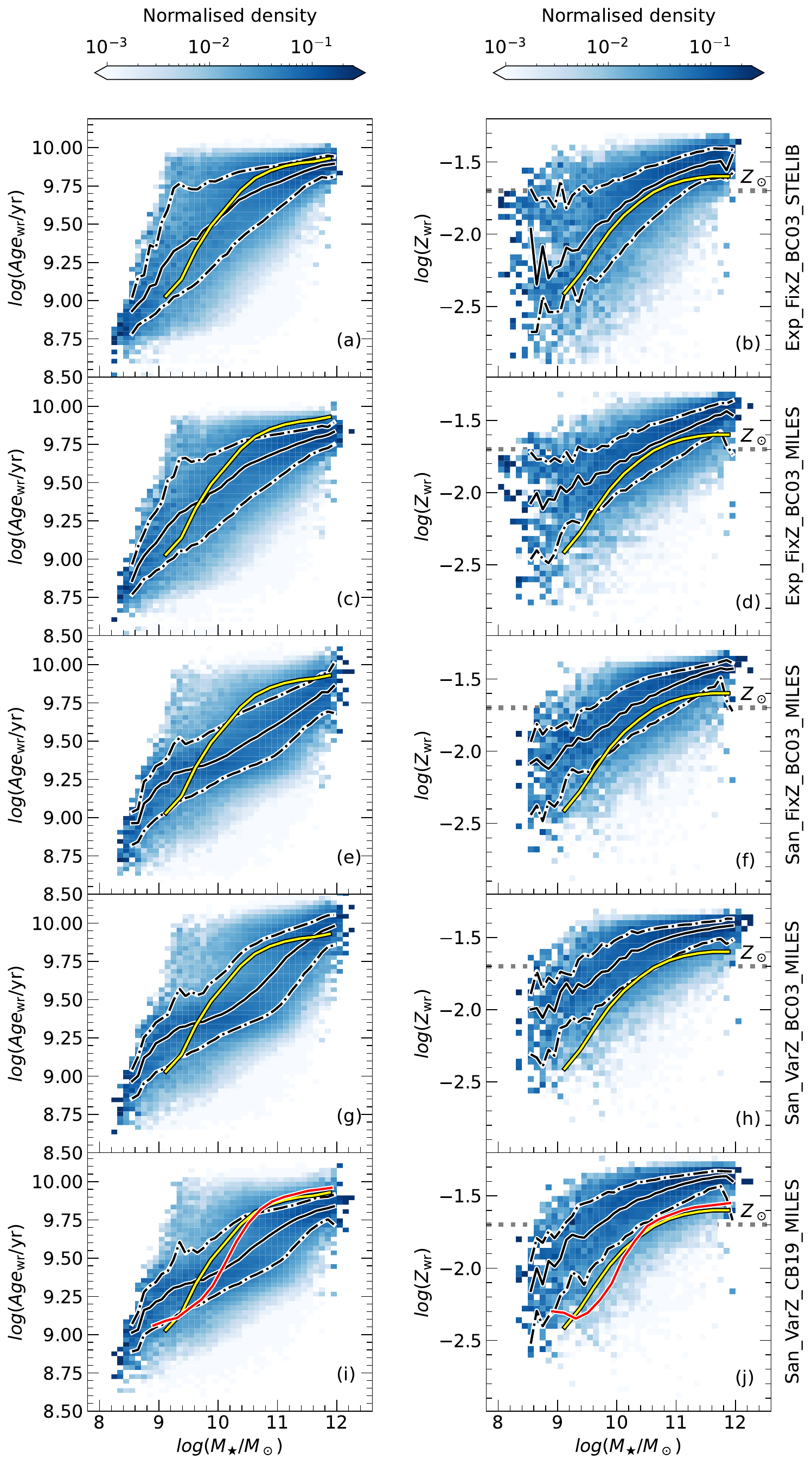}
    \caption{Density of galaxies normalised in each
  mass bin \citep[same as figure 8 from][]{Gallazzi2005} for the mass-age (left) and mass-metallicity (right) relations obtained with each SPS model in table \ref{tab:SPS_models2} (one model in each row).
  In each panel the black solid and dashed lines identify the median and $16^\mathrm{th}$ and $84^\mathrm{th}$ percentiles of the corresponding distribution in mass bins, while the yellow line identifies the median relations in the respective plane taken from \cite{Gallazzi2021}. The last row displays the distributions for the fiducial models of this work. In these panels, the red line represents the original \citetalias{Gallazzi2005} scaling relations. Note that these results are obtained \emph{without} aperture corrections, for consistency with \citetalias{Gallazzi2005} and \citetalias{Gallazzi2021}.
  The dotted horizontal lines in the mass-metallicity planes identify the solar metallicity $Z_\odot\equiv0.02$.}
    \label{fig:Comparison_models}
\end{figure*}

Fig.\ref{fig:Comparison_models} presents the mass-age (left-hand panels) and mass-metallicity (right-hand panels) relations normalised in each bin of mass.
Each row displays the relations obtained with a specific model version, identified by the label on the right side.
In each panel the yellow line represents the median relation from \citetalias{Gallazzi2021}, while the black lines identify the median (solid) and $\rm 16^{th}$ and $\rm 84^{th}$ percentiles (dashed) of the distribution obtained for the given model version.

\subsubsection{Dust treatment}\label{subsubsec:Dust_treatment}
With respect to \citetalias{Gallazzi2005} and \citetalias{Gallazzi2021} we introduced changes in the dust treatment both in the SPS modelling and in the estimation method.
CSP models in the prior distribution are created with a given dust content, whose  attenuation is obtained with the two-component model of \cite{Charlot2000}.
In our bayesian framework the photometry is used jointly with absorption indices to estimate the PDF of attenuation, as well as of the stellar populations parameters.
\citetalias{Gallazzi2005} and \citetalias{Gallazzi2021}, instead, employed dust-free models.
They estimated the colour excess from the difference between the $r-i$ colour of the galaxy and the $r-i$ colour of the dust-free model.
Then they estimated the \textit{z}-band attenuation $A_z$ assuming a single power law ($\lambda^{-0.7}$) attenuation curve \citep{Charlot2000}\footnote{
See section 2.4.3 of \cite{Gallazzi2005} for more details.
}.

To analyse the effect of this difference for the dust treatment we compare the \citetalias{Gallazzi2021} with the Exp\_FixZ\_BC03\_STELIB model version, which differs from the \citetalias{Gallazzi2005} and \citetalias{Gallazzi2021} only for the dust treatment.
We find notable differences both in the mass-age (Fig. \ref{fig:Comparison_models}, panel \emph{a}) and mass-metallicity (Fig. \ref{fig:Comparison_models}, panel \emph{b}) planes.
In the mass-age plane, for stellar masses below $10^{10}~\Msun$, the Exp\_FixZ\_BC03\_STELIB version produces higher estimates for the light-weighed ages, with differences up to $\sim0.3~\dex$.
For higher masses, the \citetalias{Gallazzi2021} estimates higher ages with maximum difference of $\sim0.15~\dex$ at $M_\star\sim10^{11}~\Msun$.
With increasing galaxy mass the differences get smaller, becoming negligible at the high-mass end.
In the mass-metallicity plane the Exp\_FixZ\_BC03\_STELIB version predicts higher stellar metallicities with respect to \citetalias{Gallazzi2021}, with differences up to $\sim0.2~\dex$ in the low-mass end.
The two relations have similar shapes, but the Exp\_FixZ\_BC03\_STELIB predicts a relation that is flatter at the low-mass end and steeper at the high-mass end.
All in all, the different dust treatment and the inclusion of photometry as an actual constraints affect the age and metallicity estimates in a mass-dependent way.

\subsubsection{Stellar spectral Libraries}\label{subsubsec:Stellar_libraries}
To assess the impact of the adopted stellar spectral libraries, we compare the outcomes of the inferences of the Exp\_FixZ\_BC03\_STELIB model, which is based on BC03 SSPs built on the STELIB stellar library \citep{LeBorgne2003}, with the results of the Exp\_FixZ\_BC03\_MILES model, based on the BC03 SSPs built on the MILES stellar library \citep{Sanchez-Blazquez2006}\footnote{Note that the BC03 MILES based SSPs are produced with the 2016 version of the BC03 code, which however has negligible differences in terms of optical indices with respect to the original BC03 code.}.

Before analysing the scaling relations, it is worth considering how the two sets of SSP tracks used to generate the CSP model libraries (dubbed ``BC03\_STELIB'' and ``BC03\_MILES'', respectively) differ in the index-index planes.
\begin{figure}
\centering
\includegraphics[width=\hsize]{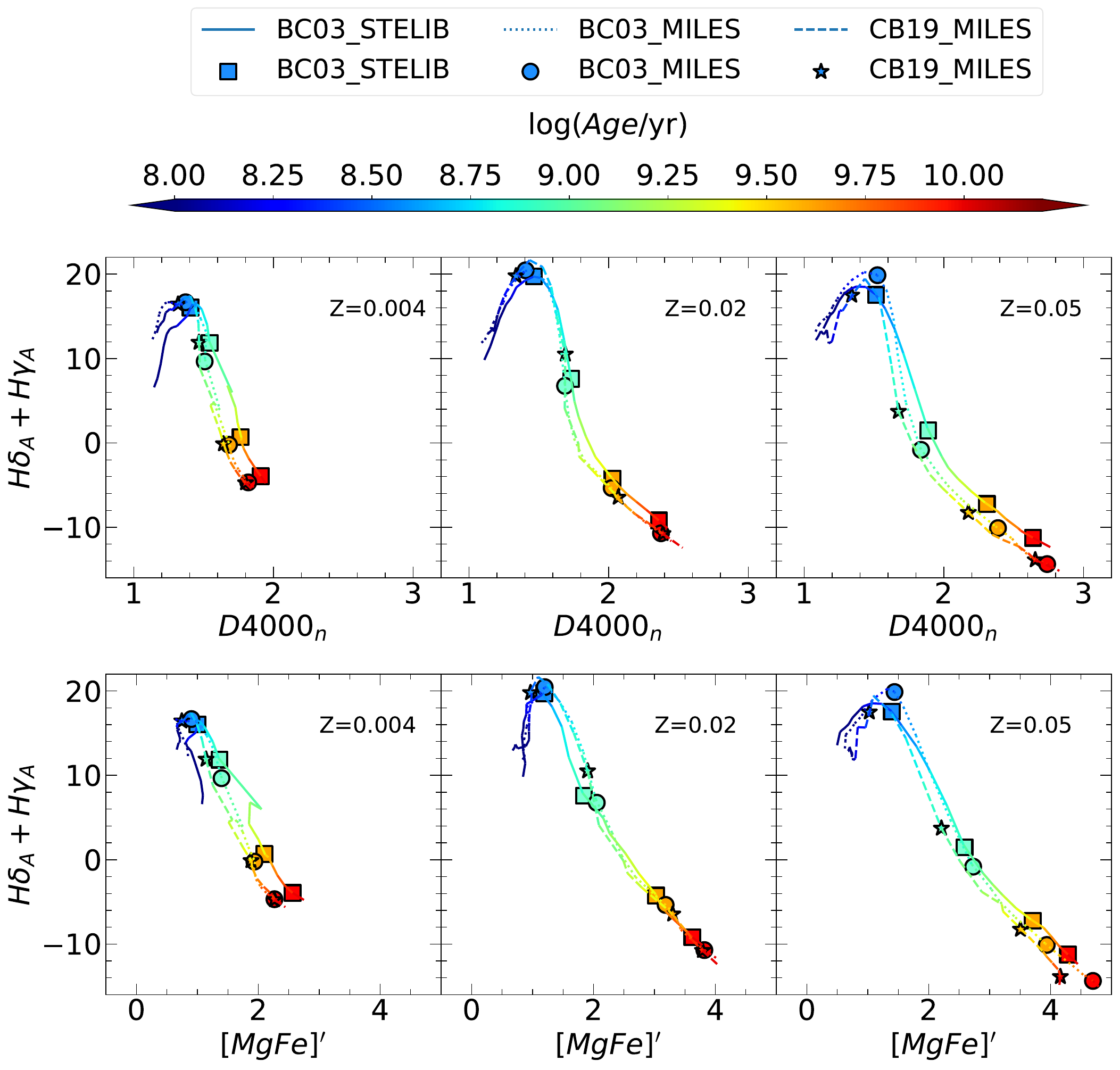}
  \caption{Evolutionary tracks in the \hdg~vs \dn~and \hdg~vs \mgfep~index planes for the BC03\_STELIB (solid), BC03\_MILES (dotted), and CB19\_MILES (dashed) SSPs (see table \ref{tab:SPS_models2} for more details).
  Left, central and right panels are associated to SSPs with different stellar metallicities: sub-solar ($Z_\star=0.004$), approximately solar ($Z_\star=0.02$), and super-solar ($Z_\star=0.05$ for BC03\_STELIB and BC03\_MILES, $Z_\star=0.05$ fro CB19\_MILES) respectively.
  The tracks are colour-coded by the age of the SSP, and the points plotted onto the tracks identify ages of $10^{8.5}~\yr$, $10^{9}~\yr$, $10^{9.5}~\yr$, and $10^{10}~\yr$, corresponding to different SSP models, squares for BC03\_STELIB, circles for BC03\_MILES, and stars for CB19\_MILES.}
    \label{fig:SSP_tracks}
\end{figure}
Figure \ref{fig:SSP_tracks} displays different SSP tracks in the \hdg~vs \dn~and \hdg~vs \mgfep~indices planes.
The BC03\_STELIB and BC03\_MILES SSP tracks are represented by the solid and dotted lines respectively.
The left, central, and right panels are associated to metallicities of $Z_\star=0.004$, $Z_\star=0.02$, and $Z_\star=0.05$ respectively.
The colour code along the SSP tracks identifies the age of the stellar population; furthermore cyan, yellow, and red symbols identify ages of $10^9~\yr$, $10^{9.5}~\yr$, and $10^{10}~\yr$, squares for BC03\_STELIB and circles for BC03\_MILES.

The BC03\_STELIB and the BC03\_MILES SSP tracks display significant differences especially at the lowest metallicities.
Furthermore, there are significant age offsets even when the tracks overlap.
Looking at the symbols as a reference, it is apparent that, at fixed index values, the BC03\_MILES SSPs predict lower ages with respect to the BC03\_STELIB ones, with stronger differences for older stellar populations.
Where the tracks do not overlap, for a given value of \hdg, the BC03\_MILES is often located at lower values of \mgfep, which are associated to more metal poor stellar populations.
Therefore, for a given observed value of the absorption indices, we expect BC03\_MILES model to deliver younger ages and higher metallicities than BC03\_STELIB.

The scaling relations normalised in mass bins obtained with the models implementing the STELIB and MILES stellar libraries show noticeable differences both in the mass-age (Fig. \ref{fig:Comparison_models}, panel \emph{c}) and mass-metallicity (Fig. \ref{fig:Comparison_models}, panel \emph{d}) planes.
The SP parameters obtained with the Exp\_FixZ\_BC03\_MILES model library display a systematic reduction in stellar age estimates for masses above $\sim10^{9.5}\,\Msun$, with respect to Exp\_FixZ\_BC03\_STELIB.
In the mass-metallicity plane we find an increase of the metallicity for masses $\mathrm{M}_\star\lesssim10^{10}\,\Msun$, with differences up to $\sim0.25~\dex$.
While systematic differences in SSPs are not trivially transposed to CSPs and, in turn, to SP parameter estimates, the most significant differences in the scaling relations mirror the most outstanding differences in the SSPs.
The largest age differences are found for stellar populations older than $\sim10^{9.3}~\yr$, while the largest metallicity differences affect metal poor populations at the low-mass end.

\subsubsection{Star formation history}\label{subsubsec:Star_formation_history}
Another substantial change we introduced in this work, with respect to \citetalias{Gallazzi2021}, concerns the Star Formation History (SFH).
Both the SFHs used in this work and in \citetalias{Gallazzi2021} are composed by a continuous component superimposed with random bursts of star formation to simulate stochasticity and rejuvenation phenomena.
\citetalias{Gallazzi2021} adopts a decaying exponential $\mathrm{SFR}(t)\propto\exp(-\gamma t)$ as continuous component, characterized by two parameters: $t_\mathrm{form}$ identifying the look-back time at which the SFH starts, and $\gamma$, the characteristic e-folding time.
In this work, the continuous component is modelled with a delayed Gaussian function \citep[][see eq. \ref{eq:Sandage_profile}, sec. \ref{subsec:Stellar_Population_Synthesis}]{Sandage1986, Gavazzi2002a}. 
In addition to purely declining SFHs, this formalism allows for a phase of increasing SFR, whose extent is parametrised by the characteristic timescale $\tau$ in relation to the formation time $t_\mathrm{form}$.
This results in more flexibility and a more extended coverage of the stellar population parameter space.

With the implementation of a new SFH we also extended the maximum $t_{\rm form,lb}$, reaching values up to $20~\Gyr$ look-back time.
This extension above the cosmologically motivated $14~\Gyr$ limit has been introduced to avoid skewing the age posterior of very old galaxies, which would have a hard boundary at the maximum $t_{\rm form,lb}$ for the old tail and practically no limitation for the young tail.

Using the Exp\_FixZ\_BC03\_MILES as a starting point, we replaced the exponential continuous component with the Sandage, while keeping all other ingredients fixed, and created a new model library called San\_FixZ\_BC03\_MILES.
With the stellar population parameters inferred with this new library, we analysed the systematic effect of the SFH relative to Exp\_FixZ\_BC03\_MILES.
In the scaling relations, the implementation of this new SFH has significant effects regarding the age estimates and little importance for stellar metallicity estimates (shown in panels \emph{e} and \emph{f} of figure \ref{fig:Comparison_models}).
The new age estimates result in lower light-weighted ages for masses between $\sim10^{9.2}~\Msun$ and $\sim10^{11}~\Msun$.
For $M_\star\gtrsim10^{10.5}~\Msun$ the median relation steepens, reaching the same median light-weighted ages of the Exp\_FixZ\_BC03\_MILES model in the high-mass end.
We also see that massive old galaxies have a broader age distribution, reaching higher light-weighted ages when compared to the estimates obtained with the previous implementation of the SFH. 
We argue that these difference could be due, at least partly, to the removal of the prior-induced bias, thanks to the extension of the maximum $t_{\rm form,lb}$ up to $20~\Gyr$.
This mass-dependent changes in the slope of the median relation is connected to the formation of a bimodal mass-age distribution, characterised by two separate sequences of young and old galaxies with a transition in the mass range $10^{10}-10^{11}~\Msun$.
The change in the SFH impacts only weakly the mass-metallicity relation, causing a shift of $\lesssim 0.1~\dex$ towards higher metallicities for masses above $10^{10}~\Msun$.

\subsubsection{Chemical enrichment history}\label{subsubsec:Chemical_enrichment_history}
The reference CSP library adopted in this work implements a parametric CEH (see eq. \ref{eq:Chemical_Enrichment_History}, sec. \ref{subsec:Stellar_Population_Synthesis}), which incorporates a dependence on the formed stellar mass, thereby establishing a connection with the SFH.
As opposed, \citetalias{Gallazzi2021} assumes a constant metallicity along the SFH, randomly drawn from a logarithmic distribution in the range $0.2-2.5~\Zsun$.
To assess the systematics induced by the implementation of this different CEH, we compare the outcome of the inference obtained with the San\_FixZ\_BC03\_MILES model and the model implementing the new parametric CEH, which will be called San\_VarZ\_BC03\_MILES and correspond to the libraries adopted in \cite{Zibetti2017,Zibetti2020,Zibetti2022}.

The implementation of the variable CEH has significant effects both in the age and metallicity estimates (as shown in panels \emph{g} and \emph{h} of figure \ref{fig:Comparison_models}).
In the mass-age plane we observe a strong increase of light-weighted ages of massive galaxies with old stellar populations.
This results in the shift of the old sequence around $Age_\mathrm{wr}\gtrsim10^{9.9}~\yr$ at stellar masses above $10^{10.5}~\Msun$.
On the other hand the galaxies populating the young sequence do not change significantly their age estimates.
These changes altogether enhance the bimodality in the mass-age distribution.
In the mass-metallicity plane, we detect an increase in stellar  metallicity for $M_\star\leq10^{11}~\Msun$ with increasing differences up to $\sim0.1~\dex$ with decreasing galaxy mass.
Consequently, these variations in metallicity estimates lead to a flattening of the MZR at lower masses.

The increase in light-weighted stellar metallicities can be understood considering the different contribution to the absorption lines of stars of different ages and the fact that the assumed CEH imposes older stars to be metal poorer. With respect to a model with fixed metallicity $Z_\star$, a model with variable metallicity having the same mean light-weighted metallicity $Z_\star$, would produce weaker absorption lines, as the cooler stars responsible for the strongest absorption are bound to be at lower metallicity than the hotter stars. Reversely, at given index strength, a model with variable metallicity would need higher mean light-weighted metallicity than a fixed-metallicity model to reproduce the absorptions.

The changes in age estimates can be seen as a side effect of the metallicity estimates caused by the classic age-metallicity degeneracy and by the CEH being linked to the SFH.

\subsubsection{Evolutionary tracks}\label{subsubsec:Evolutionary_tracks}

The latest version (CB19) of the \cite{Bruzual2003} SSP models implements the new PARSEC \citep{Bressan2012, Marigo2013a, Chen2015} stellar evolutionary tracks.
The CB19 SSPs extend the maximum metallicity, reaching up to $Z_\star=0.06$, compared to the BC03, which are limited to $Z_\star=0.05$.
Note that the PARSEC evolutionary tracks have a maximum age of $14\,\Gyr$, as opposed to $20\,\Gyr$ for the Padova94 tracks. In order for the SFHs to maintain the maximum $t_{\rm form,lb}$ to extend to $20\,\Gyr$ (see previous section), we slightly modify the synthesis of the CSPs by moving the weights of all components with age larger than $14\,\Gyr$ to the $14\,\Gyr$ component.
To assess the impact of the implementation of these updated evolutionary tracks we compare the outcome of the inference obtained with the Padova94-based San\_VarZ\_BC03\_MILES model library with the new San\_VarZ\_CB19\_MILES PARSEC-based library.

Before analysing the differences in the scaling relations, we compared the different SSPs in the indices planes shown in Fig. \ref{fig:SSP_tracks}.
The BC03\_MILES and CB19\_MILES SSPs are identified in the figure by the dotted and dashed lines, respectively.
Note that in the right panels the comparison for the SSPs at super-solar metallicity is not exact, as the same metallicity is not available for both SSPs. Therefore, we chose to show the comparison for the closest pair of super-solar metallicity, namely $Z_\star=0.05$ for BC03\_MILES, and $Z_\star=0.04$ for CB19\_MILES.

The two SSP models are in substantial agreement in terms of coverage of the index planes for $Z_\star \lesssim \mathrm{Z_\odot}$, while somewhat larger discrepancies appear at super-solar metallicity (note however the non-perfect match).
Looking at the symbols, it is notable that for stellar populations younger than $\sim10^9~\yr$, for a given combination of index values the CB19\_MILES SSPs imply higher ages by $\sim 0.2\,\dex$ with respect to the BC03\_MILES.
For older stellar populations we observe negligible age differences ($\lesssim 0.1$\,dex).
At $Z_\star \gtrsim \mathrm{Z_\odot}$ the tracks are offset with respect to each other, so in this regime we can expect systematic effects both in age and in metallicity.

To assess the systematics due to the change of the evolutionary tracks we analyse the mass-age and mass-metallicity relations obtained with the San\_VarZ\_CB19\_MILES model (shown in panels \emph{i} and \emph{j} of figure \ref{fig:Comparison_models}).
The most significant effects in the mass-age relation are mostly at the high-mass end, causing a shift of $\sim0.15~\dex$ towards lower light-weighted ages for old ($Age_\mathrm{wr}\gtrsim10^{9.75}~\yr$) and massive galaxies ($M_\star>10^{11}~\Msun$)\footnote{
Note that, because of the different age extension of the two SSP grids and the way we modified the SP synthesis (see above), the prior in light-weighted mean age of the two libraries does not match exactly, with a possible bias towards younger ages for CB19.
}
In the MZR the effects are generally negligible, except for a slight general increase in metallicity of the order of a few $0.01\,$dex and a similar reduction of the scatter at the high-mass end.

\subsection{Goodness of fits for different CSP model libraries}
\begin{figure}
  \centering
  \includegraphics[width=\hsize]{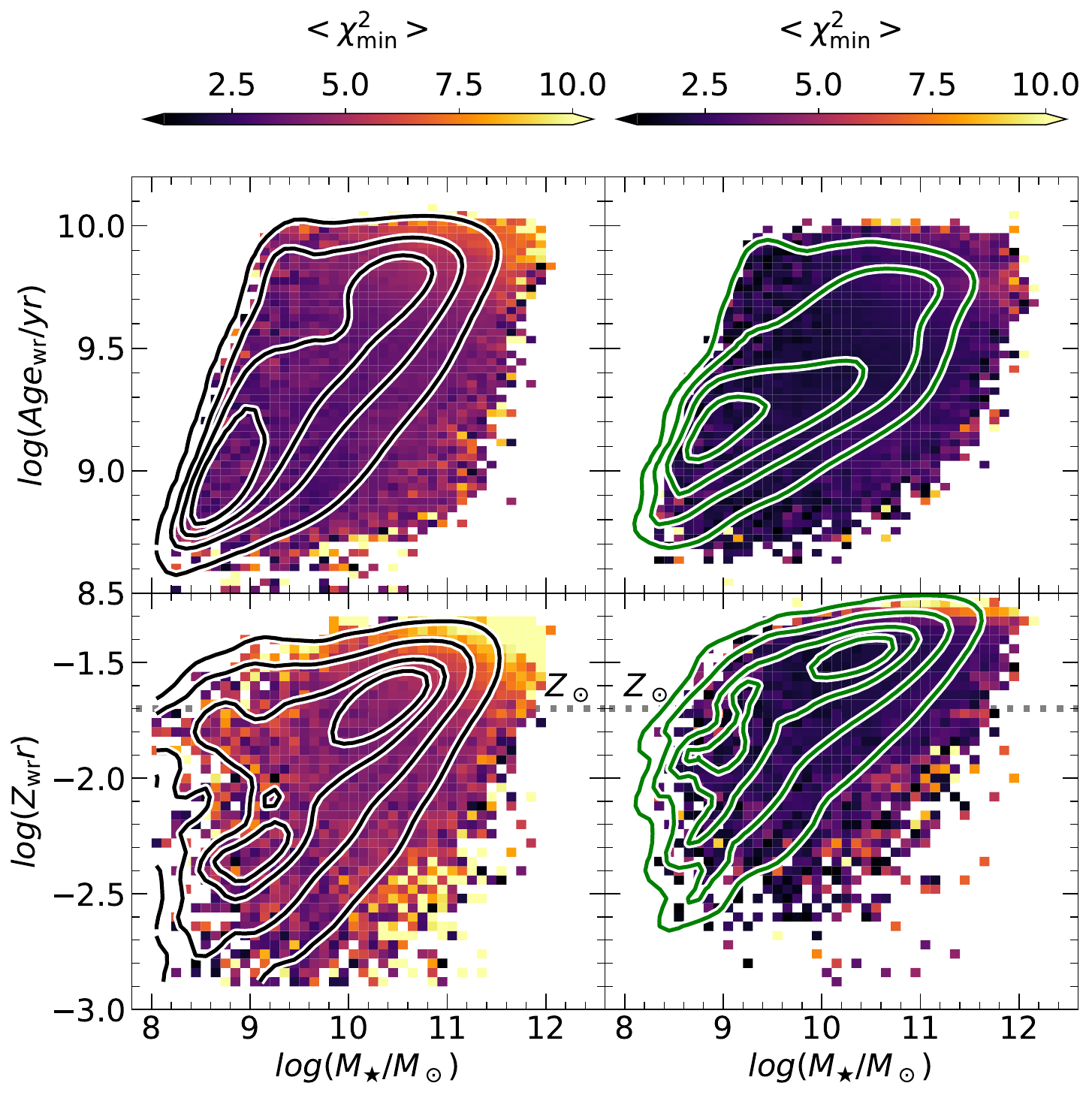}
    \caption{Mass-age and mass metallicity distributions colour-coded by the median value of the $\chi^2_\mathrm{min}$ for estimates based on the Exp\_FixZ\_BC03\_STELIB model library (same as in G05 and G21, \emph{left panels}) and the SanCB19\_MILES model library (reference models in this work, \emph{right panels}).  See Table \ref{tab:SPS_models2} for more details about the SPS models.
    The black and green lines identify the density levels enclosing $16\%$, $50\%$, $84\%$, and $97.5\%$ of the total integral of corresponding distribution.
    The dotted horizontal lines in the mass-metallicity planes identify the solar metallicity $Z_\odot\equiv0.02$.}
      \label{fig:chi2_distributions}
\end{figure}
We analyse the $\chi^2$ distribution for the fit performed with the Exp\_FixZ\_BC03\_STELIB and San\_VarZ\_CB19\_MILES models to analyse if the model changes introduced in this work allow us to fit the data more accurately.
In Fig. \ref{fig:chi2_distributions} we show the maps of the median value of $\chi^2_\mathrm{min}$ (i.e. the $\chi^2$ of the best fitting model in a maximum likelihood sense), for the Exp\_FixZ\_BC03\_STELIB (left panels) and the San\_VarZ\_CB19\_MILES (right panels) model versions, in the mass-age plane (top row) and mass-metallicity plane (bottom row)

The contours identify the isodensity levels of the galaxy sample ($\mathrm{SNR}\geq10$ and $\mathrm{SNR}\geq20$ for age and metallicity, respectively).
We see that the Exp\_FixZ\_BC03\_STELIB model produces best fits with  $\langle\chi^2_\mathrm{min}\rangle\sim4$ for the majority of galaxies, without any correlation with the physical parameters of the galaxies.
When implementing the model changes analysed in previous sections, we observe a reduction of the $\chi^2_\mathrm{min}$, obtaining $\langle\chi^2_\mathrm{min}\rangle\sim2$.
We also note that both models show an increase of the $\chi^2_\mathrm{min}$ for galaxies near the edges of the distributions, especially in the mass-metallicity plane.
However, the San\_VarZ\_CB19\_MILES models outperform the old models in terms of quality of the fit especially near the edge of the distribution, by significantly lowering the typical $\chi^2_\mathrm{min}$ from $\sim 10$ to $\sim 5$ at the high-metallicity end.

In conclusion, the new CSP model library adopted in this work result in an overall improved fit to the data.
It is worth noting that, considering nine degrees of freedom (5 from absorption indices plus 5 from photometric bands minus 1 to allow for the normalization) the obtained $\chi^2_\mathrm{min}$ values correspond to reduced $\chi^2$ well below 1.
This indicates that our models actually overfit the data.
Therefore, the uncertainties in the derived physical parameters are not driven mainly by observational errors data, rather they mostly stem from intrinsic degeneracies among the observed galaxy properties and the wide variety of CSPs that are able to reliably fit the data.

\section{Discussion}\label{sec:Discussion}
In this work we have estimated stellar masses and mean light- (and mass-) weighted ages for a sample of $354~977$ SDSS DR7 galaxies in the Local Universe characterized by a spectroscopic $\mathrm{SNR}\geq 10$. For a subsample of $89~852$ galaxies with spectroscopic $\mathrm{SNR}\geq 20$ we also estimated mean light- (and mass-) weighted stellar metallicities. We performed a careful statistical characterization of the samples, which has allowed us to compute the statistical weights to account for selection and incompleteness biases and obtain a volume-complete representation of the galaxy population. For the first time, we have implemented corrections to the spectroscopic properties of galaxies in order to mitigate the long-standing problem of SDSS fibre-aperture biases. Finally, in order to improve the accuracy in the stellar population parameter estimates, we employed more refined prescriptions for the CSP model libraries and adopted updated/state-of-the-art SSP models.

Based on this stellar population analysis, we have been able to characterize the distributions of local Universe's galaxies in the parameter space defined by stellar mass, age and metallicity, and made a preliminary exploration on their dependence on current SFR.
In this section we discuss our results in the context of our current knowledge of galaxy evolution.

\subsection{Importance of changes in SPS models}
Comparing the scaling relations derived using different SPS models we analysed the systematic in the scaling relations effects induced by variations in individual model ingredients (see Sec. \ref{sec:Models_comparison} and Fig. \ref{fig:Comparison_models}).

We found that the most significant changes in the mass-age relation arise from variations in the SFH, CEH, and evolutionary tracks.
The adoption of a more refined SFH, which allows for a rising SFR and results in a better coverage of the observable space (i.e. the index-index space), has the strongest effect.
Most notably, these changes lead to an enhancement of the mass-age bimodality, due to the presence of young galaxies for masses up to $\sim10^{11}~\Msun$, which, in turn, determines a coexistence of the young and old sequences in the mass range $10^{10}-10^{11}~\Msun$.

The updates in the evolutionary tracks and the implementation of a CEH caused significant changes in the mass-age distribution.
The estimates obtained with the new CEH prescriptions result in a further enhancement of the age bimodality.
On the other hand, with the updated PARSEC evolutionary tracks,  we infer smaller ages for massive and old galaxies. Both effects have potentially strong implications in terms of galaxy evolution, especially concerning the quenching mechanisms (connected to the age bimodality) and the evolutionary time scales of the most massive galaxies.
In the mass-metallicity plane, we found that the different ingredients implemented in the models have a significant impact and are strongly interlaced in affecting the MZR.
The combination of all these updates in the model ingredients results in a shift of the relation towards higher metallicities, which is highly relevant for our understanding of galaxy chemical enrichment.

\subsection{Importance of aperture effects corrections}
The SDSS spectroscopic data were obtained through $3''$-diameter optical fibres centred on the centre of the target galaxies, extracting spectra from a (more or less, depending on the galaxy projected size) extended inner region.
Since galaxies exhibit radial gradients in their stellar population properties, with a prevalence of older and more metal-rich stars in the centre and younger and more metal-poor stars in the outskirts \citep{Perez2013, Sanchez-Blazquez2014, GonzalezDelgado2014, GonzalezDelgado2015, Zibetti2017, Zibetti2020, Martin-Navarro2018, Goddard2017, Li2018, Johnston2022}, this selective sampling may result in a bias in the measurement of spectral properties.
This is a well-known observational bias which is generally known as ``aperture effects'' \citep{Kauffmann2003, Brinchmann2004, Gallazzi2005, Gallazzi2008, Peng2015, Trussler2020}.
To correct this bias, in this work for the first time we implemented the corrections on the observed absorption indices calculated by \cite{Zibetti2025} which were derived using IFS CALIFA observations.

The resulting reduction of both light-weighted stellar ages and metallicities (Fig. \ref{fig:Scaling_relations_comparison_fluxes}) is qualitatively consistent with the expectations from a predominance of negative gradients in stellar ages and metallicities.
Quantitatively we found that the strength of the aperture corrections varies with galaxy mass, becoming more significant at lower masses (Fig. \ref{fig:Scaling_relations_comparison_fluxes}).
This mass-dependency stems from a mixture of different causes.
On the one hand, the gradients are mass-dependent; on the other hand, they are linked also to galaxy morphology (which shows a connection with mass, see e.g. \citealp{Nair2010}).
Furthermore the amplitude of the aperture corrections depends on the relative coverage of the SDSS spectroscopic fibres which, in turn,  is determined by the galaxy apparent size \citep[see ][]{Zibetti2025}.
Since more massive galaxies are typically observed at greater distances, their higher fibre coverage fraction compared to closer low-mass galaxies may further contribute to this effect.
In addition, we observe an inverse effect in the mass-age plane for extremely young and low-mass galaxies, which increase both their light-weighted ages and stellar masses.
This is in agreement with gradients observed by \cite{Perez2013} and \cite{Johnston2022}, which revealed that lower mass galaxies can experience an outside-in growth, with younger population located in the central part of the galaxy.

Comparing the galaxy properties estimates obtained with and without the implementation of aperture corrections we find qualitatively similar scaling relations, with a bimodal mass-age relation and a unimodal MZR of increasing age and metallicity estimates with increasing galaxy mass.
However, when accounting for the aperture bias, we observe significant \emph{quantitative} changes both in the populations of the sequences of the mass-age plane (top panels of Fig. \ref{fig:Scaling_relations_comparison}), and in the slope of the MZR (bottom panels of Fig. \ref{fig:Scaling_relations_comparison}).
As we will discuss in more details in the next sections, the changes due to the implementation of the aperture corrections impact theoretical models of galaxy evolution and chemical enrichment.
In particular, the enhancement of the population of galaxies in the young sequence may suggest that the efficiency of feedback processes may be less than what implied by previous stellar population analyses.

\subsection{Constraints of galaxy quenching from the age bimodality}\label{subsec:Constraints_quenching_from_age_bimodality}
The distribution of galaxies in the mass-age plane is composed by two separate sequences of galaxies characterised by young ($Age_\mathrm{wr}\lesssim10^{9.3}\,\yr$) and old ($Age_\mathrm{wr}\sim10^{9.8}\,\yr$) stellar populations, as already shown in the Local Universe \citep[e.g.,][]{Gallazzi2005, Peng2015, Sanchez2018, Trussler2020, Sanchez2020}.
The former span the mass range from the lower completeness limit ($\sim10^9~\Msun$) to $10^{11}\,\Msun$, while the latter extends up to $10^{12}\,\Msun$.
These two sequences coexist and have similar population in the mass range $10^{10}-10^{11}\,\Msun$, defining a strongly bimodal age distribution.
This bimodality is reminiscent of and connected to well-known bimodalities already observed in other galaxy properties, such as colour \citep[e.g., ][]{Kauffmann2003a, Baldry2004, Schawinski2007, Schawinski2014}, and SFR \citep[e.g., ][]{Renzini2015}.

Our results support the idea of two distinct quenching pathways that depend on galaxy mass.
On the one hand, the predominance of old stellar populations at the high-mass end indicates the presence of a mass-dependent mechanism responsible for galaxy quenching at high masses and at high redshift.
The existence of a mass threshold beyond which galaxies evolve passively has been both pointed out observationally \citep{Kauffmann2003a, Gallazzi2005, Moustakas2013, Haines2017} and derived theoretically \citep{Dekel2006, Cattaneo2006, Peng2010}. 
This threshold is typically associated to the onset of processes that reduce or halt the inflow of gas from the IGM, ultimately shutting down star formation.
These mechanisms include inefficient accretion due to shock heating in massive halos \citep{Dekel2006, Cattaneo2006} and black hole feedback, which can either deplete the gas reservoir through AGN-driven winds, or prevent further accretion from the IGM \citep{DiMatteo2005, Croton2006a, Schawinski2007, Fabian2012, Cicone2014, Trussler2020, Bluck2020a, Bluck2020b}.
Nevertheless, the presence of a transition mass could also relate to other mechanisms capable of inducing quenching of star formation, even with the presence of a significant fraction of the molecular gas.
An example of such a mechanism is the ``morphological quenching'' \citep[e.g.,][]{Martig2009}, by which the morphological transformation of a galaxy can stabilise the disk against gas cloud fragmentation.
Moreover, resolved studies of star formation in Local Universe galaxies have shown that quenching is not a global process, but happens differently on several spatial scales (see review by \citealt{Sanchez2020} and references therein, also \citealt{Zibetti2022}).
Recent studies based on SFRs and molecular gas fractions ($f_{\rm mol}$) have shown that quenching may originate from a suppressed Star Formation Efficiency (SFE) in combination with a reduction in $f_{\rm mol}$ \citep{Saintonge2016, Saintonge2017, Colombo2020, Pan2024, Barrera2025}.

On the other hand, the coexistence of the young and old sequences in the mass-age plane at intermediate masses suggests that mass alone does not govern galaxy evolution.
This aligns with several studies in the Local Universe, which have shown that passive low-mass galaxies are predominantly satellites that experienced quenching due to environmental processes such as starvation, ram-pressure stripping, and galaxy mergers \citep{Larson1980, Balogh2000, vandenBosch2008, Pasquali2010, Schawinski2014, Gallazzi2021}.

Both theoretical and observational studies place this transition between star-forming and quenched galaxies at stellar masses of $\sim10^{10.5}\,\Msun\sim3\cdot10^{10}\,\Msun$ in the Local Universe.
In this work we obtain an estimate of the transition mass between young and old galaxies of $M_\mathrm{tr}=10^{10.65}\,\Msun$ (neglecting aperture-corrections), which is $0.15\,\dex$ (a factor of $\sim1.5$) higher than previous Local Universe estimates.
When accounting for the aperture bias the age estimates decrease for most of the galaxies in the sample.
As a result the transition mass further increases, reaching $M_\mathrm{tr}=10^{10.80}\,\Msun\sim6\cdot10^{10}\,\Msun$ (Fig. \ref{fig:Transition_mass_statistics}).
This upward shift in the transition mass serves as a crucial observational constraint for accurately modelling mass-driven galaxy quenching in theoretical frameworks of galaxy formation and evolution \citep[e.g.,][]{Kauffmann1996, DeLucia2006, Peng2010, Schaye2015, Pillepich2018}.
The fact that our aperture-corrected relations yield a higher transition mass than estimated in the past suggests that theoretical models of galaxy evolution should incorporate weaker quenching, possibly by reducing the efficiency of AGN feedback.

Our analysis of the young and old galaxy sequences shows that the old sequence is already populated at masses as low as $M_\star\sim10^{9.5}\,\Msun$, with approximately $10\%$ of all the galaxies in this mass range lying above the separation (Fig. \ref{fig:Scaling_relations_comparison} and \ref{fig:Transition_mass_statistics}).
Since most of the passive galaxies at these low masses are indeed satellites \citep{Larson1980, Balogh2000, vandenBosch2008, Pasquali2010, Schawinski2014, Gallazzi2021}, this result implies that, if environmental quenching is responsible for moving low-mass galaxies onto the old sequence, it must affect only a small fraction of the sample (less than $20\%$ for $M_\star<10^{10}\,\Msun$).
Conversely, if environmental quenching influences a significant fraction of the galaxy sample, it must have happened recently to avoid the transition of low-mass galaxies from the young to the old sequence.
This has been further investigated by analysing the mass-age relation separating passive and star-forming galaxies (Fig. \ref{fig:Mass_age_passive_starforming}).
The presence of a tail of passive low-mass galaxies with relatively young ages of $\sim10^{9.4}~\yr\sim2.5~\Gyr$ suggests that the majority of low-mass passive galaxies have experienced environment quenching recently in the past, therefore they still populate the young sequence.

\subsection{Correspondence between light-weighted ages and SFR estimates}\label{subsec:Correspondence_LWage_sSFR}
To further investigate the SFH and possible quenching paths we analysed the mass-age relations dividing the galaxy sample between star-forming and passive using the measure of the sSFR, adopting the same selection criteria of \cite{Gallazzi2021}.
We showed that the majority of star-forming and passive galaxies populate the young and old sequences respectively (Fig. \ref{fig:Mass_age_passive_starforming}).
Therefore, sSFR measures and light-weighted age estimates, although probing very different timescales ($\sim10^7\,\yr$ and $\sim1\,\Gyr$ respectively), show a good correspondence for the majority of the galaxy sample.
In accordance with similar analyses by \cite{Peng2015} and \cite{Trussler2020}, this can be interpreted as an evidence of galaxy quenching happening on timescales significantly longer than $1\,\Gyr$.
Consistent results have also been found by \cite{Corcho-Caballero2023b,Corcho-Caballero2023a}, who used the plane of observable defined by \ha~vs $\rm D4000$. 
These works have shown that the \ha$-\rm D4000$ plane can be used to investigate how star-forming galaxies evolve into passive systems.
They found that galaxies with high \ha~equivalent widths (i.e. star-forming) generally have low $\rm D4000$ (young stellar populations), while passive galaxies are generally characterised by high $\rm D4000$ (old stellar populations).
This general agreement between \ha~equivalent widths and $\rm D4000$ is interpreted as an evidence of galaxies quenching through the so called ageing process \citep[see also, ][]{Casado2015}, which refers to a gradual decrease of the star formation activity on timescales longer than $1\,\Gyr$.

Despite the general agreement between SFR estimates and light weighted ages we also find notable exceptions.
On the one hand, star-forming galaxies display a tail of high mass galaxies characterised primarily by old stellar populations.
We speculate that these may be green-valley galaxies, which display intermediate properties between star-forming and passive systems \citep{Schawinski2014, Trussler2020}.
As they undergo the transition toward quenching, they may still retain some residual star formation activity.
Alternatively, some of these galaxies could be passive systems that have recently undergone rejuvenation, making them detectable as star-forming, though without significantly altering their light-weighted ages.
On the other hand we find a tail of passive galaxies with relatively young stellar populations (ages of $\sim10^{9.4}~\yr$)\footnote{Note that the low SFR estimate for passive--young galaxies could possibly stem from the underestimation of the aperture corrections for the SFR. However, this is unlikely to be the case for the majority of these galaxies as we checked that their distribution of SFR aperture corrections is similar to the one for star-forming galaxies. Moreover, we also found that the fraction of light in fiber for passive young galaxies is typically somewhat larger than for star-forming, thus making the aperture corrections for these galaxies possibly less important.}.
We suggest that these may be recently quenched galaxies, in accordance with the analysis of the ageing diagram by \cite{Corcho-Caballero2023b,Corcho-Caballero2023a}. \footnote{
Note, however, that in \cite{Corcho-Caballero2023b,Corcho-Caballero2023a} they refer to quenching when the transition between active star formation and passive evolution happens on timescales shorter than $1~\Gyr$, and to ageing otherwise. 
In this work we do not make such differentiation and we call quenching every process which stops star formation.
}

\begin{figure}
  \centering
  \includegraphics[width=\hsize]{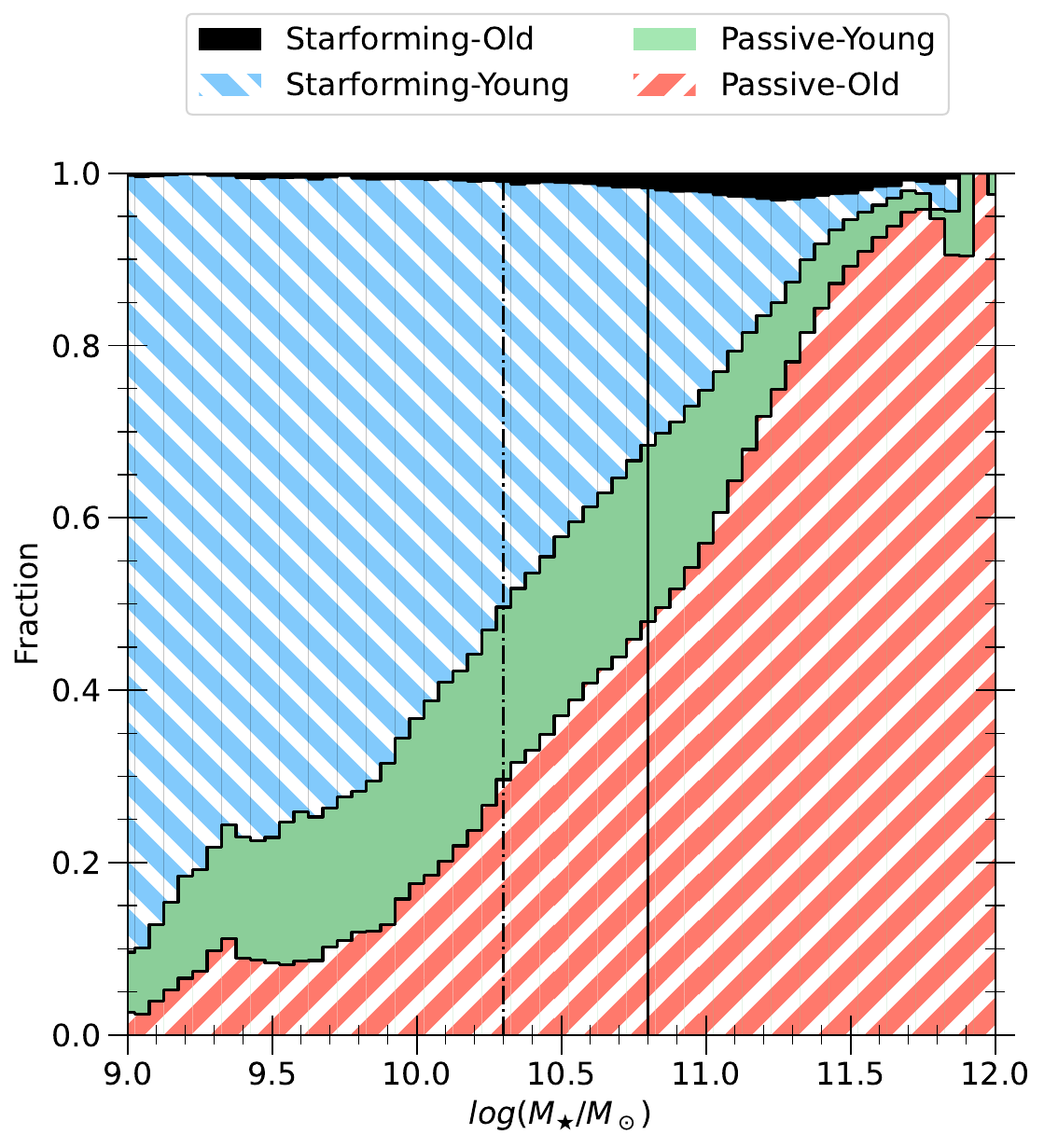}
    \caption{Fraction of galaxies in the four categories of age/SFR classification, based on the aperture-corrected physical parameter estimates: star-forming--old (black), star-forming--young (hatched blue), passive--old (hatched red), and passive--young (green)
    The vertical dot-dashed line identifies the transition mass between star-forming and passive galaxies, obtained from the mass-functions in the right panel of Fig. \ref{fig:Mass_function}.
    The vertical solid line identifies the transition mass between young and old galaxies, obtained as described in Sec. \ref{subsec:Transition_mass}.}
      \label{fig:Transition_mass_fractions}
\end{figure}

To investigate the nature of galaxy quenching we carried out a detailed analysis of the fraction of galaxies in each of the four categories defined based both on star formation activity and light-weighted ages. 
In figure \ref{fig:Transition_mass_fractions} we show the fraction of galaxies, obtained from the aperture corrected estimates, in each of the four categories: star-forming--old (black), star-forming--young (blue), passive--old (red), and passive--young (green).
We confirm that passive--young galaxies constitute a significant fraction of the passive sample at low masses.
In the low-mass end the fraction of passive--young galaxies is of $\sim20\%$ of the total sample and remains more or less constant with increasing galaxy mass, while the fraction of passive--old galaxies gradually increases.
For masses above the transition mass the fraction of passive--young galaxies decreases, becoming negligible at $M_\star\sim10^{11.5}~\Msun$.

The decrease of the fraction of passive--young galaxies above $M_\mathrm{tr}$ is connected to a more rapid increase of the passive--old fraction. This trend with $M_\star$ further supports the hypothesis of a change in the dominant quenching mechanism above the transition mass. In particular, the reduced fraction of passive--young galaxies indicates that the epoch of quenching is earlier or the timescale is shorter, or both. In addition the probability of rejuvenation might be reduced. We can interpret this as a manifestation of the mass quenching. For masses $M_\star\geq M_\mathrm{tr}$, the mass-dependent quenching becomes dominant already at $z\sim1$ \citep{Dekel2006, Cattaneo2006, Peng2010}, causing massive galaxies to evolve essentially passively.
As a result, this causes the steep increase in light-weighted ages for passive galaxies across the transition mass (see Fig. \ref{fig:Mass_age_passive_starforming}).
Below $M_\mathrm{tr}$, environmental quenching becomes more dominant \citep{Larson1980, Balogh2000, vandenBosch2008, Peng2010, Pasquali2010, Schawinski2014, Gallazzi2021}. It may either have happened recently, resulting in lower light-weighted ages, or could have happened at higher redshift, resulting in low-mass old and passive galaxies.

To further investigate galaxy quenching, in a follow-up paper we build on this work focusing on constraining timescales for galaxies SFHs, providing deeper insights into the physical processes that regulate galaxy evolution.

\subsection{Chemical evolution constraints from the mass-metallicity relation}
The mass-metallicity plane displays a clear unimodal relation between galaxy mass and stellar metallicity, with a trend of increasing mean stellar metallicity with increasing galaxy mass (Fig. \ref{fig:Scaling_relations_comparison}), in agreement with previous works in the local Universe \citep[e.g.,][]{Gallazzi2005, Gallazzi2021, Panter2008, Pasquali2010, Peng2015, Trussler2020, Zibetti2022, Looser2024} and extending towards higher redshifts \citep{Gallazzi2014, Cappellari2023, Nersesian2024, Gallazzi2025}.
At the low-mass end the relation is steep, but it becomes shallower as galaxy mass increases.
With our updated SPS modelling we obtained a shift of $\sim0.25\,\dex$ towards higher metallicities with respect to previous works from our group \citep[e.g.,][]{Gallazzi2005, Gallazzi2021}.
The implementation of corrections for the SDSS aperture effects resulted in a mass dependent decrease of stellar metallicities, with stronger effects for lower mass galaxies (Fig. \ref{fig:Scaling_relations_comparison} and \ref{fig:Scaling_relations_comparison_fluxes}).

The existence of a MZR has been interpreted as an evidence of galaxy downsizing \citep{Cowie1996, Gavazzi1996, Gallazzi2005, Panter2008, Somerville2015}, whereby higher mass galaxies have assembled at earlier cosmic epochs and more rapidly, resulting in more efficient star formation and chemical enrichment.
Another explanation for the shape of the MZR, which does not exclude downsizing, involves the presence of outflows which, depending on the galaxy gravitational potential, may cause losses of enriched gas.
In fact, observational studies of the gas MZR and supernova driven outflows in Local Universe galaxies have shown that supernovae can shape the MZR by expelling significant quantities of enriched gas \citep{Tremonti2004, Chisholm2018}.
These observational evidences suggest that more massive galaxies, thanks to their deeper gravitational wells, are less affected from metal rich outflows, reaching higher metallicities \citep{Spitoni2017, Barone2018, Barone2020, Vaughan2022}.
On the other hand, the low metallicity of gas and stellar content of low-mass galaxies may suggest that they are able to maintain a metal poor gas reservoir due to infall of low-metallicity gas from the IGM, despite the ability of supernovae-driven outflows to expel the enriched gas from the galaxy.
This is supported by the discovery of the dependence of the MZR on the star formation activity of the galaxy.
In fact several studies \citep[e.g.,][]{Ellison2008, Lara-Lopez2010, Mannucci2010} showed that more star-forming galaxies are located at lower median gas-phase metallicities with respect to less star-forming ones, defining the so-called Fundamental Metallicity Relation (FMR).  
Therefore, the normalisation and shape of the MZR result from a combination of several physical processes: metal removal via outflows; the infall of metal-poor gas from the IGM, leading to dilution of the enriched gas within galaxies; metal retention due to the gravitational potential; cosmic enrichment efficiency \citep[see review by][and references therein]{Maiolino2019}.

Our updated and aperture-corrected MZR, which exhibits significant differences in both normalisation and shape compared to previous Local Universe studies, provides crucial insights for accurately modelling galaxy evolution.
On the one hand, the shift towards higher metallicities at the high-mass end suggests that massive galaxies are more efficient at enriching the ISM than inferred from previous estimates of the stellar MZR, potentially requiring less efficient AGN driven outflows.
Conversely, the steeper slope observed at lower masses suggests that weaker chemical enrichment efficiencies are required in the low-mass regime (with significant implications on supernovae-driven outflows) and/or enhanced metal-poor gas infall are required to dilute the galaxy gas reservoir.

\subsection{Constraints on galaxy quenching from the mass metallicity relation}
As analysed in section \ref{subsec:Constraints_quenching_from_age_bimodality}, one of the possible explanation for galaxy quenching involves a reduction of
gas accretion due to the combination of shock heating of the gas accreting onto the dark matter halo \citep{Dekel2006, Cattaneo2006}, and black hole feedback \citep{DiMatteo2005, Croton2006a, Schawinski2007, Fabian2012, Cicone2014, Trussler2020}.
Lower mass galaxies may quench via starvation \citep{Larson1980, Balogh2000, vandenBosch2008}, or loose their gas reservoir due to supernovae-driven winds \citep[e.g.][]{Dalcanton2007} and ram pressure stripping \citep{Larson1980, Abadi1999}. 
Alternatively, mergers may result in morphological transformation either by disrupting the galaxy gas reservoir \citep[e.g.,][]{Schawinski2014}, or stabilising the disk against fragmentation, resulting in a lower SFE even in the presence of a significant fraction of molecular gas \citep[e.g.,][]{Martig2009}.

While both mergers and ram pressure stripping are associated to gas removal processes, starvation acts stopping the accretion of fresh gas from the IGM.
As pointed out by \cite{Peng2015}, gas removal and starvation leave different signatures in the MZR.
Analysing the MZR for passive and star-forming galaxies separately, they conclude that galaxy quenching is characterised mostly by a prolonged period of starvation.
More recently \cite{Trussler2020} supported this scenario invoking also the need of outflows to correctly reproduce the observed MZR.
On the contrary, \cite{Spitoni2017} retrieved the observed MZR by developing an analytical model implementing an exponential declining gas infall coupled with strong galactic winds, reproducing the effect of starvation.

When analysing the MZR for passive and star-forming galaxies separately, we found that star-forming and passive galaxies follow two distinct MZRs (Fig. \ref{fig:Mass_metallicity_passive_starforming}).
The two relations are separated at the low mass end by $\sim0.3\,\dex$, but get progressively closer and eventually merge for masses $M_\star>M_\mathrm{tr}$.
Nonetheless, star-forming galaxies always exhibit lower median metallicities than passive galaxies of the same mass, in agreement with \cite{Peng2015} and \cite{Trussler2020} results. The higher stellar metallicities of low-mass passive galaxies, coupled with their relatively young ages of $\sim 3$ Gyr, suggest that star formation has continued in a metal-enriched ISM for an extended period. From our MZR we also find that the scatter of the metallicity distribution for star-forming galaxies is always bigger than the one of passive galaxies of the same mass.
This may indicate that star-forming galaxies have more stochastic metallicity evolution.
Alternatively, recent results by \cite{Looser2024} suggest that the scatter may be due to the superposition of the MZR of galaxies with different sSFR.
These authors, based on MANGA observations, found that galaxies are located on different MZRs based on their sSFR, defining the so-called stellar Fundamental Metallicity Relation (stellar-FMR).
This stellar-FMR thus represent the analogous of the already observed gas-FMR \citep{Ellison2008, Lara-Lopez2010, Mannucci2010}.

In our work we also found that the MZRs of passive--young and passive--old galaxies display negligible differences, in contrast with the starkly different MZR of the star-forming ones (Fig. \ref{fig:Mass_metallicity_passive_starforming}). This shows that the chemical enrichment is more tightly connected to the current star-formation activity (timescale $\sim 10^7\,\yr$), rather than with the long-term SFH (timescale $\gtrsim 10^9\,\yr$).
A physical interpretation of this observation is not straightforward, as a direct causation appears unplausible considering that the stellar metallicity is a fossil record of the chemical enrichment over the entire evolutionary history of a galaxy. Apparently, no matter what is the timescale of the mechanism that leads to the suppression of the SFR (either starvation or abrupt/mass-driven quenching), the result of quenching is invariably an enhancement of the stellar metallicity. We will investigate this problem in further details by studying the MZR in finer bins of SFR in a forthcoming paper. Yet, one should keep in mind that observational uncertainties, particularly those related to the statistical nature of the aperture corrections and the limited size of the passive–young subsample, may result in a cross-contamination of the different subsamples by galaxies that are misclassified in terms of SFR or age, and could partly contribute to the observed trends.

\section{Summary and conclusions}\label{sec:Summary&Conclusions}
We have analysed the galaxies in the SDSS DR7 using state-of-the-art Stellar Population Synthesis modelling to estimate galaxy masses, light-weighted and mass-weighted ages and stellar metallicities.
We used statistical weights to account for biases induced by sample selections, and implemented novel corrections for SDSS aperture effects.
We summarise the main findings obtained by analysing the mass-age and mass-metallicity relations of our sample of galaxies as follows:
\begin{enumerate}
    \item In the mass-age plane, we obtained a distribution composed by two distinct sequences of young and old galaxies at ages of $\sim10^{9.3}\,\yr$ and $\sim10^{9.8}\,\yr$, respectively.
    Furthermore, such sequences show a coexistence in the mass range $10^{10}-10^{11}\,\Msun$, leading to the formation of a bimodal distribution (Fig. \ref{fig:Scaling_relations_comparison}).
    \item In the mass-metallicity plane, we obtained a clear correlation between galaxy stellar mass and stellar metallicity, as already found in several other works in the literature, in which lower-mass galaxies are characterised by lower values of stellar metallicity (Fig. \ref{fig:Scaling_relations_comparison}).
    Furthermore our metallicity estimations are shifted of $\sim0.25\,\dex$ towards higher metallicities, when compared to previous studies from the Local Universe such as \cite{Gallazzi2005, Gallazzi2021}, mainly due to our updated and refined state-of-the-art modelling for stellar population synthesis (bottom panels of Fig. \ref{fig:Comparison_models}).
    \item The implementation of novel aperture corrections to SDSS observations causes a reduction in stellar mass, age, and metallicity for most galaxies in our sample.
    In the mass-age plane, this leads to an enhancement of the population of galaxies in the young sequence at the expenses of the old one.
    In the mass-metallicity, there is a mass dependent decrease of the metallicity, with stronger effects for lower-mass galaxies, resulting in a not-uniform steepening of the relation (Fig. \ref{fig:Scaling_relations_comparison} and \ref{fig:Scaling_relations_comparison_fluxes}).
    \item We estimated a transition mass (across which the population changes from being dominated by young galaxies to being dominated by old galaxies) of $M_\mathrm{tr}=10^{10.80}\,\Msun$, which is $0.15~\dex$ higher than the value we would infer without accounting for aperture biases (Fig. \ref{fig:Transition_mass_statistics}).
    These estimates are higher than the usually quoted $10^{10.5}~\Msun$ in the literature \citep[e.g.,][]{Kauffmann2003a, Gallazzi2005, Peng2010, Haines2017, Gallazzi2021}.
    \item Analysing the mass-age scaling relation separately for passive and star-forming galaxies, we found that for the majority of the galaxy sample star-forming and passive galaxies correspond to galaxies characterised by young and old stellar population respectively (Fig. \ref{fig:Mass_age_passive_starforming}).
    When implementing aperture corrections we observe the presence of a significant fraction of passive galaxies with young stellar populations, which may have significant implication for satellite quenching.
    Similarly in the mass-metallicity plane, we confirmed results obtained in local Universe studies which showed that star-forming galaxies have lower median metallicities when compared to passive ones (Fig. \ref{fig:Mass_metallicity_passive_starforming}).
    When implementing aperture corrections the separation of the two MZRs is enhanced.
    \item Analysing each difference in the SPS ingredients we showed that each change causes noticeable effects (see sec. \ref{sec:Models_comparison}). 
    The most relevant ones for the mass-age relation can be ascribed to the SFH and CEH prescriptions, and choice of evolutionary tracks.
    For the MZR each model change acts differently, but with similar importance in the shaping of the MZR.
\end{enumerate}

Our work highlights the importance of controlling for systematic effects when comparing analyses on stellar populations conducted with different methods and models.
Our scaling relations will serve as a zero point for scaling relation studies at intermediate redshift ($0.3-1$), already ongoing \citep[e.g. LEGA-C][]{vanderWel2016} or forthcoming (e.g., WEAVE-StePS \citealp{Iovino2023a} and 4MOST-StePS \citealp{Iovino2023b}), and extending towards the cosmic noon with VLT-MOONS \citep{Cirasuolo2020a} and JWST.
As the best practice, evolutionary studies should be conducted with consistent models and methods. The stellar population scaling relations for the LEGA-C sample at $z\sim0.7$ have been already analysed in a fully consistent way with the same models and methodology as in this paper by \cite{Gallazzi2025}. We intend to extend the same approach to the forthcoming WEAVE-StePS and 4MOST-StePS and to the samples observed with VLT-MOONS in the next years.

Within this very same SPS analysis framework, we will also extend the present analysis of the aperture-corrected properties of SDSS galaxies to include more detailed information about their SFH and chemical enrichment patterns, in order to gain more insight on their evolutionary paths.

\section*{Data availability}
The table containing the observed and estimated properties of the duplicate-combined SDSS DR7 galaxies is available in electronic form at the CDS via anonymous ftp to cdsarc.u-strasbg.fr (130.79.128.5) or via \url{http://cdsweb.u-strasbg.fr/cgi-bin/qcat?J/A+A/}

\section*{Acknowledgments}
We thank the referee for useful comments which have improved the clarity of our manuscript.
This paper and related research have been conducted during and with the support of the Italian national inter-university PhD programme in Space Science and Technology.\\ S.Z. acknowledges support from the INAF-Minigrant-2023 "Enabling the study of galaxy evolution
through unresolved stellar population analysis" 1.05.23.04.01. A.R.G. acknowledges support from the INAF-Minigrant-2022 "LEGA-C" 1.05.12.04.01. L.S.D. is supported by the "Prometeus" project PID2021-123313NA-I00 of MICIN/AEI/10.13039/501100011033/FEDER, UE.

\bibliographystyle{aa}
\bibliography{Bibliografia}

\begin{thebibliography}{134}
\expandafter\ifx\csname natexlab\endcsname\relax\def\natexlab#1{#1}\fi

\bibitem[{Abadi {et~al.}(1999)Abadi, Moore, \& Bower}]{Abadi1999}
Abadi, M.~G., Moore, B., \& Bower, R.~G. 1999, MNRAS, 308, 947

\bibitem[{{Abazajian} {et~al.}(2004){Abazajian}, {Adelman-McCarthy},
  {Ag{\"u}eros}, {Allam}, {Anderson}, {Anderson}, {Annis}, {Bahcall}, {Baldry},
  {Bastian}, {Berlind}, {Bernardi}, {Blanton}, {Bochanski}, {Boroski},
  {Briggs}, {Brinkmann}, {Brunner}, {Budav{\'a}ri}, {Carey}, {Carliles},
  {Castander}, {Connolly}, {Csabai}, {Doi}, {Dong}, {Eisenstein}, {Evans},
  {Fan}, {Finkbeiner}, {Friedman}, {Frieman}, {Fukugita}, {Gal}, {Gillespie},
  {Glazebrook}, {Gray}, {Grebel}, {Gunn}, {Gurbani}, {Hall}, {Hamabe},
  {Harris}, {Harris}, {Harvanek}, {Heckman}, {Hendry}, {Hennessy}, {Hindsley},
  {Hogan}, {Hogg}, {Holmgren}, {Ichikawa}, {Ichikawa}, {Ivezi{\'c}}, {Jester},
  {Johnston}, {Jorgensen}, {Kent}, {Kleinman}, {Knapp}, {Kniazev}, {Kron},
  {Krzesinski}, {Kunszt}, {Kuropatkin}, {Lamb}, {Lampeitl}, {Lee}, {Leger},
  {Li}, {Lin}, {Loh}, {Long}, {Loveday}, {Lupton}, {Malik}, {Margon},
  {Matsubara}, {McGehee}, {McKay}, {Meiksin}, {Munn}, {Nakajima}, {Nash},
  {Neilsen}, {Newberg}, {Newman}, {Nichol}, {Nicinski}, {Nieto-Santisteban},
  {Nitta}, {Okamura}, {O'Mullane}, {Ostriker}, {Owen}, {Padmanabhan},
  {Peoples}, {Pier}, {Pope}, {Quinn}, {Richards}, {Richmond}, {Rix}, {Rockosi},
  {Schlegel}, {Schneider}, {Scranton}, {Sekiguchi}, {Seljak}, {Sergey},
  {Sesar}, {Sheldon}, {Shimasaku}, {Siegmund}, {Silvestri}, {Smith},
  {Smol{\v{c}}i{\'c}}, {Snedden}, {Stebbins}, {Stoughton}, {Strauss},
  {SubbaRao}, {Szalay}, {Szapudi}, {Szkody}, {Szokoly}, {Tegmark}, {Teodoro},
  {Thakar}, {Tremonti}, {Tucker}, {Uomoto}, {Vanden Berk}, {Vandenberg},
  {Vogeley}, {Voges}, {Vogt}, {Walkowicz}, {Wang}, {Weinberg}, {West}, {White},
  {Wilhite}, {Xu}, {Yanny}, {Yasuda}, {Yip}, {Yocum}, {York}, {Zehavi},
  {Zibetti}, \& {Zucker}}]{Abazajian2004}
{Abazajian}, K., {Adelman-McCarthy}, J.~K., {Ag{\"u}eros}, M.~A., {et~al.}
  2004, \aj, 128, 502

\bibitem[{Abazajian {et~al.}(2009)Abazajian, Adelman-McCarthy, Agüeros, Allam,
  Prieto, An, Anderson, Anderson, Annis, Bahcall, Bailer-Jones, Barentine,
  Bassett, Becker, Beers, Bell, Belokurov, Berlind, Berman, Bernardi,
  Bickerton, Bizyaev, Blakeslee, Blanton, Bochanski, Boroski, Brewington,
  Brinchmann, Brinkmann, Brunner, Budavári, Carey, Carliles, Carr, Castander,
  Cinabro, Connolly, Csabai, Cunha, Czarapata, Davenport, De~Haas, Dilday, Doi,
  Eisenstein, Evans, Evans, Fan, Friedman, Frieman, Fukugita, Gänsicke, Gates,
  Gillespie, Gilmore, Gonzalez, Gonzalez, Grebel, Gunn, Györy, Hall, Harding,
  Harris, Harvanek, Hawley, Hayes, Heckman, Hendry, Hennessy, Hindsley,
  Hoblitt, Hogan, Hogg, Holtzman, Hyde, Ichikawa, Ichikawa, Im, Ivezić,
  Jester, Jiang, Johnson, Jorgensen, Jurić, Kent, Kessler, Kleinman, Knapp,
  Konishi, Kron, Krzesinski, Kuropatkin, Lampeitl, Lebedeva, Lee, Lee, Leger,
  Lépine, Li, Lima, Lin, Long, Loomis, Loveday, Lupton, Magnier, Malanushenko,
  Malanushenko, Mandelbaum, Margon, Marriner, Martínez-Delgado, Matsubara,
  McGehee, McKay, Meiksin, Morrison, Mullally, Munn, Murphy, Nash, Nebot,
  Neilsen, Newberg, Newman, Nichol, Nicinski, Nieto-Santisteban, Nitta,
  Okamura, Oravetz, Ostriker, Owen, Padmanabhan, Pan, Park, Pauls, Peoples,
  Percival, Pier, Pope, Pourbaix, Price, Purger, Quinn, Raddick, Fiorentin,
  Richards, Richmond, Riess, Rix, Rockosi, Sako, Schlegel, Schneider, Scholz,
  Schreiber, Schwope, Seljak, Sesar, Sheldon, Shimasaku, Sibley, Simmons,
  Sivarani, Smith, Smith, Smolčić, Snedden, Stebbins, Steinmetz, Stoughton,
  Strauss, SubbaRao, Suto, Szalay, Szapudi, Szkody, Tanaka, Tegmark, Teodoro,
  Thakar, Tremonti, Tucker, Uomoto, Vanden~Berk, Vandenberg, Vidrih, Vogeley,
  Voges, Vogt, Wadadekar, Watters, Weinberg, West, White, Wilhite, Wonders,
  Yanny, Yocum, York, Zehavi, Zibetti, \& Zucker}]{Abazajian2009}
Abazajian, K.~N., Adelman-McCarthy, J.~K., Agüeros, M.~A., {et~al.} 2009,
  ApJS, 182, 543

\bibitem[{Alongi {et~al.}(1993)Alongi, Bertelli, Bressan, Chiosi, Fagotto,
  Greggio, \& Nasi}]{Alongi1993a}
Alongi, M., Bertelli, G., Bressan, A., {et~al.} 1993, A\&ASS, 97, 851

\bibitem[{{Asari} {et~al.}(2007){Asari}, {Cid Fernandes}, {Stasi{\'n}ska},
  {Torres-Papaqui}, {Mateus}, {Sodr{\'e}}, {Schoenell}, \& {Gomes}}]{Asari2007}
{Asari}, N.~V., {Cid Fernandes}, R., {Stasi{\'n}ska}, G., {et~al.} 2007,
  \mnras, 381, 263

\bibitem[{Baldry {et~al.}(2004)Baldry, Glazebrook, Brinkmann, Ivezić, Lupton,
  Nichol, \& Szalay}]{Baldry2004}
Baldry, I.~K., Glazebrook, K., Brinkmann, J., {et~al.} 2004, ApJ, 600, 681

\bibitem[{{Balogh} {et~al.}(1999){Balogh}, {Morris}, {Yee}, {Carlberg}, \&
  {Ellingson}}]{Balogh1999}
{Balogh}, M.~L., {Morris}, S.~L., {Yee}, H.~K.~C., {Carlberg}, R.~G., \&
  {Ellingson}, E. 1999, \apj, 527, 54

\bibitem[{Balogh {et~al.}(2000)Balogh, Navarro, \& Morris}]{Balogh2000}
Balogh, M.~L., Navarro, J.~F., \& Morris, S.~L. 2000, ApJ, 540, 113

\bibitem[{Barone {et~al.}(2020)Barone, D'Eugenio, Colless, \&
  Scott}]{Barone2020}
Barone, T.~M., D'Eugenio, F., Colless, M., \& Scott, N. 2020, ApJ, 898, 62

\bibitem[{Barone {et~al.}(2018)Barone, D'Eugenio, Colless, Scott, van~de Sande,
  Bland-Hawthorn, Brough, Bryant, Cortese, Croom, Foster, Goodwin,
  Konstantopoulos, Lawrence, Lorente, Medling, Owers, \& Richards}]{Barone2018}
Barone, T.~M., D'Eugenio, F., Colless, M., {et~al.} 2018, ApJ, 856, 64

\bibitem[{{Barrera-Ballesteros} {et~al.}(2025){Barrera-Ballesteros},
  {Cruz-Gonz{\'a}lez}, {Colombo}, {S{\'a}nchez}, {Levy}, {Villanueva}, {Wong},
  {Bolatto}, \& {Alonso Hern{\'a}ndez}}]{Barrera2025}
{Barrera-Ballesteros}, J.~K., {Cruz-Gonz{\'a}lez}, I., {Colombo}, D., {et~al.}
  2025, \apj, 978, 23

\bibitem[{Blanton {et~al.}(2003)Blanton, Hogg, Bahcall, Brinkmann, Britton,
  Connolly, Csabai, Fukugita, Loveday, Meiksin, Munn, Nichol, Okamura, Quinn,
  Schneider, Shimasaku, Strauss, Tegmark, Vogeley, \& Weinberg}]{Blanton2003}
Blanton, M.~R., Hogg, D.~W., Bahcall, N.~A., {et~al.} 2003, ApJ, 592, 819

\bibitem[{Blanton \& Roweis(2007)}]{Blanton2007}
Blanton, M.~R. \& Roweis, S. 2007, AJ, 133, 734

\bibitem[{Blanton {et~al.}(2005)Blanton, Schlegel, Strauss, Brinkmann,
  Finkbeiner, Fukugita, Gunn, Hogg, Ivezić, Knapp, Lupton, Munn, Schneider,
  Tegmark, \& Zehavi}]{Blanton2005}
Blanton, M.~R., Schlegel, D.~J., Strauss, M.~A., {et~al.} 2005, AJ, 129, 2562

\bibitem[{{Bluck} {et~al.}(2020{\natexlab{a}}){Bluck}, {Maiolino},
  {Piotrowska}, {Trussler}, {Ellison}, {S{\'a}nchez}, {Thorp}, {Teimoorinia},
  {Moreno}, \& {Conselice}}]{Bluck2020a}
{Bluck}, A. F.~L., {Maiolino}, R., {Piotrowska}, J.~M., {et~al.}
  2020{\natexlab{a}}, \mnras, 499, 230

\bibitem[{{Bluck} {et~al.}(2020{\natexlab{b}}){Bluck}, {Maiolino},
  {S{\'a}nchez}, {Ellison}, {Thorp}, {Piotrowska}, {Teimoorinia}, \&
  {Bundy}}]{Bluck2020b}
{Bluck}, A. F.~L., {Maiolino}, R., {S{\'a}nchez}, S.~F., {et~al.}
  2020{\natexlab{b}}, \mnras, 492, 96

\bibitem[{Bressan {et~al.}(1993)Bressan, Fagotto, Bertelli, \&
  Chiosi}]{Bressan1993}
Bressan, A., Fagotto, F., Bertelli, G., \& Chiosi, C. 1993, A\&AS, 100, 647

\bibitem[{Bressan {et~al.}(2012)Bressan, Marigo, Girardi, Salasnich, Dal~Cero,
  Rubele, \& Nanni}]{Bressan2012}
Bressan, A., Marigo, P., Girardi, L., {et~al.} 2012, MNRAS, 427, 127

\bibitem[{Brinchmann {et~al.}(2004)Brinchmann, Charlot, White, Tremonti,
  Kauffmann, Heckman, \& Brinkmann}]{Brinchmann2004}
Brinchmann, J., Charlot, S., White, S. D.~M., {et~al.} 2004, MNRAS, 351, 1151

\bibitem[{Bruzual \& Charlot(2003)}]{Bruzual2003}
Bruzual, G. \& Charlot, S. 2003, MNRAS, 344, 1000

\bibitem[{Bruzual~A.(1983)}]{BruzualA.1983a}
Bruzual~A., G. 1983, ApJ, 273, 105

\bibitem[{Bundy {et~al.}(2014)Bundy, Bershady, Law, Yan, Drory, MacDonald,
  Wake, Cherinka, Sánchez-Gallego, Weijmans, Thomas, Tremonti, Masters,
  Coccato, Diamond-Stanic, Aragón-Salamanca, Avila-Reese, Badenes,
  Falcón-Barroso, Belfiore, Bizyaev, Blanc, Bland-Hawthorn, Blanton,
  Brownstein, Byler, Cappellari, Conroy, Dutton, Emsellem, Etherington,
  Frinchaboy, Fu, Gunn, Harding, Johnston, Kauffmann, Kinemuchi, Klaene,
  Knapen, Leauthaud, Li, Lin, Maiolino, Malanushenko, Malanushenko, Mao,
  Maraston, McDermid, Merrifield, Nichol, Oravetz, Pan, Parejko, Sanchez,
  Schlegel, Simmons, Steele, Steinmetz, Thanjavur, Thompson, Tinker, Van
  Den~Bosch, Westfall, Wilkinson, Wright, Xiao, \& Zhang}]{Bundy2014a}
Bundy, K., Bershady, M.~A., Law, D.~R., {et~al.} 2014, ApJ, 798, 7

\bibitem[{{Camps-Fari{\~n}a} {et~al.}(2021){Camps-Fari{\~n}a}, {Sanchez},
  {Lacerda}, {Carigi}, {Garc{\'\i}a-Benito}, {Mast}, \&
  {Galbany}}]{Camps-Farina2021}
{Camps-Fari{\~n}a}, A., {Sanchez}, S.~F., {Lacerda}, E.~A.~D., {et~al.} 2021,
  \mnras, 504, 3478

\bibitem[{{Camps-Fari{\~n}a} {et~al.}(2022){Camps-Fari{\~n}a}, {S{\'a}nchez},
  {Mej{\'\i}a-Narv{\'a}ez}, {Lacerda}, {Carigi}, {Bruzual}, {Alvarez-Hurtado},
  {Drory}, {Lane}, {Boardman}, \& {Blanc}}]{Camps-Farina2022}
{Camps-Fari{\~n}a}, A., {S{\'a}nchez}, S.~F., {Mej{\'\i}a-Narv{\'a}ez}, A.,
  {et~al.} 2022, \apj, 933, 44

\bibitem[{{Cappellari}(2023)}]{Cappellari2023}
{Cappellari}, M. 2023, \mnras, 526, 3273

\bibitem[{{Casado} {et~al.}(2015){Casado}, {Ascasibar}, {Gavil{\'a}n},
  {Terlevich}, {Terlevich}, {Hoyos}, \& {D{\'\i}az}}]{Casado2015}
{Casado}, J., {Ascasibar}, Y., {Gavil{\'a}n}, M., {et~al.} 2015, \mnras, 451,
  888

\bibitem[{Cattaneo {et~al.}(2006)Cattaneo, Dekel, Devriendt, Guiderdoni, \&
  Blaizot}]{Cattaneo2006}
Cattaneo, A., Dekel, A., Devriendt, J., Guiderdoni, B., \& Blaizot, J. 2006,
  MNRAS, 370, 1651

\bibitem[{Chabrier(2003)}]{Chabrier2003}
Chabrier, G. 2003, PASP, 115, 763

\bibitem[{Charlot \& Fall(2000)}]{Charlot2000}
Charlot, S. \& Fall, S.~M. 2000, ApJ, 539, 718

\bibitem[{Chen {et~al.}(2015)Chen, Bressan, Girardi, Marigo, Kong, \&
  Lanza}]{Chen2015}
Chen, Y., Bressan, A., Girardi, L., {et~al.} 2015, MNRAS, 452, 1068

\bibitem[{Chisholm {et~al.}(2018)Chisholm, Tremonti, \&
  Leitherer}]{Chisholm2018}
Chisholm, J., Tremonti, C., \& Leitherer, C. 2018, MNRAS, 481, 1690

\bibitem[{Cicone {et~al.}(2014)Cicone, Maiolino, Sturm, Graciá-Carpio,
  Feruglio, Neri, Aalto, Davies, Fiore, Fischer, García-Burillo,
  González-Alfonso, Hailey-Dunsheath, Piconcelli, \& Veilleux}]{Cicone2014}
Cicone, C., Maiolino, R., Sturm, E., {et~al.} 2014, A\&A, 562, A21

\bibitem[{{Cid Fernandes} {et~al.}(2007){Cid Fernandes}, {Asari}, {Sodr{\'e}},
  {Stasi{\'n}ska}, {Mateus}, {Torres-Papaqui}, \&
  {Schoenell}}]{Cid-Fernandes2007}
{Cid Fernandes}, R., {Asari}, N.~V., {Sodr{\'e}}, L., {et~al.} 2007, \mnras,
  375, L16

\bibitem[{Cirasuolo {et~al.}(2020)Cirasuolo, Fairley, Rees, Gonzalez, Taylor,
  Maiolino, Afonso, Evans, Flores, Lilly, Oliva, Paltani, Vanzi, Abreu,
  Accardo, Adams, Álvarez Méndez, Amans, Amarantidis, Atek, Atkinson,
  Banerji, Barrett, Barrientos, Bauer, Beard, Béchet, Belfiore, Bellazzini,
  Benoist, Best, Biazzo, Black, Boettger, Bonifacio, Bowler, Bragaglia,
  Brierley, Brinchmann, Brinkmann, Buat, Buitrago, Burgarella, Burningham,
  Buscher, Cabral, Caffau, Cardoso, Carnall, Carollo, Castillo, Castignani,
  Catelan, Cicone, Cimatti, Cioni, Clementini, Cochrane, Coelho, Colling,
  Contini, Contreras, Conzelmann, Cresci, Cropper, Cucciati, Cullen, Cumani,
  Curti, Da~Silva, Daddi, Dalessandro, Dalessio, Dauvin, Davidson, de~Laverny,
  Delplancke-Ströbele, De~Lucia, Del~Vecchio, Dessauges-Zavadsky, Di~Matteo,
  Dole, Drass, Dunlop, Dünner, Eales, Ellis, Enriques, Fasola, Ferguson,
  Ferruzzi, Fisher, Flores, Fontana, Forchi, Francois, Franzetti, Gargiulo,
  Garilli, Gaudemard, Gieles, Gilmore, Ginolfi, Gomes, Guinouard, Gutierrez,
  Haigron, Hammer, Hammersley, Haniff, Harrison, Haywood, Hill, Hubin,
  Humphrey, Ibata, Infante, Ives, Ivison, Iwert, Jablonka, Jakob, Jarvis, King,
  Kneib, Laporte, Lawrence, Lee, Li~Causi, Lorenzoni, Lucatello, Luco, Macleod,
  Magliocchetti, Magrini, Mainieri, Maire, Mannucci, Martin, Matute,
  Maurogordato, McGee, Mcleod, McLure, McMahon, Melse, Messias, Mucciarelli,
  Nisini, Nix, Norberg, Oesch, Oliveira, Origlia, Padilla, Palsa, Pancino,
  Papaderos, Pappalardo, Parry, Pasquini, Peacock, Pedichini, Pello, Peng,
  Pentericci, Pfuhl, Piazzesi, Popovic, Pozzetti, Puech, Puzia, Raichoor,
  Randich, Recio-Blanco, Reis, Reix, Renzini, Rodrigues, Rojas,
  Rojas-Arriagada, Rota, Royer, Sacco, Sanchez-Janssen, Sanna, Santos, Sarzi,
  Schaerer, Schiavon, Schnell, Schultheis, Scodeggio, Serjeant, Shen, Simmonds,
  Smoker, Sobral, Sordet, \& Spérone}]{Cirasuolo2020a}
Cirasuolo, M., Fairley, A., Rees, P., {et~al.} 2020, The Messenger, 180, 10

\bibitem[{{Colombo} {et~al.}(2020){Colombo}, {Sanchez}, {Bolatto}, {Kalinova},
  {Wei{\ss}}, {Wong}, {Rosolowsky}, {Vogel}, {Barrera-Ballesteros},
  {Dannerbauer}, {Cao}, {Levy}, {Utomo}, \& {Blitz}}]{Colombo2020}
{Colombo}, D., {Sanchez}, S.~F., {Bolatto}, A.~D., {et~al.} 2020, \aap, 644,
  A97

\bibitem[{{Corcho-Caballero} {et~al.}(2023{\natexlab{a}}){Corcho-Caballero},
  {Ascasibar}, {Cortese}, {S{\'a}nchez}, {L{\'o}pez-S{\'a}nchez},
  {Fraser-McKelvie}, \& {Zafar}}]{Corcho-Caballero2023b}
{Corcho-Caballero}, P., {Ascasibar}, Y., {Cortese}, L., {et~al.}
  2023{\natexlab{a}}, \mnras, 524, 3692

\bibitem[{{Corcho-Caballero} {et~al.}(2023{\natexlab{b}}){Corcho-Caballero},
  {Ascasibar}, {S{\'a}nchez}, \&
  {L{\'o}pez-S{\'a}nchez}}]{Corcho-Caballero2023a}
{Corcho-Caballero}, P., {Ascasibar}, Y., {S{\'a}nchez}, S.~F., \&
  {L{\'o}pez-S{\'a}nchez}, {\'A}.~R. 2023{\natexlab{b}}, \mnras, 520, 193

\bibitem[{Cowie {et~al.}(1996)Cowie, Songaila, Hu, \& Cohen}]{Cowie1996}
Cowie, L.~L., Songaila, A., Hu, E.~M., \& Cohen, J.~G. 1996, AJ, 112, 839

\bibitem[{Croton {et~al.}(2006)Croton, Springel, White, De~Lucia, Frenk, Gao,
  Jenkins, Kauffmann, Navarro, \& Yoshida}]{Croton2006a}
Croton, D.~J., Springel, V., White, S. D.~M., {et~al.} 2006, MNRAS, 365, 11

\bibitem[{da~Cunha {et~al.}(2008)da~Cunha, Charlot, \& Elbaz}]{daCunha2008}
da~Cunha, E., Charlot, S., \& Elbaz, D. 2008, MNRAS, 388, 1595

\bibitem[{{Dalcanton}(2007)}]{Dalcanton2007}
{Dalcanton}, J.~J. 2007, \apj, 658, 941

\bibitem[{De~Lucia {et~al.}(2006)De~Lucia, Springel, White, Croton, \&
  Kauffmann}]{DeLucia2006}
De~Lucia, G., Springel, V., White, S. D.~M., Croton, D., \& Kauffmann, G. 2006,
  MNRAS, 366, 499

\bibitem[{Dekel \& Birnboim(2006)}]{Dekel2006}
Dekel, A. \& Birnboim, Y. 2006, MNRAS, 368, 2

\bibitem[{Di~Matteo {et~al.}(2005)Di~Matteo, Springel, \&
  Hernquist}]{DiMatteo2005}
Di~Matteo, T., Springel, V., \& Hernquist, L. 2005, Nature, 433, 604

\bibitem[{Ellison {et~al.}(2008)Ellison, Patton, Simard, \&
  McConnachie}]{Ellison2008}
Ellison, S.~L., Patton, D.~R., Simard, L., \& McConnachie, A.~W. 2008, ApJ,
  672, L107

\bibitem[{{Erb}(2008)}]{Erb2008}
{Erb}, D.~K. 2008, \apj, 674, 151

\bibitem[{Fabian(2012)}]{Fabian2012}
Fabian, A.~C. 2012, Annual Review of A\&A, 50, 455

\bibitem[{Fagotto {et~al.}(1994{\natexlab{a}})Fagotto, Bressan, Bertelli, \&
  Chiosi}]{Fagotto1994}
Fagotto, F., Bressan, A., Bertelli, G., \& Chiosi, C. 1994{\natexlab{a}},
  A\&AS, 104, 365

\bibitem[{Fagotto {et~al.}(1994{\natexlab{b}})Fagotto, Bressan, Bertelli, \&
  Chiosi}]{Fagotto1994a}
Fagotto, F., Bressan, A., Bertelli, G., \& Chiosi, C. 1994{\natexlab{b}},
  A\&AS, 105, 29

\bibitem[{{Fontanot} {et~al.}(2009){Fontanot}, {De Lucia}, {Monaco},
  {Somerville}, \& {Santini}}]{Fontanot2009}
{Fontanot}, F., {De Lucia}, G., {Monaco}, P., {Somerville}, R.~S., \&
  {Santini}, P. 2009, \mnras, 397, 1776

\bibitem[{Fukugita {et~al.}(1996)Fukugita, Ichikawa, Gunn, Doi, Shimasaku, \&
  Schneider}]{Fukugita1996}
Fukugita, M., Ichikawa, T., Gunn, J.~E., {et~al.} 1996, AJ, 111, 1748

\bibitem[{Gallazzi {et~al.}(2014)Gallazzi, Bell, Zibetti, Brinchmann, \&
  Kelson}]{Gallazzi2014}
Gallazzi, A., Bell, E.~F., Zibetti, S., Brinchmann, J., \& Kelson, D.~D. 2014,
  ApJ, 788, 72

\bibitem[{Gallazzi {et~al.}(2008)Gallazzi, Brinchmann, Charlot, \&
  White}]{Gallazzi2008}
Gallazzi, A., Brinchmann, J., Charlot, S., \& White, S. D.~M. 2008, MNRAS, 383,
  1439

\bibitem[{Gallazzi {et~al.}(2005)Gallazzi, Charlot, Brinchmann, White, \&
  Tremonti}]{Gallazzi2005}
Gallazzi, A., Charlot, S., Brinchmann, J., White, S. D.~M., \& Tremonti, C.~A.
  2005, MNRAS, 362, 41

\bibitem[{{Gallazzi} {et~al.}(2025){Gallazzi}, {Zibetti}, \& {van der
  Wel}}]{Gallazzi2025}
{Gallazzi}, A., {Zibetti}, S., \& {van der Wel}, A. 2025, \aap

\bibitem[{Gallazzi {et~al.}(2021)Gallazzi, Pasquali, Zibetti, \&
  Barbera}]{Gallazzi2021}
Gallazzi, A.~R., Pasquali, A., Zibetti, S., \& Barbera, F.~L. 2021, MNRAS, 502,
  4457

\bibitem[{Gavazzi {et~al.}(2002)Gavazzi, Bonfanti, Sanvito, Boselli, \&
  Scodeggio}]{Gavazzi2002a}
Gavazzi, G., Bonfanti, C., Sanvito, G., Boselli, A., \& Scodeggio, M. 2002,
  ApJ, 576, 135

\bibitem[{Gavazzi \& Scodeggio(1996)}]{Gavazzi1996}
Gavazzi, G. \& Scodeggio, M. 1996, A\&A, 312, L29

\bibitem[{Girardi {et~al.}(1996)Girardi, Bressan, Chiosi, Bertelli, \&
  Nasi}]{Girardi1996}
Girardi, L., Bressan, A., Chiosi, C., Bertelli, G., \& Nasi, E. 1996, A\&AS,
  117, 113

\bibitem[{Goddard {et~al.}(2017)Goddard, Thomas, Maraston, Westfall,
  Etherington, Riffel, Mallmann, Zheng, Argudo-Fernández, Lian, Bershady,
  Bundy, Drory, Law, Yan, Wake, Weijmans, Bizyaev, Brownstein, Lane, Maiolino,
  Masters, Merrifield, Nitschelm, Pan, Roman-Lopes, Storchi-Bergmann, \&
  Schneider}]{Goddard2017}
Goddard, D., Thomas, D., Maraston, C., {et~al.} 2017, MNRAS, 466, 4731

\bibitem[{Gonzalez {et~al.}(1993)Gonzalez, Faber, \& Worthey}]{Gonzalez1993}
Gonzalez, J.~J., Faber, S.~M., \& Worthey, G. 1993, AASM, 183, 42.06

\bibitem[{González~Delgado {et~al.}(2015)González~Delgado, García-Benito,
  Pérez, Cid~Fernandes, de~Amorim, Cortijo-Ferrero, Lacerda,
  López~Fernández, Vale-Asari, Sánchez, Mollá, Ruiz-Lara,
  Sánchez-Blázquez, Walcher, Alves, Aguerri, Bekeraité, Bland-Hawthorn,
  Galbany, Gallazzi, Husemann, Iglesias-Páramo, Kalinova, López-Sánchez,
  Marino, Márquez, Masegosa, Mast, Méndez-Abreu, Mendoza, del Olmo, Pérez,
  Quirrenbach, \& Zibetti}]{GonzalezDelgado2015}
González~Delgado, R.~M., García-Benito, R., Pérez, E., {et~al.} 2015, A\&A,
  581, A103

\bibitem[{González~Delgado {et~al.}(2014)González~Delgado, Pérez,
  Cid~Fernandes, García-Benito, de~Amorim, Sánchez, Husemann,
  Cortijo-Ferrero, López~Fernández, Sánchez-Blázquez, Bekeraite, Walcher,
  Falcón-Barroso, Gallazzi, van~de Ven, Alves, Bland-Hawthorn, Kennicutt,
  Kupko, Lyubenova, Mast, Mollá, Marino, Quirrenbach, Vílchez, \&
  Wisotzki}]{GonzalezDelgado2014}
González~Delgado, R.~M., Pérez, E., Cid~Fernandes, R., {et~al.} 2014, A\&A,
  562, A47

\bibitem[{Gunn {et~al.}(1998)Gunn, Carr, Rockosi, Sekiguchi, Berry, Elms,
  de~Haas, Ivezić, Knapp, Lupton, Pauls, Simcoe, Hirsch, Sanford, Wang, York,
  Harris, Annis, Bartozek, Boroski, Bakken, Haldeman, Kent, Holm, Holmgren,
  Petravick, Prosapio, Rechenmacher, Doi, Fukugita, Shimasaku, Okada, Hull,
  Siegmund, Mannery, Blouke, Heidtman, Schneider, Lucinio, \&
  Brinkman}]{Gunn1998}
Gunn, J.~E., Carr, M., Rockosi, C., {et~al.} 1998, AJ, 116, 3040

\bibitem[{Haines {et~al.}(2017)Haines, Iovino, Krywult, Guzzo, Davidzon,
  Bolzonella, Garilli, Scodeggio, Granett, de~la Torre, De~Lucia, Abbas, Adami,
  Arnouts, Bottini, Cappi, Cucciati, Franzetti, Fritz, Gargiulo, Le~Brun,
  Le~Fèvre, Maccagni, Małek, Marulli, Moutard, Polletta, Pollo, Tasca,
  Tojeiro, Vergani, Zanichelli, Zamorani, Bel, Branchini, Coupon, Ilbert,
  Moscardini, Peacock, \& Siudek}]{Haines2017}
Haines, C.~P., Iovino, A., Krywult, J., {et~al.} 2017, A\&A, 605, A4

\bibitem[{Iovino {et~al.}(2023{\natexlab{a}})Iovino, Mercurio, Gallazzi,
  La~Barbera, Longhetti, Tortora, Zibetti, Belfiore, Bianconi, Busarello,
  Corsini, Costantin, De~Lucia, De~Propris, D’Eugenio, Fontanot,
  García-Benito, Hirschmann, Haines, Mannucci, McGee, Merluzzi, Morelli,
  Moretti, Pasquali, Poggianti, Pozzetti, Rodighiero, Sánchez-Blázquez, Van
  Der~Wel, Vazdekis, Vulcani, Zanella, Annunziatella, Concas, Cassarà, Cresci,
  Curti, De~Lorenzo-Cáceres, Mateu, Delgado, Mancini, Pacifici, Perez-Montero,
  Pizzella, Perez-Gonzalez, Trager, \& Vergani}]{Iovino2023b}
Iovino, A., Mercurio, A., Gallazzi, A.~R., {et~al.} 2023{\natexlab{a}}, The
  Messenger, pp. 22-24, 3 pages

\bibitem[{Iovino {et~al.}(2023{\natexlab{b}})Iovino, Poggianti, Mercurio,
  Longhetti, Bolzonella, Busarello, Gullieuszik, La~Barbera, Merluzzi, Morelli,
  Tortora, Vergani, Zibetti, Haines, Costantin, Ditrani, Pozzetti, Angthopo,
  Balcells, Bardelli, Benn, Bianconi, Cassarà, Corsini, Cucciati, Dalton,
  Ferré-Mateu, Fossati, Gallazzi, García-Benito, Granett, González~Delgado,
  Ikhsanova, Iodice, Jin, Knapen, McGee, Moretti, Murphy, Peralta De~Arriba,
  Pizzella, Sánchez-Blázquez, Spiniello, Talia, Trager, Vazdekis, Vulcani, \&
  Zucca}]{Iovino2023a}
Iovino, A., Poggianti, B.~M., Mercurio, A., {et~al.} 2023{\natexlab{b}}, A\&A,
  672, A87

\bibitem[{Johnston {et~al.}(2022)Johnston, Häußler, Jegatheesan,
  Fraser-McKelvie, Coccato, Cortesi, Jaffé, Galaz, Mora, \&
  Ordenes-Briceño}]{Johnston2022}
Johnston, E.~J., Häußler, B., Jegatheesan, K., {et~al.} 2022, MNRAS, 514,
  6141

\bibitem[{Kauffmann(1996)}]{Kauffmann1996}
Kauffmann, G. 1996, MNRAS, 281, 487

\bibitem[{Kauffmann {et~al.}(2003{\natexlab{a}})Kauffmann, Heckman, White,
  Charlot, Tremonti, Brinchmann, Bruzual, Peng, Seibert, Bernardi, Blanton,
  Brinkmann, Castander, Csábai, Fukugita, Ivezic, Munn, Nichol, Padmanabhan,
  Thakar, Weinberg, \& York}]{Kauffmann2003}
Kauffmann, G., Heckman, T.~M., White, S. D.~M., {et~al.} 2003{\natexlab{a}},
  MNRAS, 341, 33

\bibitem[{Kauffmann {et~al.}(2003{\natexlab{b}})Kauffmann, Heckman, White,
  Charlot, Tremonti, Peng, Seibert, Brinkmann, Nichol, SubbaRao, \&
  York}]{Kauffmann2003a}
Kauffmann, G., Heckman, T.~M., White, S. D.~M., {et~al.} 2003{\natexlab{b}},
  MNRAS, 341, 54

\bibitem[{Lara-López {et~al.}(2010)Lara-López, Cepa, Bongiovanni,
  Pérez~García, Ederoclite, Castañeda, Fernández~Lorenzo, Pović, \&
  Sánchez-Portal}]{Lara-Lopez2010}
Lara-López, M.~A., Cepa, J., Bongiovanni, A., {et~al.} 2010, A\&A, 521, L53

\bibitem[{Larson {et~al.}(1980)Larson, Tinsley, \& Caldwell}]{Larson1980}
Larson, R.~B., Tinsley, B.~M., \& Caldwell, C.~N. 1980, ApJ, 237, 692

\bibitem[{Le~Borgne {et~al.}(2003)Le~Borgne, Bruzual, Pelló, Lançon,
  Rocca-Volmerange, Sanahuja, Schaerer, Soubiran, \&
  Vílchez-Gómez}]{LeBorgne2003}
Le~Borgne, J.-F., Bruzual, G., Pelló, R., {et~al.} 2003, A\&A, 402, 433

\bibitem[{Li {et~al.}(2018)Li, Mao, Cappellari, Ge, Long, Li, Mo, Li, Zheng,
  Bundy, Thomas, Brownstein, Roman~Lopes, Law, \& Drory}]{Li2018}
Li, H., Mao, S., Cappellari, M., {et~al.} 2018, MNRAS, 476, 1765

\bibitem[{Looser {et~al.}(2024)Looser, D'Eugenio, Piotrowska, Belfiore,
  Maiolino, Cappellari, Baker, \& Tacchella}]{Looser2024}
Looser, T.~J., D'Eugenio, F., Piotrowska, J.~M., {et~al.} 2024, MNRAS, 532,
  2832

\bibitem[{Lupton {et~al.}(1999)Lupton, Gunn, \& Szalay}]{Lupton1999a}
Lupton, R.~H., Gunn, J.~E., \& Szalay, A.~S. 1999, AJ, 118, 1406

\bibitem[{Maiolino \& Mannucci(2019)}]{Maiolino2019}
Maiolino, R. \& Mannucci, F. 2019, A\&AR, 27, 3

\bibitem[{Mannucci {et~al.}(2010)Mannucci, Cresci, Maiolino, Marconi, \&
  Gnerucci}]{Mannucci2010}
Mannucci, F., Cresci, G., Maiolino, R., Marconi, A., \& Gnerucci, A. 2010,
  MNRAS, 408, 2115

\bibitem[{Marigo {et~al.}(2013)Marigo, Bressan, Nanni, Girardi, \&
  Pumo}]{Marigo2013a}
Marigo, P., Bressan, A., Nanni, A., Girardi, L., \& Pumo, M.~L. 2013, MNRAS,
  434, 488

\bibitem[{{Martig} {et~al.}(2009){Martig}, {Bournaud}, {Teyssier}, \&
  {Dekel}}]{Martig2009}
{Martig}, M., {Bournaud}, F., {Teyssier}, R., \& {Dekel}, A. 2009, \apj, 707,
  250

\bibitem[{Martín-Navarro {et~al.}(2018)Martín-Navarro, Vazdekis,
  Falcón-Barroso, La~Barbera, Yıldırım, \& van~de Ven}]{Martin-Navarro2018}
Martín-Navarro, I., Vazdekis, A., Falcón-Barroso, J., {et~al.} 2018, MNRAS,
  475, 3700

\bibitem[{{Mateus} {et~al.}(2006){Mateus}, {Sodr{\'e}}, {Cid Fernandes},
  {Stasi{\'n}ska}, {Schoenell}, \& {Gomes}}]{Mateus2006}
{Mateus}, A., {Sodr{\'e}}, L., {Cid Fernandes}, R., {et~al.} 2006, \mnras, 370,
  721

\bibitem[{Moustakas {et~al.}(2013)Moustakas, Coil, Aird, Blanton, Cool,
  Eisenstein, Mendez, Wong, Zhu, \& Arnouts}]{Moustakas2013}
Moustakas, J., Coil, A.~L., Aird, J., {et~al.} 2013, ApJ, 767, 50

\bibitem[{Nair \& Abraham(2010)}]{Nair2010}
Nair, P.~B. \& Abraham, R.~G. 2010, ApJS, 186, 427

\bibitem[{{Nersesian} {et~al.}(2024){Nersesian}, {van der Wel}, {Gallazzi},
  {Leja}, {Bezanson}, {Bell}, {D'Eugenio}, {de Graaff}, {Kaushal}, {Martorano},
  {Maseda}, \& {Zibetti}}]{Nersesian2024}
{Nersesian}, A., {van der Wel}, A., {Gallazzi}, A., {et~al.} 2024, \aap, 681,
  A94

\bibitem[{{Pacifici} {et~al.}(2016){Pacifici}, {Kassin}, {Weiner}, {Holden},
  {Gardner}, {Faber}, {Ferguson}, {Koo}, {Primack}, {Bell}, {Dekel}, {Gawiser},
  {Giavalisco}, {Rafelski}, {Simons}, {Barro}, {Croton}, {Dav{\'e}}, {Fontana},
  {Grogin}, {Koekemoer}, {Lee}, {Salmon}, {Somerville}, \&
  {Behroozi}}]{Pacifici2016}
{Pacifici}, C., {Kassin}, S.~A., {Weiner}, B.~J., {et~al.} 2016, \apj, 832, 79

\bibitem[{{Pan} {et~al.}(2024){Pan}, {Lin}, {Ellison}, {Thorp}, {S{\'a}nchez},
  {Bluck}, {Belfiore}, {Piotrowska}, {Scudder}, \& {Baker}}]{Pan2024}
{Pan}, H.-A., {Lin}, L., {Ellison}, S.~L., {et~al.} 2024, \apj, 964, 120

\bibitem[{{Panter} {et~al.}(2003){Panter}, {Heavens}, \&
  {Jimenez}}]{Panter2003}
{Panter}, B., {Heavens}, A.~F., \& {Jimenez}, R. 2003, \mnras, 343, 1145

\bibitem[{{Panter} {et~al.}(2008){Panter}, Jimenez, Heavens, \&
  Charlot}]{Panter2008}
{Panter}, B., Jimenez, R., Heavens, A.~F., \& Charlot, S. 2008, MNRAS, 391,
  1117

\bibitem[{Pasquali {et~al.}(2010)Pasquali, Gallazzi, Fontanot, van~den Bosch,
  De~Lucia, Mo, \& Yang}]{Pasquali2010}
Pasquali, A., Gallazzi, A., Fontanot, F., {et~al.} 2010, MNRAS, 407, 937

\bibitem[{Peng {et~al.}(2015)Peng, Maiolino, \& Cochrane}]{Peng2015}
Peng, Y., Maiolino, R., \& Cochrane, R. 2015, Nature, 521, 192

\bibitem[{Peng {et~al.}(2010)Peng, Lilly, Kovač, Bolzonella, Pozzetti,
  Renzini, Zamorani, Ilbert, Knobel, Iovino, Maier, Cucciati, Tasca, Carollo,
  Silverman, Kampczyk, de~Ravel, Sanders, Scoville, Contini, Mainieri,
  Scodeggio, Kneib, Le~Fèvre, Bardelli, Bongiorno, Caputi, Coppa, de~la Torre,
  Franzetti, Garilli, Lamareille, Le~Borgne, Le~Brun, Mignoli, Perez~Montero,
  Pello, Ricciardelli, Tanaka, Tresse, Vergani, Welikala, Zucca, Oesch, Abbas,
  Barnes, Bordoloi, Bottini, Cappi, Cassata, Cimatti, Fumana, Hasinger,
  Koekemoer, Leauthaud, Maccagni, Marinoni, McCracken, Memeo, Meneux, Nair,
  Porciani, Presotto, \& Scaramella}]{Peng2010}
Peng, Y.-j., Lilly, S.~J., Kovač, K., {et~al.} 2010, ApJ, 721, 193

\bibitem[{Pillepich {et~al.}(2018)Pillepich, Springel, Nelson, Genel, Naiman,
  Pakmor, Hernquist, Torrey, Vogelsberger, Weinberger, \&
  Marinacci}]{Pillepich2018}
Pillepich, A., Springel, V., Nelson, D., {et~al.} 2018, MNRAS, 473, 4077

\bibitem[{Pérez {et~al.}(2013)Pérez, Cid~Fernandes, González~Delgado,
  García-Benito, Sánchez, Husemann, Mast, Rodón, Kupko, Backsmann,
  de~Amorim, van~de Ven, Walcher, Wisotzki, Cortijo-Ferrero, \&
  Collaboration}]{Perez2013}
Pérez, E., Cid~Fernandes, R., González~Delgado, R.~M., {et~al.} 2013, ApJ,
  764, L1

\bibitem[{Renzini \& Peng(2015)}]{Renzini2015}
Renzini, A. \& Peng, Y.-j. 2015, ApJ, 801, L29

\bibitem[{{Rossi}(2025)}]{Rossi2025}
{Rossi}, E. 2025, arXiv e-prints, arXiv:2507.06006

\bibitem[{{Saintonge} {et~al.}(2016){Saintonge}, {Catinella}, {Cortese},
  {Genzel}, {Giovanelli}, {Haynes}, {Janowiecki}, {Kramer}, {Lutz},
  {Schiminovich}, {Tacconi}, {Wuyts}, \& {Accurso}}]{Saintonge2016}
{Saintonge}, A., {Catinella}, B., {Cortese}, L., {et~al.} 2016, \mnras, 462,
  1749

\bibitem[{{Saintonge} {et~al.}(2017){Saintonge}, {Catinella}, {Tacconi},
  {Kauffmann}, {Genzel}, {Cortese}, {Dav{\'e}}, {Fletcher},
  {Graci{\'a}-Carpio}, {Kramer}, {Heckman}, {Janowiecki}, {Lutz}, {Rosario},
  {Schiminovich}, {Schuster}, {Wang}, {Wuyts}, {Borthakur}, {Lamperti}, \&
  {Roberts-Borsani}}]{Saintonge2017}
{Saintonge}, A., {Catinella}, B., {Tacconi}, L.~J., {et~al.} 2017, \apjs, 233,
  22

\bibitem[{Salim {et~al.}(2007)Salim, Rich, Charlot, Brinchmann, Johnson,
  Schiminovich, Seibert, Mallery, Heckman, Forster, Friedman, Martin,
  Morrissey, Neff, Small, Wyder, Bianchi, Donas, Lee, Madore, Milliard, Szalay,
  Welsh, \& Yi}]{Salim2007}
Salim, S., Rich, R.~M., Charlot, S., {et~al.} 2007, ApJS, 173, 267

\bibitem[{{S{\'a}nchez}(2020)}]{Sanchez2020}
{S{\'a}nchez}, S.~F. 2020, \araa, 58, 99

\bibitem[{{S{\'a}nchez} {et~al.}(2018){S{\'a}nchez}, {Avila-Reese},
  {Hernandez-Toledo}, {Cortes-Su{\'a}rez}, {Rodr{\'\i}guez-Puebla},
  {Ibarra-Medel}, {Cano-D{\'\i}az}, {Barrera-Ballesteros}, {Negrete},
  {Calette}, {de Lorenzo-C{\'a}ceres}, {Ortega-Minakata}, {Aquino},
  {Valenzuela}, {Clemente}, {Storchi-Bergmann}, {Riffel}, {Schimoia}, {Riffel},
  {Rembold}, {Brownstein}, {Pan}, {Yates}, {Mallmann}, \&
  {Bitsakis}}]{Sanchez2018}
{S{\'a}nchez}, S.~F., {Avila-Reese}, V., {Hernandez-Toledo}, H., {et~al.} 2018,
  \rmxaa, 54, 217

\bibitem[{Sandage(1986)}]{Sandage1986}
Sandage, A. 1986, A\&A, 161, 89

\bibitem[{Schawinski {et~al.}(2007)Schawinski, Thomas, Sarzi, Maraston,
  Kaviraj, Joo, Yi, \& Silk}]{Schawinski2007}
Schawinski, K., Thomas, D., Sarzi, M., {et~al.} 2007, MNRAS, 382, 1415

\bibitem[{Schawinski {et~al.}(2014)Schawinski, Urry, Simmons, Fortson, Kaviraj,
  Keel, Lintott, Masters, Nichol, Sarzi, Skibba, Treister, Willett, Wong, \&
  Yi}]{Schawinski2014}
Schawinski, K., Urry, C.~M., Simmons, B.~D., {et~al.} 2014, MNRAS, 440, 889

\bibitem[{Schaye {et~al.}(2015)Schaye, Crain, Bower, Furlong, Schaller, Theuns,
  Dalla~Vecchia, Frenk, McCarthy, Helly, Jenkins, Rosas-Guevara, White, Baes,
  Booth, Camps, Navarro, Qu, Rahmati, Sawala, Thomas, \& Trayford}]{Schaye2015}
Schaye, J., Crain, R.~A., Bower, R.~G., {et~al.} 2015, MNRAS, 446, 521

\bibitem[{Schmidt(1968)}]{Schmidt1968}
Schmidt, M. 1968, ApJS, 151, 393

\bibitem[{Serra \& Trager(2007)}]{Serra2007a}
Serra, P. \& Trager, S.~C. 2007, MNRAS, 374, 769

\bibitem[{Somerville \& Davé(2015)}]{Somerville2015}
Somerville, R.~S. \& Davé, R. 2015, ARAA, 53, 51

\bibitem[{Spitoni {et~al.}(2017)Spitoni, Vincenzo, \& Matteucci}]{Spitoni2017}
Spitoni, E., Vincenzo, F., \& Matteucci, F. 2017, A\&A, 599, A6

\bibitem[{Stoughton {et~al.}(2002)Stoughton, Lupton, Bernardi, Blanton, Burles,
  Castander, Connolly, Eisenstein, Frieman, Hennessy, Hindsley, Ivezić, Kent,
  Kunszt, Lee, Meiksin, Munn, Newberg, Nichol, Nicinski, Pier, Richards,
  Richmond, Schlegel, Smith, Strauss, SubbaRao, Szalay, Thakar, Tucker,
  Vanden~Berk, Yanny, Adelman, Anderson, Anderson, Annis, Bahcall, Bakken,
  Bartelmann, Bastian, Bauer, Berman, Böhringer, Boroski, Bracker, Briegel,
  Briggs, Brinkmann, Brunner, Carey, Carr, Chen, Christian, Colestock, Crocker,
  Csabai, Czarapata, Dalcanton, Davidsen, Davis, Dehnen, Dodelson, Doi,
  Dombeck, Donahue, Ellman, Elms, Evans, Eyer, Fan, Federwitz, Friedman,
  Fukugita, Gal, Gillespie, Glazebrook, Gray, Grebel, Greenawalt, Greene, Gunn,
  de~Haas, Haiman, Haldeman, Hall, Hamabe, Hansen, Harris, Harris, Harvanek,
  Hawley, Hayes, Heckman, Helmi, Henden, Hogan, Hogg, Holmgren, Holtzman,
  Huang, Hull, Ichikawa, Ichikawa, Johnston, Kauffmann, Kim, Kimball, Kinney,
  Klaene, Kleinman, Klypin, Knapp, Korienek, Krolik, Kron, Krzesiński, Lamb,
  Leger, Limmongkol, Lindenmeyer, Long, Loomis, Loveday, MacKinnon, Mannery,
  Mantsch, Margon, McGehee, McKay, McLean, Menou, Merelli, Mo, Monet, Nakamura,
  Narayanan, Nash, Neilsen, Newman, Nitta, Odenkirchen, Okada, Okamura,
  Ostriker, Owen, Pauls, Peoples, Peterson, Petravick, Pope, Pordes, Postman,
  Prosapio, Quinn, Rechenmacher, Rivetta, Rix, Rockosi, Rosner, Ruthmansdorfer,
  Sandford, Schneider, Scranton, Sekiguchi, Sergey, Sheth, Shimasaku, Smee,
  Snedden, Stebbins, Stubbs, Szapudi, Szkody, Szokoly, Tabachnik, Tsvetanov,
  Uomoto, Vogeley, Voges, Waddell, Walterbos, Wang, Watanabe, Weinberg, White,
  White, Wilhite, Wolfe, Yasuda, York, Zehavi, \& Zheng}]{Stoughton2002}
Stoughton, C., Lupton, R.~H., Bernardi, M., {et~al.} 2002, AJ, 123, 485

\bibitem[{Strauss {et~al.}(2002)Strauss, Weinberg, Lupton, Narayanan, Annis,
  Bernardi, Blanton, Burles, Connolly, Dalcanton, Doi, Eisenstein, Frieman,
  Fukugita, Gunn, Ivezić, Kent, Kim, Knapp, Kron, Munn, Newberg, Nichol,
  Okamura, Quinn, Richmond, Schlegel, Shimasaku, SubbaRao, Szalay, Vanden~Berk,
  Vogeley, Yanny, Yasuda, York, \& Zehavi}]{Strauss2002}
Strauss, M.~A., Weinberg, D.~H., Lupton, R.~H., {et~al.} 2002, AJ, 124, 1810

\bibitem[{Sánchez {et~al.}(2022)Sánchez, Barrera-Ballesteros, Lacerda,
  Mejía-Narvaez, Camps-Fariña, Bruzual, Espinosa-Ponce, Rodríguez-Puebla,
  Calette, Ibarra-Medel, Avila-Reese, Hernandez-Toledo, Bershady, Cano-Diaz, \&
  Munguia-Cordova}]{Sanchez2022a}
Sánchez, S.~F., Barrera-Ballesteros, J.~K., Lacerda, E., {et~al.} 2022, ApJSS,
  262, 36

\bibitem[{Sánchez {et~al.}(2012)Sánchez, Kennicutt, Gil~de Paz, van~de Ven,
  Vílchez, Wisotzki, Walcher, Mast, Aguerri, Albiol-Pérez, Alonso-Herrero,
  Alves, Bakos, Bartáková, Bland-Hawthorn, Boselli, Bomans, Castillo-Morales,
  Cortijo-Ferrero, de~Lorenzo-Cáceres, Del~Olmo, Dettmar, Díaz, Ellis,
  Falcón-Barroso, Flores, Gallazzi, García-Lorenzo, González~Delgado, Gruel,
  Haines, Hao, Husemann, Iglésias-Páramo, Jahnke, Johnson, Jungwiert,
  Kalinova, Kehrig, Kupko, López-Sánchez, Lyubenova, Marino,
  Mármol-Queraltó, Márquez, Masegosa, Meidt, Mendez-Abreu, Monreal-Ibero,
  Montijo, Mourão, Palacios-Navarro, Papaderos, Pasquali, Peletier, Pérez,
  Pérez, Quirrenbach, Relaño, Rosales-Ortega, Roth, Ruiz-Lara,
  Sánchez-Blázquez, Sengupta, Singh, Stanishev, Trager, Vazdekis, Viironen,
  Wild, Zibetti, \& Ziegler}]{Sanchez2012}
Sánchez, S.~F., Kennicutt, R.~C., Gil~de Paz, A., {et~al.} 2012, A\&A, 538,
  A8, aDS Bibcode: 2012A\&A...538A...8S

\bibitem[{Sánchez-Blázquez {et~al.}(2006)Sánchez-Blázquez, Peletier,
  Jimenez-Vicente, Cardiel, Cenarro, Falcon-Barroso, Gorgas, Selam, \&
  Vazdekis}]{Sanchez-Blazquez2006}
Sánchez-Blázquez, P., Peletier, R.~F., Jimenez-Vicente, J., {et~al.} 2006,
  MNRAS, 371, 703

\bibitem[{Sánchez-Blázquez {et~al.}(2014)Sánchez-Blázquez, Rosales-Ortega,
  Méndez-Abreu, Pérez, Sánchez, Zibetti, Aguerri, Bland-Hawthorn,
  Catalán-Torrecilla, Cid~Fernandes, de~Amorim, de~Lorenzo-Caceres,
  Falcón-Barroso, Galazzi, García~Benito, Gil~de Paz, González~Delgado,
  Husemann, Iglesias-Páramo, Jungwiert, Marino, Márquez, Mast, Mendoza,
  Mollá, Papaderos, Ruiz-Lara, van~de Ven, Walcher, \&
  Wisotzki}]{Sanchez-Blazquez2014}
Sánchez-Blázquez, P., Rosales-Ortega, F.~F., Méndez-Abreu, J., {et~al.}
  2014, A\&A, 570, A6

\bibitem[{Thomas {et~al.}(2003)Thomas, Maraston, \& Bender}]{Thomas2003}
Thomas, D., Maraston, C., \& Bender, R. 2003, MNRAS, 339, 897

\bibitem[{Thomas {et~al.}(2010)Thomas, Maraston, Schawinski, Sarzi, \&
  Silk}]{Thomas2010}
Thomas, D., Maraston, C., Schawinski, K., Sarzi, M., \& Silk, J. 2010, MNRAS,
  404, 1775

\bibitem[{Tinsley(1980)}]{Tinsley1980a}
Tinsley, B.~M. 1980, Fundamentals of Cosmic Physics, 5, 287

\bibitem[{Tinsley \& Gunn(1976)}]{Tinsley1976a}
Tinsley, B.~M. \& Gunn, J.~E. 1976, ApJ, 203, 52

\bibitem[{Tremonti {et~al.}(2004)Tremonti, Heckman, Kauffmann, Brinchmann,
  Charlot, White, Seibert, Peng, Schlegel, Uomoto, Fukugita, \&
  Brinkmann}]{Tremonti2004}
Tremonti, C.~A., Heckman, T.~M., Kauffmann, G., {et~al.} 2004, ApJ, 613, 898

\bibitem[{Trussler {et~al.}(2020)Trussler, Maiolino, Maraston, Peng, Thomas,
  Goddard, \& Lian}]{Trussler2020}
Trussler, J., Maiolino, R., Maraston, C., {et~al.} 2020, MNRAS, 491, 5406

\bibitem[{{Vale Asari} {et~al.}(2009){Vale Asari}, {Stasi{\'n}ska}, {Cid
  Fernandes}, {Gomes}, {Schlickmann}, {Mateus}, \& {Schoenell}}]{Asari2009}
{Vale Asari}, N., {Stasi{\'n}ska}, G., {Cid Fernandes}, R., {et~al.} 2009,
  \mnras, 396, L71

\bibitem[{van~den Bosch {et~al.}(2008)van~den Bosch, Aquino, Yang, Mo,
  Pasquali, McIntosh, Weinmann, \& Kang}]{vandenBosch2008}
van~den Bosch, F.~C., Aquino, D., Yang, X., {et~al.} 2008, MNRAS, 387, 79

\bibitem[{van~der Wel {et~al.}(2016)van~der Wel, Noeske, Bezanson, Pacifici,
  Gallazzi, Franx, Muñoz-Mateos, Bell, Brammer, Charlot, Chauké, Labbé,
  Maseda, Muzzin, Rix, Sobral, van~de Sande, van Dokkum, Wild, \&
  Wolf}]{vanderWel2016}
van~der Wel, A., Noeske, K., Bezanson, R., {et~al.} 2016, ApJSS, 223, 29

\bibitem[{Vaughan {et~al.}(2022)Vaughan, Barone, Croom, Cortese, D'Eugenio,
  Brough, Colless, McDermid, van~de Sande, Scott, Bland-Hawthorn, Bryant,
  Lawrence, López-Sánchez, Lorente, Owers, \& Richards}]{Vaughan2022}
Vaughan, S.~P., Barone, T.~M., Croom, S.~M., {et~al.} 2022, MNRAS, 516, 2971

\bibitem[{Worthey {et~al.}(1994)Worthey, Faber, Gonzalez, \&
  Burstein}]{Worthey1994}
Worthey, G., Faber, S.~M., Gonzalez, J.~J., \& Burstein, D. 1994, ApJS, 94, 687

\bibitem[{Worthey \& Ottaviani(1997)}]{Worthey1997}
Worthey, G. \& Ottaviani, D.~L. 1997, ApJS, 111, 377

\bibitem[{York {et~al.}(2000)York, Adelman, Anderson, Anderson, Annis, Bahcall,
  Bakken, Barkhouser, Bastian, Berman, Boroski, Bracker, Briegel, Briggs,
  Brinkmann, Brunner, Burles, Carey, Carr, Castander, Chen, Colestock,
  Connolly, Crocker, Csabai, Czarapata, Davis, Doi, Dombeck, Eisenstein,
  Ellman, Elms, Evans, Fan, Federwitz, Fiscelli, Friedman, Frieman, Fukugita,
  Gillespie, Gunn, Gurbani, de~Haas, Haldeman, Harris, Hayes, Heckman,
  Hennessy, Hindsley, Holm, Holmgren, Huang, Hull, Husby, Ichikawa, Ichikawa,
  Ivezić, Kent, Kim, Kinney, Klaene, Kleinman, Kleinman, Knapp, Korienek,
  Kron, Kunszt, Lamb, Lee, Leger, Limmongkol, Lindenmeyer, Long, Loomis,
  Loveday, Lucinio, Lupton, MacKinnon, Mannery, Mantsch, Margon, McGehee,
  McKay, Meiksin, Merelli, Monet, Munn, Narayanan, Nash, Neilsen, Neswold,
  Newberg, Nichol, Nicinski, Nonino, Okada, Okamura, Ostriker, Owen, Pauls,
  Peoples, Peterson, Petravick, Pier, Pope, Pordes, Prosapio, Rechenmacher,
  Quinn, Richards, Richmond, Rivetta, Rockosi, Ruthmansdorfer, Sandford,
  Schlegel, Schneider, Sekiguchi, Sergey, Shimasaku, Siegmund, Smee, Smith,
  Snedden, Stone, Stoughton, Strauss, Stubbs, SubbaRao, Szalay, Szapudi,
  Szokoly, Thakar, Tremonti, Tucker, Uomoto, Vanden~Berk, Vogeley, Waddell,
  Wang, Watanabe, Weinberg, Yanny, Yasuda, \& Collaboration}]{York2000}
York, D.~G., Adelman, J., Anderson, John~E., J., {et~al.} 2000, AJ, 120, 1579

\bibitem[{Zibetti \& Gallazzi(2022)}]{Zibetti2022}
Zibetti, S. \& Gallazzi, A.~R. 2022, MNRAS, 512, 1415

\bibitem[{Zibetti {et~al.}(2017)Zibetti, Gallazzi, Ascasibar, Charlot, Galbany,
  García~Benito, Kehrig, de~Lorenzo-Cáceres, Lyubenova, Marino, Márquez,
  Sánchez, van~de Ven, Walcher, \& Wisotzki}]{Zibetti2017}
Zibetti, S., Gallazzi, A.~R., Ascasibar, Y., {et~al.} 2017, MNRAS, 468, 1902

\bibitem[{Zibetti {et~al.}(2020)Zibetti, Gallazzi, Hirschmann, Consolandi,
  Falcón-Barroso, van~de Ven, \& Lyubenova}]{Zibetti2020}
Zibetti, S., Gallazzi, A.~R., Hirschmann, M., {et~al.} 2020, MNRAS, 491, 3562

\bibitem[{Zibetti {et~al.}(2025)Zibetti, Pratesi, Gallazzi, Mattolini, \&
  Scholz-Díaz}]{Zibetti2025}
Zibetti, S., Pratesi, J., Gallazzi, A.~R., Mattolini, D., \& Scholz-Díaz, L.
  2025 [\eprint{Arxiv:2508.19462v1}]

\bibitem[{Zibetti {et~al.}(2024)Zibetti, Rossi, \& Gallazzi}]{Zibetti2024a}
Zibetti, S., Rossi, E., \& Gallazzi, A.~R. 2024, MNRAS, 528, 2790

\end{thebibliography}
\begin{appendix}
\section{Generating the CSP model library}\label{App:Model_library}
In this appendix we detail our standard procedure to generate the stellar population library that we use for the inference of the stellar population parameters (referred to in the text as San\_VarZ\_CB19\_MILES). This was introduced by \cite{Zibetti2017} and adopted in the subsequent works based on BaStA (\citealp{Zibetti2020, Zibetti2022}; this work; \citealp{Gallazzi2025}).

In order to generate our library of Composite Stellar Population (CSP) models, we first produce the continuous, secular component and then the stochastic burst component. Then we apply dust attenuation on the resulting spectra. As a final step, we equalize the distribution of models in an observable parameter space, in order to optimize the sampling of the posterior PDF.

\begin{table*}[!htbp]
    \centering
    \begin{tabular}{p{0.10\textwidth}|p{0.40\textwidth}|p{0.45\textwidth}}
         {\small Parameters} & {\small Description} & {\small Prior PDF}  \\
         \noalign{\vspace{3pt}}
         \hline \hline
         \noalign{\vspace{3pt}}
         \multicolumn{3}{c}{Secular SFH: $\mli{SFR}_\tau(t)\propto\frac{(t-t_{\mathrm{form}})}{\tau}\exp\left(-\frac{(t-t_{\mathrm{form}})^2}{\tau^2}\right)$}\\
         \noalign{\vspace{3pt}}
         \hline\hline
         \noalign{\vspace{3pt}}
          $t_\mathrm{form,lb}$ \\
          $\equiv t_0-t_\mathrm{form}$ & Time elapsed since the beginning of the SFH & Log-uniform between $5\cdot10^8$ and $2\cdot10^{10}$\,yr  \\
          \noalign{\vspace{3pt}}
          \hline
          \noalign{\vspace{3pt}}
          {$\tau$} & Time scale/peak & From log-uniform of $\tau/t_\mathrm{form,lb}$ between $1/20$ and $2$  \\
          \noalign{\vspace{3pt}}
          \hline\hline
          \noalign{\vspace{3pt}}
         \multicolumn{3}{c}{Secular Chemical Enrichment History:}\\
         \noalign{\vspace{3pt}}
         \multicolumn{3}{c}{$Z_*(t)=Z_*\left(M(t)\right)=Z_{*\,\text{final}}-\left(Z_{*\,\text{final}}-Z_{*\,\text{0}}\right)\left(1-\frac{M(t)}{M_\text{final}}\right)^\alpha$}\\
         \noalign{\vspace{3pt}}
         \hline \hline
         \noalign{\vspace{3pt}}
         $\left<Z_{*\,\rm{fm-w}}\right>$ & Formed-mass-weighted metallicity (non-generative, constraint for generative pars.)& $\tanh$ distribution of $P(\log \left<Z_{*\,\rm{fm-w}}\right>)$: $P(1/50\, \mathrm{Z_\odot})=0$, $P(0.1\, \mathrm{Z_\odot})=0.9$, $P(3.8\, \mathrm{Z_\odot})=1$, $\left<Z_{*\,\rm{fm-w}}\right> <3.8\, \mathrm{Z_\odot}$ \\
         \noalign{\vspace{3pt}}
         \hline
         \noalign{\vspace{3pt}}
         {$\alpha$} & Swiftness of chemical enrichment & $0.25 \leq \alpha \leq 19$,  from a uniform distribution in $\beta\equiv\frac{1}{1+\alpha},\, 0.05 \leq \beta \leq 0.80$  \\
         \noalign{\vspace{3pt}}
         \hline
         \noalign{\vspace{3pt}}
         $Z_{*\,\text{0}}$ & Initial metallicity & Log-uniform between $\log (Z_{\rm{min}}/Z_\odot) \equiv \log(1/50)$ and $\min(\log Z_{\rm{0\,max}}/Z_\odot \equiv \log(0.05), \left<Z_{*\,\rm{fm-w}}\right>)$ \\
         \noalign{\vspace{3pt}}
         \hline
         \noalign{\vspace{3pt}}
         $Z_{*\,\text{final}}$ & Final metallicity & Derived from  $Z_{\rm{min}}$, $\alpha$, and $\left<Z_{*\,\rm{fm-w}}\right>$, with the constraint $Z_{*\,\text{final}} \leq 4 Z_{*\,\odot}$ \\
         \noalign{\vspace{3pt}}
         \hline \hline
         \noalign{\vspace{3pt}}
         \multicolumn{3}{c}{Bursts}\\
         \noalign{\vspace{3pt}}
         \hline \hline
         \noalign{\vspace{3pt}}
         $N_\mathrm{burst}$ & Number of bursts & $1/3$ of models without bursts; for the remaining $2/3$: $N_\mathrm{burst}\leq 6$, $P(N_\mathrm{burst})\propto e^{-N_\mathrm{burst}}$\\
         \noalign{\vspace{3pt}}
         \hline
         \noalign{\vspace{3pt}}
         $t_{\mathrm{burst},i}$ & Age of the $i$th burst (instantaneous) & Log-uniform distribution between $10^5$\, yr and the $t_\mathrm{form}$ of the secular component\\
         \noalign{\vspace{3pt}}
         \hline
         \noalign{\vspace{3pt}}
         $Z_{*\,\mathrm{burst},i}$ & Metallicity of the $i$th burst & Gaussian distribution in log-$Z$, with mean equal to the metallicity value of the secular component at $t=t_{\mathrm{burst},i}$ and width $\sigma_{Z_{*}\,\mathrm{burst}} = 0.2$\,dex\\
         \noalign{\vspace{3pt}}
         \hline
         \noalign{\vspace{3pt}}
         $\mathcal{M}_{\mathrm{burst},i}$ & Mass fraction formed in the $i$th burst relative to the mass formed in secular mode & Log-uniform distribution between $10^{-3}$ and $2$. More restrictive upper limits applied for $t_{\mathrm{burst},i} < 10^8$\,yr \\
         \noalign{\vspace{3pt}}
         \hline \hline
         \noalign{\vspace{3pt}}
         \multicolumn{3}{c}{Dust attenuation \citep{Charlot2000}}\\
         \noalign{\vspace{3pt}}
         \hline
         \noalign{\vspace{3pt}}
         $\tau_V$ & Total optical depth due to diffuse ISM and birth cloud (affects stars younger than $10^7$\,yr) & $P(\tau_V)\propto1-\tanh(1.5\tau_V-6.7)$, constant at low values, exponential drop to $0$ between $\tau_V=4$ and $\tau_V=6$ \\
         \noalign{\vspace{3pt}}
         \hline
         \noalign{\vspace{3pt}}
         $\mu$ & Fraction of total optical depth in diffuse ISM & $P(\mu)\propto1-\tanh(8\mu-6)$, exponential drop to $0$ between $\mu=0.5$ and $\mu=1$\\
         \noalign{\vspace{3pt}}
         \hline \hline
    \end{tabular}
    \caption{Summary table of the parameters used to generate the CSP library.}
    \label{tab:SFHinpars}
\end{table*}

\subsection{Secular component of the SFH}
The star-formation history of the secular component is described by a delayed Gaussian function, following \cite{Sandage1986} and \cite{Gavazzi2002a}, according to the following equation:
\begin{equation}\label{eq:sandage_SFH}
    \mli{SFR}_\tau(t)\propto\frac{(t-t_{\mathrm{form}})}{\tau}\exp\left(-\frac{(t-t_{\mathrm{form}})^2}{2\tau^2}\right)
\end{equation}
$t_{\mathrm{form}}$ is the time corresponding to the beginning of the SFH and is randomly generated from a log-uniform variate of $t_\mathrm{form,lb}\equiv t_0 - t_\mathrm{form}$ between $5\cdot10^8$ and $2\cdot10^{10}$\, yr, with $t_0$ being the cosmic time of the observation. 
Note, however, that the Bayesian inference for SDSS galaxies is based only on models with $t_{\rm{form,lb}}\geq1.5\cdot10^{9}\,\rm yr$.
$\tau$\, represents the time scale of the SFH and corresponds to the time elapsed between $t_{\mathrm{form}}$ and the peak of the SFH, at which the SFR starts to decline as a quadratic exponential after the first phase of nearly-linear increase. In order to generate $\tau$, we generated a random distribution of $\tau/t_{\mathrm{form,lb}}$ drawn from a log-uniform variate between $1/20$ and $2$. In this way we cover from almost instantaneous bursts to linearly increasing SFHs. Note that defining $\tau$ in relation to $t_{\mathrm{form,lb}}$ rather than in absolute terms is motivated by the fact that stellar population properties are sensitive to the relative time duration of the SFH. The lower limit of $1/20$ is dictated by the limited time resolution achievable in SP analysis \citep{Zibetti2024a}.

The Chemical Enrichment History (CEH) of the secular component is described by the following equation:
\begin{equation}\label{eq:CEH}
Z_*(t)=Z_*\left(M(t)\right)=Z_{*\,\text{final}}-\left(Z_{*\,\text{final}}-Z_{*\,\text{0}}\right)\left(1-\frac{M(t)}{M_\text{final}}\right)^\alpha,  \alpha > 0
\end{equation}
In order to generate the parameters that govern the CEH ($Z_{*\,\text{0}}$, $Z_{*\,\text{final}}$ and $\alpha$), we take care that the resulting prior distribution in mean formed-mass-weighted metallicity $\log \left<Z_{*\,\rm{fm-w}}\right>$ is as smooth as possible over a range relevant to a broad variety of galaxies. Therefore, we initially generate this parameter sampling from a hyperbolic-tangent variate:
\begin{equation}
    \frac{dP}{d\log Z}=\kappa\,\tanh(\gamma(\log Z-\log Z_{\mathrm{min}}))
\end{equation}
We adopt as a minimum value for the metallicity $Z_{\mathrm{min}} = 1/50\, \mathrm{Z_\odot}$. The shape parameters $\kappa$ and $\gamma$ are determined in order to satisfy the following conditions:
\begin{align}
    \frac{dP}{d\log Z}(Z_\mathrm{max}\equiv 3.8\, \mathrm{Z_\odot}) =1 \\
    \frac{dP}{d\log Z}(Z_\mathrm{0.9}\equiv 0.1\,\mathrm{Z_\odot}) = 0.9
\end{align}

This ensures that the distribution of $\left<Z_{*\,\rm{fm-w}}\right>$ is almost uniform in log between $0.1\,\mathrm{Z_\odot}$ and $3.8\, \mathrm{Z_\odot}$, while gradually dropping to 0 when going down to the minimum value of $1/50\, \mathrm{Z_\odot}$.

For each model, once $\log \left<Z_{*\,\rm{fm-w}}\right>$ is assigned, we generate the parameters that describe the CEH according to Eq. \ref{eq:CEH}.
$\alpha$ is a shape parameter that describes how quickly the enrichment occurs, from 
delayed/instantaneous ($\alpha=0$) to early/rapid ($\alpha \gg 1$). It is randomly generated from a uniform distribution in
$\beta\equiv\frac{1}{1+\alpha}$ between $0.05$ and $0.80$, which corresponds to a minimum and maximum $\alpha = 0.25$ and $19$, respectively.

The initial metallicity $\log Z_{*\,\text{0}}$ is drawn from a uniform distribution between $\log (Z_{\rm{min}}/Z_\odot) \equiv \log(1/50)$ and $\log Z_{\rm{0\,max}}/Z_\odot \equiv \log(0.05)$. If the generated value of $\log Z_{*\,\text{0}}$ is larger than $\log \left<Z_{*\,\rm{fm-w}}\right>$, $\log \left<Z_{*\,\rm{fm-w}}\right>$ is taken instead. 

The final metallicity $Z_{*\,\text{final}}$ is obtained by enforcing that the $\log \left<Z_{*\,\rm{fm-w}}\right>$ as derived by integrating Eq. \ref{eq:CEH}, using the actual values of $\alpha$ and $\log Z_{*\,\text{0}}$. 
After simple calculation, one obtains $\log \left<Z_{*\,\rm{fm-w}}\right> \equiv \int_0^{M_{\rm final}} dM\, Z_*\left(M\right) = Z_{*\,\text{final}}-\frac{Z_{*\,\text{final}}-Z_{*\,\text{0}}}{\alpha+1}$, hence $Z_{*\,\text{final}} = \frac{(\alpha+1)\left<Z_{*,\rm{fm-w}}\right>-Z_{*\,\text{0}}}{\alpha}$. In case $Z_{*\,\text{final}}$ exceeds the maximum metallicity value of $Z_{*\,\text{final},{\text {max}}}\equiv 0.6 \approx 4 Z_{*\,\odot}$, such maximum value is assumed  as $Z_{*\,\text{final}}$ instead.

As a consequence of the small adjustments of $Z_{*\,\text{0}}$ and $Z_{*\,\text{final}}$ to fit the allowed parameter range, the final distribution of $\left<Z_{*\,\rm{fm-w}}\right>$ is slightly different from the original tanh distribution.

Note that the parameters of the CEH adopted in \cite{Zibetti2017, Zibetti2022} slightly differ from the present ones, due to the less extended metallicity range of the BC03 models based on Padova 1994 evolutionary tracks adopted in those works with respect to the PARSEC tracks used in the present one.

\subsection{Burst component of the SFH}
Each model in the library may have a number of bursts between 0 and 6 superimposed on the top of the secular component of the SFH.
We leave a fraction $f_{\mathrm{no\,burst}}=1/3$ of models without burst, with a pure continuous secular SFH. For the remaining $2/3$ of the models, the probability of having a number of bursts $N_\mathrm{burst}$ between 1 and 6 is proportional to $\exp(-N_\mathrm{burst})$.
Each burst is modeled as a single-age, single-metallicity stellar population and it is thus characterized by three parameters:
the look-back time at which the burst occurs $t_{\mathrm{burst\, lb},i}$, which is randomly drawn from a log-uniform distribution between $10^5$\, yr and the $t_\mathrm{form\,lb}$ of the secular component; the metallicity of the burst $Z_{*\,\mathrm{burst},i}$, which is randomly generated from a Gaussian distribution in log-$Z$, with mean at the metallicity value of the secular component at $t=t_{\mathrm{burst},i}$ and a width $\sigma_{Z_{*}\,\mathrm{burst}} = 0.2$\,dex; the fraction of mass formed in the burst relative to the integrated mass formed in the secular component, $\mathcal{M}_{\mathrm{burst},i}$ randomly generated from a log-uniform distribution between $10^{-3}$ and $2$. In this way we cover from very minor burst resulting a tiny ``contamination'' of the secular component, up to strong burst dominating the SFH. However, In order to avoid unrealistic recent bursts that completely outshine the rest of the stars (which are indeed already represented by low $t_\mathrm{form lb}$ and short $\tau$), resulting in most of the stellar mass to be formed in less than $100~\mathrm{Myr}$, for $t_\mathrm{burst\,lb}<10^8$\,yr we have introduced an upper limit to $\mathcal{M}_{\mathrm{burst}}$ defined by the parabolic function going through the points $(\log(t_\mathrm{burst\, lb}),\log(\mathcal{M}_{\mathrm{burst}}))= (5,-2.5),(7,-1),(8,\log(2))$.

\subsection{Dust attenuation treatment}
For the treatment of dust attenuation we adopt the two-component \cite{Charlot2000} formalism, whereby all stars older than the birth-cloud (BC) dissipation timescale $t_\mathrm{BC}\equiv 10^7$\,yr are attenuated by the diffuse ISM only, while stars younger than $t_\mathrm{BC}$ are additionally subject to the birth-cloud attenuation. In the synthesis of the stellar population spectra we first compute the spectra of the stars younger and older than $t_\mathrm{BC}$, and apply the attenuation curves separately, before summing the two attenuated spectra. Following \cite{Charlot2000} we adopt an effective attenuation curve with optical depth proportional to $\lambda^{-0.7}$ and $\lambda^{-1.3}$, for the diffuse ISM and the BC, respectively. In the reference $V$ band, a total optical depth $\tau_V$  applies to the young stars, given by the sum of the BC component $\tau_V^{\mathrm{BC}}\equiv(1-\mu)\cdot\tau_V$ and of the diffuse ISM component $\tau_V^{\mathrm{ISM}}\equiv\mu\cdot\tau_V$, while $\tau_V^{\mathrm{ISM}}$ applies to the old stars.
The CSP model library is composed by $\sim25\%$ dust free models. 
For the remaining $75\%$ models each $\tau_V$ and $\mu$ are randomly generated using the same prior probability distributions of \cite{daCunha2008}, as follows: $P(\tau_V)\propto1-\tanh(1.5\tau_V-6.7)$, which is constant at low values and drops exponentially to $0$ between $\tau_V=4$ and $\tau_V=6$; $P(\mu)\propto1-\tanh(8\mu-6)$, which drops exponentially to $0$ between $\mu=0.5$ and $\mu=1$.

\subsection{Library equalization}
The library as described so far produces a prior distribution that is smooth and nearly flat for most of the $\log \mathrm{Age}-\log \mathrm{Z_*}$ plane. Looking at the distribution of the models in the observable space, chiefly in the classical Balmer plane defined by $\mathrm{H}\delta_{\rm A}+\mathrm{H}\gamma_{\rm A}$ vs $\mathrm{D4000_n}$, we notice several overdense features and, vice versa, many regions that are severely under-populated. These regions correspond to young and/or bursty SFH, where indices span large dynamical ranges because of their sensitive response to small changes in SFH and metallicity. In order to sample the posterior PDF with a sufficient number of models and avoid biases towards overdense regions in the observable parameter space, it is indeed desirable to have the distribution of models in such space as smooth and uniform as possible, while retaining a distribution in the space of the physical parameters to be retrieved which is as smooth as possible on the scales of the expected uncertainties. Thus we implemented an equalization strategy as follows. We first generate a library with 10 times more models than actually used in the inferences. Then we resample the library in order to obtain a uniform distribution over the $\mathrm{H}\delta_{\rm A}+\mathrm{H}\gamma_{\rm A}$ vs $\mathrm{D4000_n}$ plane, retaining only the desired number of models for the inference. Extensive testing of the BaStA algorithm shows that such an equalization procedure does not produce significant biases in the retrieved posterior PDF for the main parameters of interest, such as mean age and metallicity \citep[see also][]{Rossi2025}. 

\section{Mass metallicity relation for young and old galaxies}\label{App:MZR_young_old}
\begin{figure}
  \centering
  \includegraphics[width=\hsize]{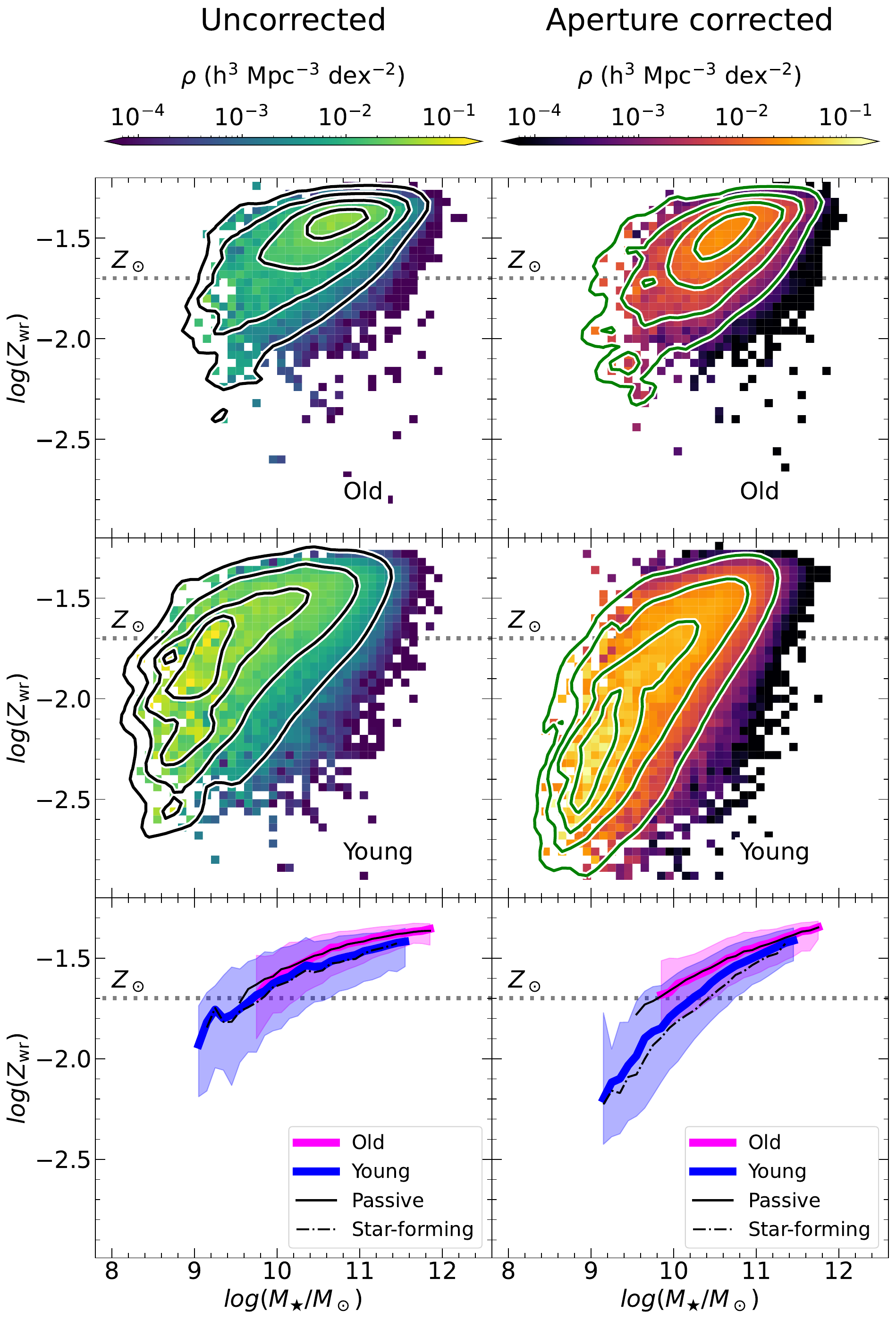}
    \caption{Number density of young (\emph{top row}) and old (\emph{middle row}) galaxies in the mass-metallicity plane, obtained using both the volume and completeness statistical corrections. 
  Following the same colour-code of Fig. \ref{fig:Scaling_relations_comparison}, the \emph{left panels} show the distributions obtained without the use of the aperture corrections, and the \emph{right panels} show the distributions obtained implementing aperture corrections.
  The contours identify the levels containing $16\%$, $50\%$, $84\%$, and $97.5\%$ of the total integral of the distributions.
  The bottom panels display the median relations (solid lines) and the respective $16^\mathrm{th}$ and $84^\mathrm{th}$ percentiles (shaded area) as a function of mass, for old (red) and young (blue) galaxies.
  The solid and dot-dashed black lines identify the median MZRs obtained with the passive and star-forming sample, respectively.
  The dotted horizontal lines in the mass-metallicity planes identify the solar metallicity $Z_\odot\equiv0.02$.}
      \label{fig:Mass_metallicity_young_old}
\end{figure}
In this appendix we analyse the MZRs obtained dividing our galaxy sample between old and young galaxies using the divisions defined in sec. \ref{subsec:Transition_mass}.
Top and central panels of figure \ref{fig:Mass_metallicity_young_old} show the MZRs obtained from the old and young subsamples, respectively.
Following the same structure and colour-coding of Fig. \ref{fig:Mass_metallicity_passive_starforming}, the left panels display the uncorrected distributions and the right panels display the aperture-corrected ones.
In both figures the black and green contours identify the isodensity levels.
The bottom panels display the median (solid line) and $16^\mathrm{th}-84^\mathrm{th}$ percentile ranges (shaded area) of the relations of young (blue) and old (red) galaxies.
The solid and dot-dashed black lines identify the median relations for passive and star-forming galaxies, respectively.

From Fig. \ref{fig:Mass_metallicity_young_old} we see that young and old galaxies have MZRs similar to those of star-forming and passive galaxies, respectively.
This is expected given the overall agreement between light-weighted ages and SFR estimates analysed in sec. \ref{subsec:Passive_vs_starforming}.
Nevertheless, looking at the aperture-corrected relations, we also find some differences, in that
the MZR of young galaxies is located at higher values than the one of star-forming galaxies.
This is expected based on what we showed in sec. \ref{subsec:Passive_vs_starforming} concerning the shape and normalisation of the MZR being mainly set by the star formation activity.
As the young population contains a fraction of passive galaxies, which are located at higher metallicities (see Fig. \ref{fig:Mass_age_passive_starforming} and \ref{fig:Mass_metallicity_passive_starforming} for more details), this results in a shift of the median and percentiles of the relation of young galaxies towards higher metallicities.

\end{appendix}
\end{document}